\documentclass[aip,jap,amsmath,amssymb,reprint,floatfix,citeautoscript]{revtex4-2}
\usepackage{times,amsfonts,bm,braket,siunitx,graphicx}
\usepackage{hyperref}
\usepackage[utf8]{inputenc}
\usepackage[T1]{fontenc}
\usepackage[capitalize]{cleveref}
\crefname{subsection}{Section}{Sections}
\hypersetup{pagebackref=true,colorlinks=true,urlcolor=blue,linkcolor=blue,citecolor=blue}
\DeclareMathOperator{\sgn}{sgn}
\DeclareMathOperator{\tr}{tr}
\DeclareMathOperator{\pf}{pf}
\DeclareMathOperator{\diag}{diag}
\renewcommand\Im{\operatorname{Im}}
\renewcommand\Re{\operatorname{Re}}
\newcommand{\ii}{\mathrm{i}}
\newcommand{\ee}{\mathrm{e}}
\newcommand{\dd}{\mathrm{d}}
\newcommand{\id}{\openone}
\newcommand{\up}{{\uparrow}}
\newcommand{\down}{{\downarrow}}
\newcommand{\updown}{{\uparrow\!\downarrow}}
\newcommand{\nostar}{{\phantom{*}}}

\begin{document}

\title{Majorana nanowires for topological quantum computation}
\author{Pasquale Marra}
\email{pmarra@ms.u-tokyo.ac.jp}
\affiliation{Graduate School of Mathematical Sciences, The University of Tokyo, 3-8-1 Komaba, Meguro, Tokyo, 153-8914, Japan}
\affiliation{Department of Physics, and Research and Education Center for Natural Sciences, Keio University, 4-1-1 Hiyoshi, Yokohama, Kanagawa, 223-8521, Japan}
\date{\today}

\begin{abstract}
Majorana bound states are quasiparticle excitations localized at the boundaries of a topologically nontrivial superconductor.
They are zero-energy, charge-neutral, particle-hole symmetric, and spatially-separated end modes which are topologically protected by the particle-hole symmetry of the superconducting state.
Due to their topological nature, they are robust against local perturbations and, in an ideal environment, free from decoherence.
Furthermore, unlike ordinary fermions and bosons, the adiabatic exchange of Majorana modes is noncommutative, i.e., the outcome of exchanging two or more Majorana modes depends on the order in which exchanges are performed.
These properties make them ideal candidates for the realization of topological quantum computers.
In this tutorial, I will present a pedagogical review of 1D topological superconductors and Majorana modes in quantum nanowires.
I will give an overview of the Kitaev model and the more realistic Oreg-Lutchyn model, discuss the experimental signatures of Majorana modes, and highlight their relevance in the field of topological quantum computation.
This tutorial may serve as a pedagogical and relatively self-contained introduction for graduate students and researchers new to the field, as well as an overview of the current state-of-the-art of the field and a reference guide to specialists.
\end{abstract}

\maketitle

\section{Introduction}

A Majorana fermion~\cite{majorana_teoria_1937} is a charge-neutral particle satisfying the Dirac equation and which, contrarily to ordinary Dirac fermions~\cite{dirac_the-quantum_1928,dirac_a-theory_1930}, coincides with its own antiparticle.
In high-energy physics, all fermionic particles in the standard model are Dirac fermions with perhaps one exception, neutrinos, which might be instead Majorana fermions~\cite{balantekin_properties_2018}.
In condensed matter, Dirac and Majorana fermions emerge as low-energy quasiparticle excitations~\cite{wehling_dirac_2014,shen_topological_2017} realized in, e.g., superfluid helium-3~\cite{leggett_a-theoretical_1975,volovik_the-universe_2009}, graphene~\cite{castro-neto_the-electronic_2009}, topological insulators and superconductors~\cite{hasan_colloquium:_2010,qi_topological_2011,bernevig_topological_2013,ando_topological_2013,sato_majorana_2016,asboth_a-short_2016,sato_topological_2017,shen_topological_2017}.
In particular, pointlike (0D) Majorana modes are spatially-separated, zero-energy quasiparticle excitations localized at the vortices of a two-dimensional (2D) topological superconductor or superfluid~\cite{volovik_fermion_1999,read_paired_2000,das-sarma_proposal_2006,tewari_quantum_2007,zhang_pxipy_2008,fujimoto_topological_2008,zhang_pxipy_2008,fu_superconducting_2008} or at the boundaries of the nontrivial phase of a one-dimensional (1D) topological superconductor~\cite{kitaev_unpaired_2001,lutchyn_majorana_2010,oreg_helical_2010}.
Majorana fermions in condensed matter physics and Majorana fermions in high-energy physics are connected by the fact that the effective theory describing collective excitations in superconductors is analogous to the Dirac-Majorana equation describing a Majorana fermion in 
quantum field theory, with the vacuum replaced by the superconducting condensate and particle-antiparticle (charge) symmetry replaced by particle-hole symmetry~\cite{chamon_quantizing_2010}.

The existence of spatially-separated Majorana modes in topological superconductors can be explained in terms of topology.
Topology studies the properties of geometrical objects which are invariant under continuous transformations.
Under the lens of topology, for instance, 
spheres and cylinders are equivalent since they can be transformed one into the other via a continuous transformation.
Conversely, spheres and toruses are distinct, inequivalent objects since they cannot be transformed one into the other via a continuous transformation.
This approach can be applied to condensed matter physics:
A topological superconductor is a phase of matter which is \emph{nontrivial}, i.e., topologically distinct from an ordinary superconductor and from the vacuum.
A very general statement underlying the physics of topological phases of matter is that a topologically nontrivial phase must exhibit localized modes at its boundaries~\cite{hatsugai_chern_1993,ryu_topological_2002,teo_topological_2010}.
Indeed, a topologically nontrivial phase cannot be connected by a smooth transformation to a trivial phase (e.g., the vacuum) unless the energy gap closes at their shared boundary.
In other words, the gap must vanish at the boundary between trivial and nontrivial phases, allowing the topological invariant to change.
This mandates the existence of localized modes at the boundary or domain wall between topologically inequivalent phases.
These modes are exponentially localized and spatially separated, i.e., their mutual overlap vanishes in the limit of infinite system size, being exponentially small in their mutual distance.

Intuitively, a Majorana mode corresponds to ``half'' a fermion, in the sense that any fermionic operator $c$ can be formally recast in terms of two Majorana operators as
\begin{equation}\label{eq:sillymajo}
	c=\frac12(\gamma_A+\ii\gamma_B),
\end{equation}
where the operators $\gamma_A$ and $\gamma_B$ are self-adjoint, idempotent, and obey the fermionic anticommutation rules, i.e.,
\begin{equation}\label{eq:Majorana-Properties-Intro}
	\gamma_{\alpha}^\dag=\gamma_{\alpha},
	\quad
	\gamma_{\alpha}^2=\gamma^\dag_{\alpha}\gamma_{\alpha}=1,
	\quad
	\{\gamma_\alpha,\gamma_\beta\}=2\delta_{\alpha\beta}.
\end{equation}
The Majorana operators 
$\gamma_A=c^\dag+c$ and $\gamma_B=\ii(c^\dag-c)$ 
correspond respectively to the ``real part'' and ``imaginary part'' or, more precisely, to the hermitian and antihermitian parts of the fermion operator $c$.
Like fermions, Majorana operators anticommute but, unlike ordinary fermions, Majorana modes are their own antiparticle.
Like bosons, creating two identical Majorana modes is not forbidden by the Pauli principle but, unlike bosons, this operation has no effect since Majorana operators are idempotent.

The seemingly harmless transformation in \cref{eq:sillymajo} has profound consequences in the case where the two Majorana modes are spatially separated.
Spatial separation requires their wavefunctions to decay exponentially as $\ee^{-x/\xi_\mathrm{M}}$ and their mutual distance to be much larger than their localization length $L\gg\xi_\mathrm{M}$, i.e., the limit where their distance becomes effectively infinite $L\to\infty$.
Indeed, any spatially-separated mode in a superconductor exhibits very distinctive properties.
Consider, e.g., a semi-infinite wire with one mode localized at its end.
Due to particle-hole symmetry, every mode with finite positive energy (particle-like) must correspond to a mode with opposite energy (hole-like).
However, particle-like and hole-like modes must coincide since there is only one mode at the boundary.
For this reason, spatially-separated modes must have zero energy, be self-adjoint, and invariant under the particle-hole symmetry enforced by the superconducting pairing, i.e., must coincide with their own antiparticles.
This also means that local perturbations cannot lift the energy of a lonely spatially-separated mode from zero as long as the particle-hole gap remains open.
Indeed, spatially-separated modes can be removed from the boundary only by going through a quantum phase transition which closes the particle-hole gap.
Moreover, being an equal superposition of particles and holes, they are necessarily charge-neutral and insensitive to electric and magnetic fields.
Furthermore, time-reversal symmetry must be broken to lift the Kramers degeneracy between spin up and down electrons and allow the existence of only one zero-energy mode per boundary.
Hence, this mode must be spinless, otherwise its energy would be lifted by the broken time-reversal symmetry (e.g., by a finite magnetic field).
Spatially-separated zero-energy 0D Majorana modes in 1D systems are usually referred to as Majorana zero modes, Majorana end modes, or Majorana edge modes.
A nonlocal fermionic mode $c$ with zero energy which can be decomposed into two spatially-separated Majorana zero modes as in \cref{eq:sillymajo} and with
\begin{equation}
	[H,\gamma_{\alpha}]=0,
\end{equation}
is usually referred to as a Majorana bound state.
The terms Majorana bound states, Majorana zero modes, Majorana end modes, or Majorana edge modes are usually considered near-synonyms in the literature, with the term ``mode'' emphasizing the separation of two Majorana modes as distinct entities, and the term ``state'' emphasizing the collective existence of two Majorana modes as a nonlocal fermionic state.
The splitting of a single fermionic mode into two spatially-separated Majorana modes is an example of ``fractionalization''~\cite{semenoff_stretching_2006,semenoff_stretched_2007}, i.e., the separation of an elementary excitation into distinct quasiparticles.

Notice that perfectly spatially-separated zero-energy modes can be achieved only in infinite-size systems.
Unluckily, in the real world, everything is finite: 
In this case, Majorana modes localized at a finite distance $L$ become hybridized due to the finite overlap between their wavefunctions.
This hybridization results in a fermionic mode with a small but finite energy $E_\mathrm{M}>0$.
In particular, since Majorana modes are exponentially localized, this energy is exponentially small, i.e., it decreases exponentially with the mutual distance $L$, being
\begin{equation}
	[H,c^\dag]=E_\mathrm{M}c^\dag,\qquad E_\mathrm{M}\propto
	\ee^{-L/\xi_\mathrm{M}},
	\end{equation}
where $\xi_\mathrm{M}$ is the Majorana localization length, which depends on the system parameters and diverges at the quantum phase transition between the topologically trivial and nontrivial phases.
This exponential scaling is a distinctive fingerprint of topological states of matter, analogous to the quantization of the quantum Hall conductance up to exponentially small corrections~\cite{klitzing_new-method_1980,laughlin_quantized_1981,thouless_quantized_1982,niu_quantum_1987,thouless_topological_1998_s73}.

There is a fundamental reason why realizing spatially-separated Majorana modes is a big deal: 
these Majorana modes exhibit nonabelian exchange statistics, which make them ideal building blocks for the realization of topological quantum computers~\cite{read_paired_2000,ivanov_non-abelian_2001,kitaev_fault-tolerant_2003,stern_geometric_2004,nayak_non-abelian_2008,alicea_non-abelian_2011}.
A quantum many-body state has a well-defined symmetry under the exchange of identical fermions or bosons. 
Furthermore, the exchange is an abelian (i.e., commutative) operation, i.e., when applying several exchanges one after another, the final outcome does not depend on the order in which they are performed.
Majorana modes are different.
The exchange (braiding) of two or more Majorana modes is nonabelian (i.e., noncommutative)~\cite{moore_nonabelions_1991,nayak_2nquasihole_1996,read_paired_2000,ivanov_non-abelian_2001,stern_geometric_2004,kitaev_anyons_2006}, i.e.,
the final outcome does depend on the order in which exchanges are performed.
Mathematically, these exchange operations are described by unitary matrices acting on a degenerate many-body groundstate:
Applying two or more exchange operations consecutively corresponds to multiplying the groundstate by two or more unitary matrices consecutively.
Changing the order of the exchange operations changes the order of the matrix multiplications and thus may change the final outcome, since matrix multiplication is noncommutative. 
All this requires the existence of a degenerate many-body groundstate:
Indeed, the groundstate of a system of $2N$ Majorana modes is $2^N$-fold degenerate, with exchange operations corresponding to adiabatic unitary evolutions within the groundstate manifold.
In a topological quantum computer, the quantum information is encoded nonlocally in the spatially-separated Majorana modes, and the quantum information is manipulated via quantum gates realized via exchange operations.
A fundamental advantage is that these exchange operations are independent of the fine details of the adiabatic evolution and only depend on the order in which the exchanges are performed.
Moreover, being charge-neutral, nonlocal, and topologically protected, Majorana modes are robust, i.e., their existence remains unscathed as long as the energy scale of perturbations does not exceed the particle-hole energy gap.
Indeed, small local perturbations (having a length scale smaller than the distance between the Majorana modes) cannot remove a Majorana mode from one end of a 1D system without concomitantly removing the Majorana mode at the opposite end.
Consequently, the only way to destroy Majorana modes is by a perturbation strong enough to close the particle-hole gap, move them close to each other, make their wavefunctions overlap, and split their energy.
In ideal conditions, Majorana modes are also free from decoherence since their decay is protected by fermion-parity conservation and by the particle-hole gap~\cite{kitaev_unpaired_2001,kitaev_fault-tolerant_2003}.
These properties may give them an edge over other possible implementations of quantum computing~\cite{kitaev_unpaired_2001,kitaev_fault-tolerant_2003,nayak_non-abelian_2008}.
On the other hand, these same properties are somewhat a hindrance to the experimental detection and manipulation of Majorana modes and to the readout of the quantum information they encode.

Theoretically, a topological superconductor supporting spatially-separated Majorana end modes can be realized in a 1D lattice of spinless fermions with $p$-wave superconducting pairing~\cite{kitaev_unpaired_2001}.
Despite its simplicity, this model is rather unrealistic since spinless fermions do not exist, superconductivity is not stable against quantum fluctuations in 1D~\cite{giamarchi_quantum_2003}, and spin-triplet $p$-wave superconductors are rare in nature~\cite{kallin_chiral_2016,sato_majorana_2016,sato_topological_2017}.
A more realistic proposal is to consider a heterostructure between a topological insulator and a conventional superconductor~\cite{fu_superconducting_2008,fu_josephson_2009,fu_probing_2009,akhmerov_electrically_2009,linder_unconventional_2010,hosur_majorana_2011,mi_proposal_2013}.
In this system, the 1D helical edge modes of the topological insulator~\cite{bernevig_quantum_2006a,bernevig_quantum_2006b,fu_topological_2007} are gapped by the proximitized superconducting pairing, becoming effectively equivalent to a spinless superconductor.
However, it was soon understood that a much simpler recipe to engineer nontrivial superconducting phases includes only three simple ingredients~\cite{sato_non-abelian_2009,sato_non-abelian_2010,lutchyn_majorana_2010,oreg_helical_2010,sau_generic_2010,alicea_majorana_2010,sau_non-abelian_2010,sau_robustness_2010,akhmerov_quantized_2011,stanescu_majorana_2011}, i.e., conventional $s$-wave superconductivity (induced by proximity), spin-orbit coupling, and broken time-reversal symmetry (e.g., magnetic field).
Indeed, since Majorana modes are equal superpositions of particle and hole excitations, one needs a superconducting state to enforce the particle-hole symmetry.
In addition, one needs to break time-reversal symmetry to remove Kramers degeneracy between opposite spin channels, otherwise spin-degenerate modes would simply hybridize in the presence of local perturbations and lift their energy.
Finally, spin-orbit coupling is needed to tilt the spin of electrons with the same energy and opposite momenta in different directions:
This tilting allows them to form a Cooper pair via proximization with a conventional spin-singlet superconductor and, therefore, \emph{effectively} realize a spin-triplet $p$-wave superconducting phase.

Perhaps the simplest proposal to realize this physics is that of a semiconducting nanowire with strong spin-orbit coupling and proximitized by a conventional superconductor in an external magnetic field~\cite{lutchyn_majorana_2010,oreg_helical_2010,sau_non-abelian_2010,akhmerov_quantized_2011,stanescu_majorana_2011}.
These systems, usually dubbed Majorana nanowires, are so far the most actively studied platform to realize Majorana bound states, both theoretically and experimentally.
Other nanowire-based setups are possible, e.g., semiconducting nanowires in the presence of magnetic textures~\cite{mohanta_electrical_2019}, nuclear magnetic order~\cite{braunecker_interplay_2013,klinovaja_topological_2013}, periodic arrays of nanomagnets~\cite{kjaergaard_majorana_2012,klinovaja_transition_2012}, or helical-magnetic superconducting compounds~\cite{martin_majorana_2012}, semiconducting nanowires fully coated with a superconducting shell (full-shell nanowires)~\cite{vaitiekenas_flux-induced_2020,penaranda_even-odd_2020}, 
or covered by both superconducting and ferromagnetic layers~\cite{maiani_topological_2021,khindanov_topological_2021,langbehn_topological_2021,escribano_tunable_2021,liu_electronic_2021,woods_electrostatic_2021}, metallic nanowires~\cite{potter_multichannel_2010,potter_topological_2012}, topological-insulator nanowires~\cite{cook_majorana_2011,potter_engineering_2011,cook_stability_2012,legg_majorana_2021,breunig_opportunities_2022}, or second-order topological-insulator nanowires at zero fields~\cite{hsu_majorana_2018}.
Furthermore, many other 1D platforms have been proposed, including 1D stripes in 2D semiconductor-superconductor planar heterostructures (e.g., planar Josephson junctions)~\cite{shabani_two-dimensional_2016,hell_two-dimensional_2017,hell_coupling_2017,pientka_topological_2017,haim_benefits_2019,setiawan_topological_2019,melo_supercurrent-induced_2019,stern_fractional_2019,zhou_phase_2020,laeven_enhanced_2020,mohanta_skyrmion_2021,paudel_enhanced_2021} or in topological insulator-superconductor heterostructures~\cite{hegde_a-topological_2020}, thin films of proximitized 3D topological insulators covered by a strip of magnetic insulators~\cite{papaj_creating_2021}, arrays of superconducting quantum dots~\cite{leijnse_parity_2012,sau_realizing_2012,fulga_adaptive_2013}, arrays of magnetic atoms deposited on a conventional superconductor substrate~\cite{choy_majorana_2011,nadj-perge_proposal_2013,pientka_topological_2013,vazifeh_self-organized_2013,braunecker_interplay_2013,klinovaja_topological_2013,pientka_unconventional_2014,kim_helical_2014,heimes_majorana_2014,li_topological_2014,peng_strong_2015,hui_majorana_2015,brydon_topological_2015,li_manipulating_2016,schecter_self-organized_2016,marra_controlling_2017}, carbon-based setups such as proximitized carbon nanotubes~\cite{egger_emerging_2012,klinovaja_electric-field-induced_2012,desjardins_synthetic_2019}, graphene nanoribbons~\cite{klinovaja_giant_2013,san-jose_majorana_2015}, 1D channels obtained via electrostatic confinement in bilayer graphene~\cite{klinovaja_helical_2012}, or in armchair edge states of encapsulated bilayer graphene~\cite{penaranda_majorana_2022}, and optically-trapped 1D lattices of ultracold fermionic atoms in a Zeeman field with intrinsic attractive interactions or coupled to a surrounding molecular BEC cloud~\cite{jiang_majorana_2011,diehl_topology_2011,nascimbene_realizing_2013,ruhman_topological_2015,ptok_quantum_2018}.
Moreover, dispersive 1D or 2D Majorana modes may be realized as edge or surface modes of 2D or 3D topological superconductors~\cite{sato_topological_2009,tanaka_manipulation_2009,qi_time-reversal-invariant_2009,qi_chiral_2010,sato_topological_2010,linder_unconventional_2010,nakosai_topological_2012,he_platform_2019,he_optical_2021} (see Refs.~\onlinecite{sato_majorana_2016,sato_topological_2017,haim_time-reversal-invariant_2019} for a review) or, alternatively, as the result of the hybridization of arrays of 0D Majorana modes~\cite{neupert_chain_2010,rex_majorana_2020,marra_1d-majorana_2022} or in the presence of additional synthetic dimensions~\cite{marra_topologically_2019}.

In principle, Majorana modes should exhibit distinctive experimental signatures in the tunneling conductance~\cite{sengupta_midgap_2001,law_majorana_2009,flensberg_tunneling_2010,stanescu_majorana_2011,prada_transport_2012,prada_measuring_2017}, ballistic point contact conductance~\cite{wimmer_quantum_2011}, Josephson current~\cite{kwon_fractional_2004,fu_superconducting_2008,fu_josephson_2009,lutchyn_majorana_2010,oreg_helical_2010,ioselevich_anomalous_2011,van-heck_coulomb_2011,law_robustness_2011,jiang_unconventional_2011,dominguez_dynamical_2012,pikulin_phenomenology_2012,san-jose_ac-josephson_2012,badiane_ac-josephson_2013,houzet_dynamics_2013,virtanen_microwave_2013,vayrynen_microwave_2015,peng_signatures_2016,sau_detecting_2017,dominguez_josephson_2017}, and Coulomb blockade spectroscopy~\cite{fu_electron_2010,zazunov_coulomb_2011,hutzen_majorana_2012,van-heck_conductance_2016,chiu_conductance_2017,cao_decays_2019}.
Besides, due to their topological nature, these signatures should persist in a large parameter range.
Indeed, there are already many experimental observations compatible with the presence of spatially-separated and topologically-protected Majorana modes in Majorana nanowires, such as zero-bias conductance peaks~\cite{mourik_signatures_2012,das_zero-bias_2012,deng_anomalous_2012,finck_anomalous_2013,churchill_superconductor-nanowire_2013,lee_spin-resolved_2014,deng_majorana_2016,chen_experimental_2017,deng_nonlocality_2018,gul_ballistic_2018,bommer_spin-orbit_2019,grivnin_concomitant_2019,menard_conductance-matrix_2020,puglia_closing_2021,heedt_shadow-wall_2021,zhang_large_2021,zhang_suppressing_2022}, fractional Josephson current~\cite{rokhinson_the-fractional_2012,laroche_observation_2019}, and Coulomb blockade experiments~\cite{deng_parity_2014,higginbotham_parity_2015,deng_majorana_2016,albrecht_exponential_2016,albrecht_transport_2017,shen_parity_2018,van-zanten_photon-assisted_2020}.
Nevertheless, there is still no consensus on the interpretation of these signatures, which may indeed have alternative explanations~\cite{kells_near-zero-energy_2012,lee_zero-bias_2012,liu_zero-bias_2012,pikulin_a-zero-voltage_2012,bagrets_class_2012}.
Furthermore, some concerns were raised about some of these experiments (see Refs.~\onlinecite{frolov_quantum_2021,castelvecchi_evidence_2021}).

There are several popular accounts~\cite{wilczek_majorana_2009,franz_majoranas_2013,brouwer_enter_2012,wilczek_majorana_2012,alicea_majorana_2013,beenakker_a-road_2016,aguado_majorana_2020} and reviews on Majorana modes and topological superconductors~\cite{alicea_new-directions_2012,tanaka_symmetry_2012,beenakker_search_2013,elliott_colloquium:_2015,beenakker_random-matrix_2015,sato_majorana_2016,sato_topological_2017,aguado_majorana_2017,haim_time-reversal-invariant_2019,sau_from_2021,lesser_majorana_2022}, mainly focusing on Majorana modes in 1D systems~\cite{alicea_new-directions_2012,leijnse_introduction_2012,stanescu_majorana_2013,oppen_topological_2017,aguado_majorana_2017,pawlak_majorana_2019,prada_from_2020,flensberg_engineered_2021,laubscher_majorana_2021,sau_from_2021,lesser_majorana_2022}, their experimental realization~\cite{lutchyn_majorana_2018,zhang_next_2019,li_exploring_2019,frolov_topological_2020,flensberg_engineered_2021,jack_detecting_2021,fu_experimental_2021,cao_recent_2022}, their applications in the field of topological quantum computation~\cite{nayak_non-abelian_2008,stern_non-abelian_2010,pachos_introduction_2012,stern_topological_2013,beenakker_search_2013,das-sarma_majorana_2015,roy_topological_2017,lahtinen_a-short_2017,stanescu_introduction_2017,beenakker_search_2020,oreg_majorana_2020,aguado_a-perspective_2020,sau_from_2021}, and their connection to other topological states of matter~\cite{hasan_colloquium:_2010,qi_topological_2011,bernevig_topological_2013,shen_topological_2017}.
In this tutorial, I will present a pedagogical review on Majorana modes and topological superconductivity in Majorana nanowires and their relevance for topological quantum computation.
First, I will show how Majorana modes emerge as topologically-protected end modes of a spinless superconductor (\cref{sec:pwave}) and give an overview of their nonabelian braiding statistics and their relevance in the field of topological quantum computation (\cref{sec:TQC}).
I will then introduce the Oreg-Lutchyn model~\cite{lutchyn_majorana_2010,oreg_helical_2010}, which offers a more realistic description of Majorana nanowires (\cref{sec:swave}), review the theoretical predictions regarding the experimental signatures of topological superconductivity (\cref{sec:experiments}), and shortly give an outlook on the recent development of the field (\cref{sec:outlook}).

\section{Majorana zero modes\label{sec:pwave}}

\subsection{Spinless topological superconductors\label{sec:pwavecontinuous}}

\begin{figure*}[t]
\includegraphics[width=\textwidth]{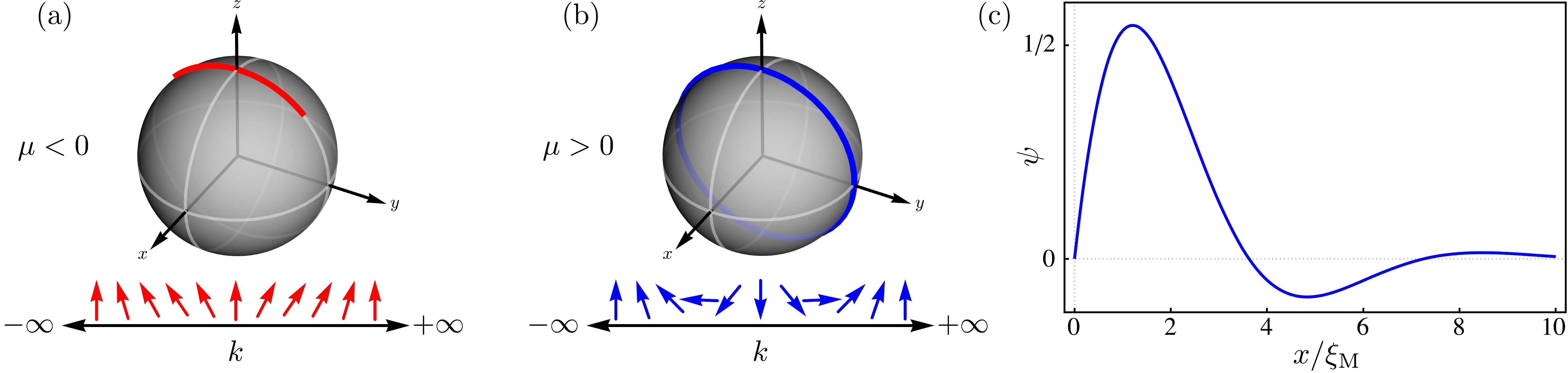}
\caption{
The path followed by the unit vector $\mathbf{u}(k)=\mathbf{h}(k)/|\mathbf{h}(k)|$ on the $yz$ plane of the unit sphere when $k$ runs from $-\infty$ to $+\infty$, and the Majorana zero mode in the topologically nontrivial phase of a spinless topological superconductor. 
(a) For $\mu<0$, the unit vector follows a circular arc that does not encircle the origin, having winding number $W=0$ corresponding to a topologically trivial phase.
(b) For $\mu>0$, the unit vector describes a full circle on the $yz$ plane, having winding number $W=\pm1$ (with the sign 
equal to 
the sign of the superconducting pairing $\Delta$) corresponding to a topologically nontrivial phase.
(c) In the nontrivial phase, a Majorana zero mode localizes at the boundary, decaying exponentially from the origin at $x=0$.
}
\label{fig:JR}
\end{figure*}

The simplest model, although somewhat unrealistic, of a 1D topological superconductor that exhibits spatially-separated Majorana modes is given by a $p$-wave spinless superconductor described by the Hamiltonian
\begin{align}
	\mathcal{H}=\int\dd{x}\Big[&
	\Psi^\dag(x)\left(\frac{p_x^2}{2m}-\mu\right)\Psi(x)
	+\nonumber\\
+
	&
	\left(\frac{\ii\Delta}{2\hbar} 
	\Psi(x)p_x\Psi(x)+\text{h.c.}\right)
	\Big],
\end{align}
where $\Psi(x)$ is the real-space fermion field, $p_x=-\ii\hbar\partial_x$ the momentum operator, $m$ the effective mass, $\mu$ the chemical potential, and $\Delta$ the superconducting pairing.
For simplicity, we choose $\Delta$ to be real.
Introducing the Nambu spinor $\bm\Psi^\dag(x)=[\Psi^\dag(x),\Psi(x)]$, 
and by formally substituting $\Psi^\dag(x)\Psi(x)\to(\Psi^\dag(x)\Psi(x)-\Psi(x)\Psi^\dag(x))/2$,
the continuum Hamiltonian can be rewritten in the Bogoliubov-de~Gennes (BdG) form as $\mathcal{H}=\frac12\int\dd{x}\bm\Psi^\dag(x)H(x)\bm\Psi(x)$, with Hamiltonian density
\begin{equation}\label{eq:H-pwaveC}
	H(x)=\left(\frac{p_x^2}{2m}-\mu\right)\tau_z
+
	\frac\Delta\hbar p_x\,\tau_y,
\end{equation}
where $\tau_{xyz}$ are the Pauli matrices. 
Notice that, with the assumption $\Delta\in\mathbb{R}$, the Hamiltonian is real.
In the momentum basis the Hamiltonian becomes $\mathcal{H}=\frac12\int\dd{k}\bm\Psi^\dag(k)H(k)\bm\Psi(k)$
with 
\begin{equation}\label{eq:H-pwaveCkBdG}
	H(k)=\left(\frac{\hbar^2k^2}{2m}-\mu\right)\tau_z
+
	\Delta k\,\tau_y,
\end{equation}
where the Nambu spinor in momentum space is $\bm\Psi^\dag(k)=[\Psi^\dag(k),\Psi(-k)]$.
The energy dispersion is given by a particle $E_k\ge0$ and hole $-E_k\le0$ branches with
\begin{equation}\label{eq:kEnergyCont}
	E_k=\sqrt{\varepsilon_k^2+\Delta^2 k^2},
\end{equation}
where $\varepsilon_k=\hbar^2k^2/2m-\mu$.
Contrarily to the case of a conventional $s$-wave superconductor which is always gapped, the energy dispersion becomes gapless at $k=0$ for $\mu=0$, i.e., when the chemical potential is fine-tuned to the bottom of the conduction band.
This is a direct consequence of the odd-parity symmetry of the superconducting pairing $\Delta k$, which is an odd function of the momentum:
The particle-hole gap can close only at $k=0$ if the bare electron dispersion $\varepsilon_k$ vanishes altogether.
The closing of the particle-hole gap at $\mu=0$ coincides with a topological phase transition between topologically distinct phases for $\mu<0$ and $\mu>0$.

Notice that the phase with $\mu<0$ remains gapped for $\Delta=0$ and, consequently, can be continuously deformed into a trivial insulator by switching off the superconducting pairing without closing the particle-hole gap.
Hence, this phase is topologically trivial.
On the other hand, the gapped phase with $\mu>0$ becomes gapless for $\Delta=0$ and thus cannot be continuously deformed into a trivial insulator.
This phase is topologically distinct from the trivial phase.
Topologically distinct phases correspond to topologically distinct mappings between the momentum $k$ and the Hamiltonian density.
To see this, let us rewrite the Hamiltonian density as
\begin{equation}
	H(k)=\mathbf{h}(k)\cdot\bm\tau,
\end{equation}
with
\begin{equation}
	\mathbf{h}(k)=[0,
	\Delta k,
	\varepsilon_k], 
	\qquad
	|\mathbf{h}(k)|=E_k,
\end{equation}
and consider the mapping between the momentum $k\in[-\infty,+\infty]$ to the unit vector $\mathbf{u}(k)=\mathbf{h}(k)/|\mathbf{h}(k)|$ moving on the $yz$ plane of the unit sphere.
The $y$ component of the unit vector goes from 
negative to positive values from $k=-\infty$ to $k=\infty$ for $\Delta>0$ (and from positive to negative for $\Delta<0$).
For $\mu<0$, one has $\varepsilon_k>0$ for all momenta, and thus the $z$ component of the unit vector is always positive with $\mathbf{u}(k)=[0,0,1]$ for $k=0$ and $k\to\pm\infty$.
In this case, the unit vector describes a circular arc as the momentum varies from $k=-\infty$ to $+\infty$, as shown in \cref{fig:JR}(a).
For $\mu>0$ instead, $\varepsilon_k$ changes its sign from positive at $k=-\infty$ to negative at $k=0$ and back to positive values at $k=+\infty$.
Thus, the $z$ component is positive at large momenta and negative at small momenta, 
with $\mathbf{u}(k)=[0,0,1]$ for $k=\pm\infty$ and $\mathbf{u}(k)=[0,0,-1]$ for $k=0$.
In this case, the unit vector describes a full circle as the momentum varies from $k=-\infty$ to $+\infty$, as shown in \cref{fig:JR}(b).
The trajectories followed by the unit vector are topologically distinct: a circular arc for $\mu<0$ and a full circle for $\mu>0$, characterized by a different winding number.
The winding number $W$ is defined as the total number of turns that the unit vector $\mathbf{u}(k)$ makes around the origin in the $yz$ plane as the momentum varies from $k=-\infty$ to $+\infty$, with positive or negative signs corresponding to 
anticlockwise or clockwise 
directions, respectively.
Hence, the winding number is $W=0$ for $\mu<0$, while it is $W=\pm1$ for $\mu>0$ with the sign 
equal to 
the sign of the superconducting pairing $\Delta$.
The phase with winding number $W=0$ is topologically trivial, whereas the phases with $W=\pm1$ are topologically nontrivial.
The trivial and nontrivial phases correspond to the so-called strong-pairing and weak-pairing, as I will discuss in \cref{sec:extendedbulk}.

The Hamiltonian in \cref{eq:H-pwaveC,eq:H-pwaveCkBdG} is equivalent to a 1D Dirac Hamiltonian with a Dirac mass $M=-\mu/\Delta^2$ and a quadratic correction $\propto k^2$ (assuming $\Delta>0$).
The quantum phase transition between the topologically trivial $M>0$ (i.e., $\mu<0$) and nontrivial $M<0$ (i.e., $\mu>0$) phase coincide with the inversion of the gap $M$~\cite{shen_topological_2011,shen_topological_2017}.
The BdG Hamiltonian describing quasiparticle states in a superconductor is analogous to the Dirac equation of a Majorana fermion, where the vacuum is replaced by the superconducting condensate and the role of particle-antiparticle symmetry by particle-hole symmetry (see also Refs.~\onlinecite{shen_topological_2011,elliott_colloquium:_2015,shen_topological_2017,aguado_majorana_2017}).
The connection to the Dirac equation (with quadratic corrections) and the identification of the sign of the Dirac mass with the topological invariant offers a unifying framework for understanding topological phases of matter, in particular topological insulators and superconductors, as highlighted in Refs.~\onlinecite{shen_topological_2011,shen_topological_2017}.
In this framework, the emergence of boundary modes can be seen as a direct consequence of the change of the sign of the Dirac mass.
Since phases with opposite Dirac masses are topologically distinct, they cannot be connected by a smooth transformation unless the Dirac mass becomes zero at their shared boundary.
In other words, the energy gap must close at the boundary between topologically distinct phases, allowing the topological invariant to change.
This mandates the existence of localized modes closing the energy gap at the boundary or domain wall between topologically inequivalent phases.
This direct correspondence between the topological invariant and the existence of boundary modes is called the bulk-boundary correspondence~\cite{hatsugai_chern_1993,ryu_topological_2002,teo_topological_2010}.
With a slight abuse of language, one may say that the inversion of the energy gap (i.e., the sign change of the Dirac mass) requires the existence of a gapless boundary.

Due to the bulk-boundary correspondence~\cite{hatsugai_chern_1993,ryu_topological_2002,teo_topological_2010}, Majorana modes localize at the boundary between topologically trivial and nontrivial phases or, equivalently, at the boundary between the nontrivial phase and the vacuum.
To verify this directly, let us consider a semi-infinite 1D superconductor $x\ge0$ and look for the zero-energy modes of the Hamiltonian in \cref{eq:H-pwaveC} localized at $x=0$. 
The eigenfunction of such zero-energy mode must satisfy
\begin{equation}
	\left[\left(-\eta\partial_x^2-\mu\right)\tau_z
	-
	\ii\Delta\partial_x\,\tau_y \right]\bm\psi(x)=0,
\end{equation}
where $\eta={\hbar^2}/{2m}$.
Since we are looking for the localized modes, we require the wavefunction $\bm\psi(x)$ to vanish at the boundary and at infinity, i.e., $\bm\psi(0)=\bm\psi(+\infty)=0$. 
Multiplying the equation above by $\ii\tau_y$ yields
\begin{equation}\label{eq:DiffEq}
	\left[\left(\eta\partial_x^2+\mu\right)\tau_x
	+
	\Delta\partial_x \right]\bm\psi(x)=0.
\end{equation}
Since the above equation contains only 
one Pauli
matrix, the wavefunction $\bm\psi(x)$ must be an eigenstate of $\tau_x$ with eigenvalue $s=\pm1$.
Moreover, since the wavefunction must vanish at infinity, we take $\bm\psi(x)\propto\ee^{-\lambda x}$ with $\Re(\lambda)>0$.
Thus, our trial solution is either
\begin{equation}
	\bm\psi(x)\propto
	\begin{bmatrix} 1 \\ 1 \\ \end{bmatrix}
	\ee^{-\lambda x},
	\text{\quad or \quad}
	\bm\psi(x)\propto
	\begin{bmatrix} 1 \\ -1 \\ \end{bmatrix}
	\ee^{-\lambda x}.
\end{equation} 
Substituting into \cref{eq:DiffEq} yields the algebraic equations
\begin{equation}
	\eta\lambda^2
	-
	\Delta\lambda+\mu=0,
	\qquad
	\eta\lambda^2
	+
	\Delta\lambda+\mu=0.
\end{equation}
Assuming $\Delta>0$, only the 
	first
equation has both roots with $\Re(\lambda)>0$ for $\mu>0$, given by
\begin{equation}
	\lambda_\pm=\frac1{2\eta}\left(\Delta\pm\sqrt{\Delta^2-4\eta\mu}\right).
\end{equation}
Hence, the boundary conditions can be fulfilled for $\mu>0$ by a zero-energy mode with a purely real wavefunction
\begin{equation}\label{eq:JRlike}
	\bm\psi(x)= \frac1{Z} \begin{bmatrix} 1 \\ 
	1
	\\ \end{bmatrix} \left(\ee^{-\lambda_+ x}-\ee^{-\lambda_- x}\right),
\end{equation} 
with $1/Z$ a normalization factor where 
$Z^2=|1/\lambda_{+}+1/\lambda_{-}-4/(\lambda_{+}+\lambda_{-})|$.
This wavefunction represents a self-adjoint Majorana zero mode which decays exponentially away from the boundary $x=0$, with oscillating behavior in the case where $\Im(\lambda_\pm)\neq0$, as shown in \cref{fig:JR}(c).
The decay rate is determined by the values of $|\lambda_\pm|$, which correspond to the characteristic lengths $\xi_\pm=1/|\lambda_\pm|$.
The localization length of the Majorana mode is given by the largest of these two, i.e., $\xi_\mathrm{M}=\max(1/|\lambda_+|,1/|\lambda_-|)$.
The Majorana mode at the origin corresponds to a topologically-protected end mode localized at the boundary between the topologically nontrivial phase and the vacuum.
For example, in the case of a nonuniform chemical potential such that $\mu(x)\lessgtr0$ for $x\lessgtr0$, a Majorana mode localizes at the origin $x=0$.
The zero-energy mode obtained is analogous to Jackiw-Rebbi solitons, which are solutions of the Dirac equation localized at the interface between regions with positive and negative masses~\cite{jackiw_solitons_1976,jackiw_solitons_1981}.
In particular, the wavefunction in \cref{eq:JRlike} reduces to the Jackiw-Rebbi soliton $\propto\ee^{-(\mu/\Delta) x}$ when $\eta\to0$ (i.e., $m\to\infty$), that is, when the quadratic term in the Hamiltonian in \cref{eq:H-pwaveC,eq:H-pwaveCkBdG} becomes negligible.
See Refs.~\onlinecite{elliott_colloquium:_2015,aguado_majorana_2017,shen_topological_2017} for a more thorough discussion on the connection between topologically nontrivial phases and the Dirac equation.

\subsection{The Kitaev model}

Historically, spatially-separated, pointlike Majorana modes in a 1D topological superconductor were first considered in a discrete model introduced by Kitaev~\cite{kitaev_unpaired_2001}.
The Kitaev model is a tight-binding Hamiltonian describing spinless electrons on a 1D lattice with $p$-wave superconducting pairing, which reads 
\begin{align}
	\mathcal{H}=
	-\sum_{n=1}^N & \mu c^\dag_n c_n -\sum_{n=1}^{N-1}\left(t c^\dag_n c_{n+1}+\text{h.c.}\right)+
	\nonumber\\\label{eq:H-pwave}
	+
	\sum_{n=1}^{N-1} & \left(\Delta\ee^{\ii\phi}c_n c_{n+1}+\text{h.c.}\right),
\end{align}
where $c^\dag_n$ and $c_n$ are respectively the creation and annihilation operators on each lattice site, $\mu$ is the chemical potential, $t>0$ the hopping parameter, $\Delta\ee^{\ii\phi}$ the superconducting pairing, and $N$ the number of lattice sites.
Notice that the phase $\phi$ of the superconducting pairing can always be absorbed by a unitary transformation $c_n\to\ee^{-\ii\phi/2}c_n$.
Hence, without loss of generality and for the sake of simplicity, let us assume that $\phi=0$ hereafter.
Notice that the superconducting term couples only electron states on neighboring sites: 
This is because spinless electrons cannot doubly occupy the same lattice site due to the Pauli exclusion principle.
Notice also that the superconducting term does not conserve the number of fermions: 
Electrons can be added or removed from the system.
However, it still conserves the fermion parity, i.e., the number of fermions modulo 2, since the superconducting term can only create or annihilate 2 fermions at a time.
Of course, particle conservation is not violated at a fundamental level.
The superconducting pairing term is, in fact, a mean-field description of the quasiparticle excitations of a fermionic condensate, which is the many-body groundstate described by the Bardeen-Cooper-Schrieffer (BCS) theory~\cite{bardeen_microscopic_1957,bardeen_theory_1957}.
Therefore, added and removed electrons are just coming in and out of the BCS condensate, which acts as a reservoir.

It is useful to rewrite the Hamiltonian in the BdG form~\cite{de-gennes_superconductivity_1999_c51}.
This can be achieved straightforwardly by using the fermion anticommutation relations $\{c_n,c_m\}=0$ and $\{c_n^\dag,c_m\}=\delta_{nm}$ which give $c_{n}c_{m}=(c_{n}c_{m}-c_{m}c_{n})/2$ and $c_n^\dag{c_m}=(\delta_{nm}+{c_n^\dag}c_m-c_m{c_n^\dag})/2$.
Using these simple identities, one can rewrite \cref{eq:H-pwave} as
\begin{align}
\mathcal{H}=
\frac12 \sum_{n=1}^N&
\begin{bmatrix} c_n^\dag , & c_n \\ \end{bmatrix}
\!\cdot\!
\begin{bmatrix}
-
\mu & 0 \\
0 & \mu \\
\end{bmatrix}
\!\cdot\!
\begin{bmatrix}
c_n \\
c_n^\dag \\
\end{bmatrix}
+
\nonumber\\
+
\frac12 \sum_{n=1}^{N-1}&
\begin{bmatrix} c_n^\dag , & c_n \\ \end{bmatrix}
\!\cdot\!
\begin{bmatrix}
-
t & -\Delta \\
\Delta & t \\
\end{bmatrix}
\!\cdot\!
\begin{bmatrix}
c_{n+1} \\
c_{n+1}^\dag \\
\end{bmatrix}
+\text{h.c.},
\label{eq:H-pwave-BdG}
\end{align}
up to a constant term.
The Hamiltonian above is identical to the Hamiltonian in \cref{eq:H-pwave} (up to a constant term).
Notice that, in the gauge $\phi=0$, the Hamiltonian is purely real, and therefore its eigenstates can be written in terms of real wavefunctions.

In the BdG formalism, the fermion degrees of freedom are doubled.
Each fermionic ``particle'' state (with creation operator $c^\dag$) is mirrored by a fermionic ``hole'' state (with creation operator $\tilde{c}^\dag=c$).
Indeed, the particle creation operator coincides with the hole annihilation operator.
In other words, creating a particle with energy $E$ is equivalent to annihilating a hole with energy $-E$.
The groundstate of the BdG Hamiltonian is the state where all quasiparticle hole states are occupied.
On the other hand, the excited states coincide with the creation of a particle with energy $E>0$, or equivalently the annihilation of a hole with energy $-E<0$.
This particle-hole symmetry can be formally described as an antiunitary symmetry that anticommutes with the Hamiltonian, as I will discuss in \cref{sec:topology}.

The Kitaev model in \cref{eq:H-pwave} corresponds to the discrete version of the continuum Hamiltonian of a spinless topological superconductor in \cref{eq:H-pwaveC}, obtained via \cref{eq:discretization} with $t=\eta/a^2=\hbar^2/2ma^2$ the hopping parameter, and by redefining the superconducting pairing $\widetilde\Delta=\Delta/2a$ and rescaling the chemical potential $\widetilde\mu=\mu-2t$ (see \cref{sec:discretization}).

\subsection{Extended bulk states\label{sec:extendedbulk}}

Let us first consider the bulk properties of the Kitaev model.
Imposing periodic boundary conditions to the 1D chain, i.e., by adding the lattice site $n=0$ and imposing $c_0=c_N$, via Fourier transform $c_n=(1/\sqrt{N})\sum_k\ee^{\ii kna}c_k$, the Hamiltonian in \cref{eq:H-pwave} becomes in momentum space
\begin{equation}\label{eq:H-pwave-kspace}
	\mathcal{H}=
	\sum_k \left(
	\varepsilon_k c^\dag_k c_k +
	\frac12 \Delta_k c_{-k} c_{k} +
	\frac12 \Delta_k^* c^\dag_{k} c^\dag_{-k}
	\right),
\end{equation}
with $\varepsilon_k=-(\mu+2t\cos{(ka)})$, 
$\Delta_k
=
2\ii\Delta\sin{(ka)}$, and where the momenta are quantized as $k=2\pi m/N$ with $m=1,\ldots,N$ integer.
The Hamiltonian above can be rewritten using the fermion anticommutation relations
in the more compact BdG form as
\begin{equation}
	\mathcal{H}=\frac12\sum_k
	\begin{bmatrix} c_k^\dag , & c_{
-
	k} \\ \end{bmatrix}
	\cdot H(k) \cdot
	\begin{bmatrix} c_k \\ c_{-k}^\dag \\ \end{bmatrix},
\end{equation}
with
\begin{equation}\label{eq:H-pwave-kspace-BdG}
	H(k)=
	\begin{bmatrix}
	\varepsilon_k & \Delta_k^* \\
	\Delta_k & -\varepsilon_k
	\end{bmatrix}
	=\varepsilon_k\tau_z + \Im(\Delta_k)\tau_y,
\end{equation}
up to a constant term.
The energy levels $E_k$ are the solutions of the characteristic equation $\det\left(H(k)-E_k\right)=0$,
which yields two dispersion branches $\pm E_k$ with 
\begin{align}
	E_k=&\sqrt{\varepsilon_k^2+|\Delta_k|^2}=
	\nonumber\\
	=&\sqrt{(\mu+2t\cos{(ka)})^2+4\Delta^2\sin^2{(ka)}},
	\label{eq:kEnergy}
\end{align}
where positive and negative energy levels correspond respectively to particle and hole states.
The continuum Hamiltonian in \cref{eq:H-pwaveCkBdG} and its discrete counterpart in \cref{eq:H-pwave-kspace-BdG} are approximately equivalent at small momenta $k\approx0$, as one can see by expanding \cref{eq:H-pwave-kspace-BdG} at the second order in $k$.
However, the energy dispersion has a finite energy width in the discrete cases, whereas it is unbounded in the continuum case, being parabolic $\propto k^2$ [see \cref{eq:kEnergyCont}].
In practice, this difference does not have any physical consequence as long as one is interested in the low-energy properties of the model.

\begin{figure}[t]
\includegraphics[width=\columnwidth]{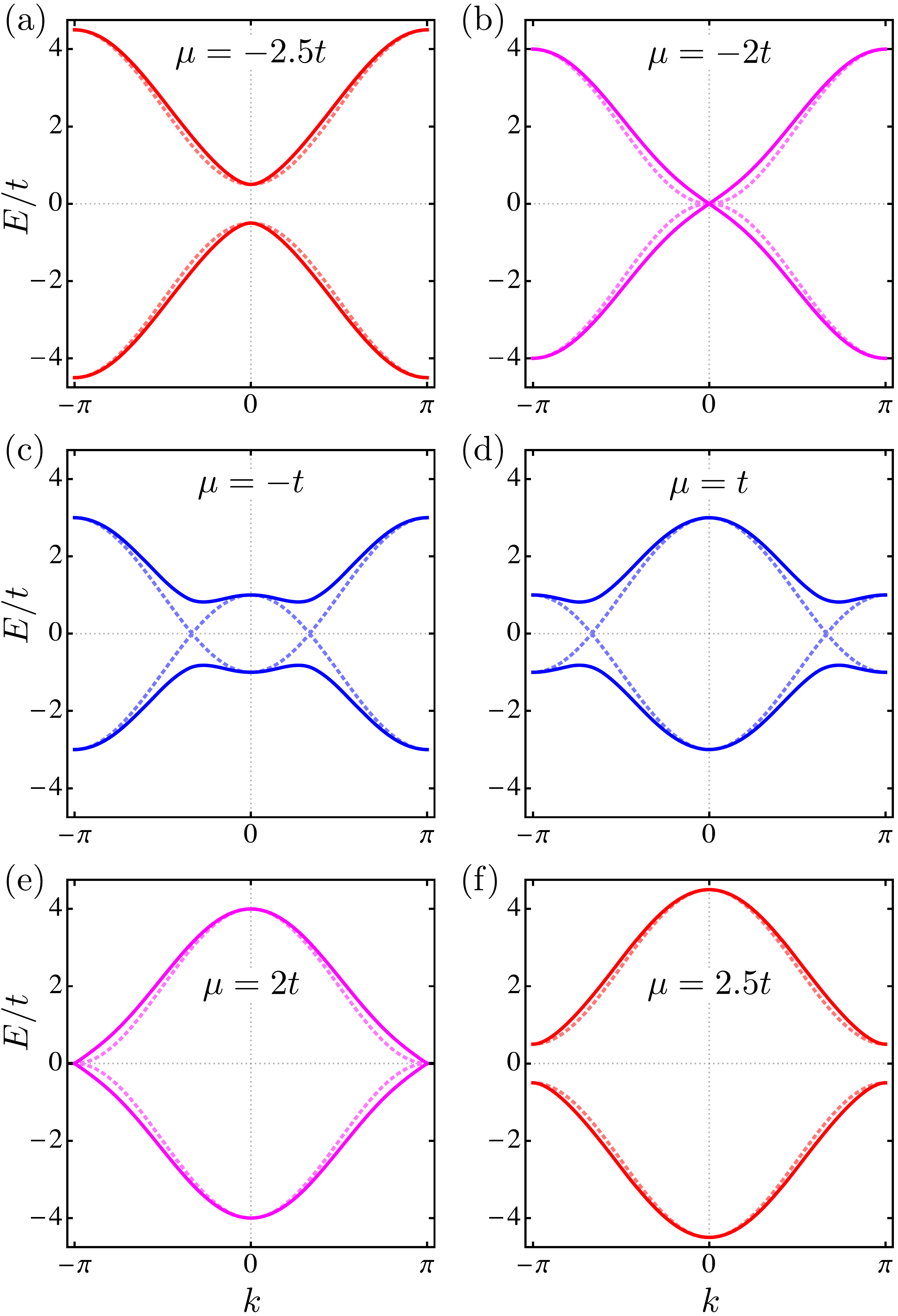}
\caption{
Energy dispersion $E_k$ (continuous lines) of the superconducting quasiparticles compared with the bare electron dispersion $\varepsilon_k$ (dotted lines) of the Kitaev model for different choices of the chemical potential $\mu$ and with $\Delta=t/2$, as in \cref{eq:kEnergy}.
Positive and negative branches correspond respectively to the particle and hole levels.
The bare electron dispersion (dotted lines) is gapped for $\mu<-2t$ (a) and $\mu>2t$ (f), whereas it is metallic for $-2t<\mu<2t$, (c) and (d). 
The particle-hole gap closes at $k=0$ for $\mu=-2t$ (b) and at $k=\pi$ for $\mu=2t$ (e).
}
\label{fig:bulk}
\end{figure}

\Cref{fig:bulk} shows the energy dispersion $E_k$ of superconducting electrons and the bare electron dispersion $\varepsilon_k$ for different choices of the chemical potential $\mu$.
The bare electron dispersion is gapped for $|\mu|>2t$ [see \cref{fig:bulk}(a) and~\ref{fig:bulk}(f)], whereas it is gapless for $|\mu|<2t$ with a Fermi momentum $k_\mathrm{F}$ determined by $2t\cos{(k_\mathrm{F}a)}=-\mu$ [see \cref{fig:bulk}(c) and~\ref{fig:bulk}(d)]. 
The energy dispersion of superconducting electrons instead 
has a finite gap $\Delta E=|2t-\mu|$
for $2t\neq|\mu|$ but becomes gapless at the time-reversal symmetry points $k=0,\pm\pi$ for $\mu=\pm2t$, i.e., when the chemical potential is fine-tuned to the top or the bottom of the conduction band [see \cref{fig:bulk}(b) and~\ref{fig:bulk}(e)].
As mentioned, this is a direct consequence of the odd-parity symmetry of the superconducting pairing (i.e., $\Delta_k=\Delta_{-k}$), which can vanish only at $k=0$ and $k=\pm\pi$.
Thus, the particle-hole gap closes only if the bare electron dispersion $\varepsilon_k$ vanishes altogether, which is indeed what happens for $\mu=-2t$ and $\mu=2t$ where $\varepsilon_k=0$ respectively at $k=0$ and $k=\pm\pi$.
The Hamiltonian is diagonalized by a Bogoliubov transformation 
\begin{equation}\label{eq:Bogoliubov1}
\begin{bmatrix}
d_k \\
d_{-k}^\dag\\
\end{bmatrix}=
U_k
\cdot
\begin{bmatrix}
c_k \\
c_{-k}^\dag\\
\end{bmatrix},
\qquad
U_k=
\begin{bmatrix}
u_k & v_k \\
v_{-k}^* & u_{-k}^* \\
\end{bmatrix},
\end{equation}
with 
\begin{subequations}\label{eq:absukvk}\begin{align}
|u_k|^2=&\frac12 + \frac{\varepsilon_k}{2 E_k},
\\
|v_k|^2=&\frac12 - \frac{\varepsilon_k}{2 E_k},
\\
v_k/u_k=&\frac{E_k-\varepsilon_k}{\Delta_k},
\end{align}\end{subequations}
which fully determines the Bogoliubov coefficients up to an arbitrary phase.
This transformation defines a new set of fermion operators, i.e., the Bogoliubov quasiparticles $d_k=u_{k}c_{k}+v_{k}c_{-k}^\dag$, mixing creation and annihilation operators (particle and hole states) with opposite momenta.
In the gauge $\phi=0$, the two Bogoliubov coefficients can be chosen to be real numbers, such that the eigenstates are described by real wavefunctions.
The groundstate of the Hamiltonian is the vacuum of the Bogoliubov quasiparticles, i.e., the state $\ket{0}$ annihilated by all fermion operators $d_k$, i.e., $d_k\ket{0}=0$ for all momenta $k$.
The excitations of the system are the states $d_k^\dag\ket{0}$, which can be interpreted equivalently as the creation of a particle with energy $E_k$, or the annihilation of a hole with energy $-E_k$.
Creating a state with positive energy (particle) is equivalent to annihilating a state with negative energy (hole).

\begin{figure}[t]
\includegraphics[width=\columnwidth]{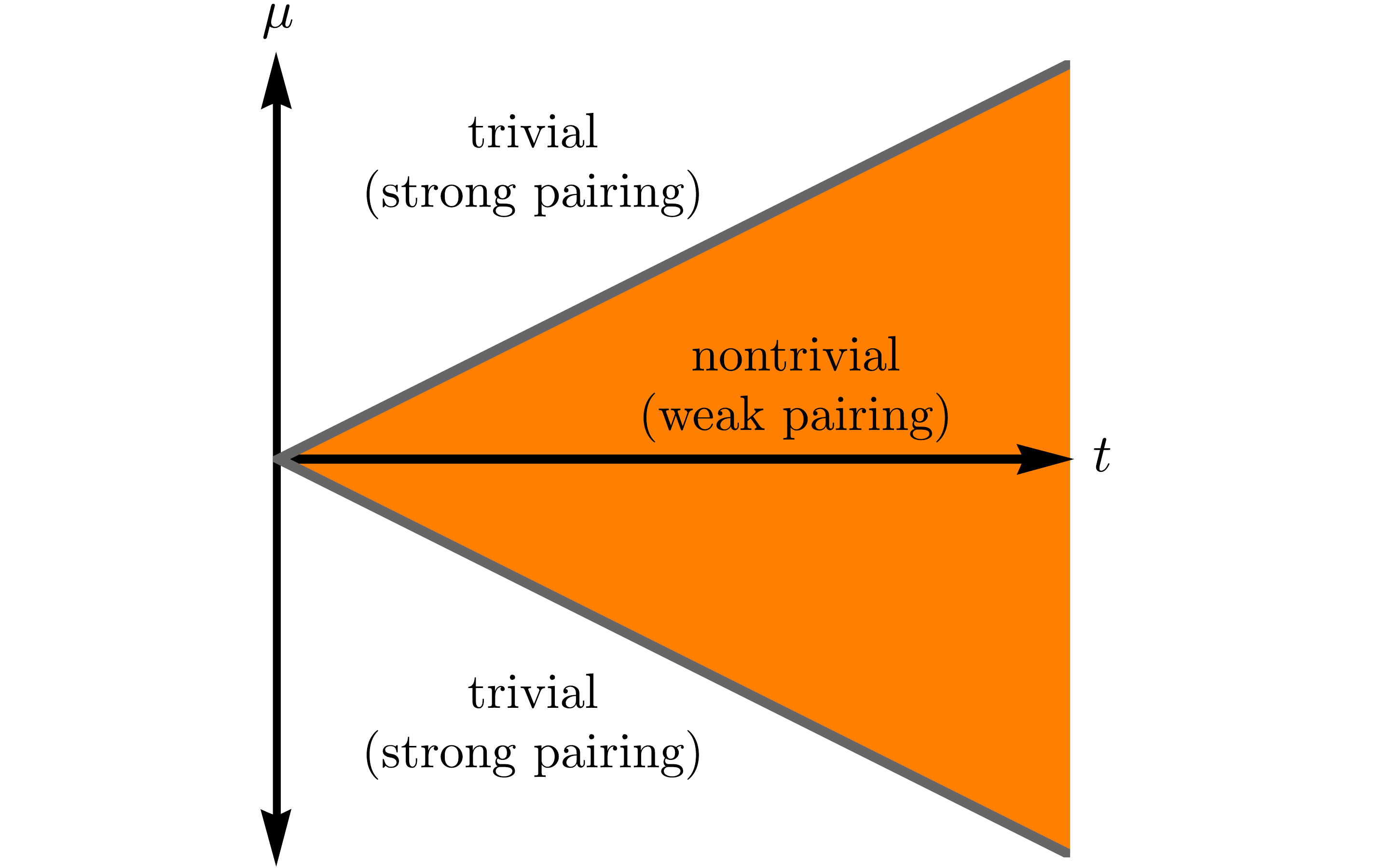}
\caption{
Phase space of the Kitaev model as a function of the hopping parameter $t$ and chemical potential $\mu$.
In the strong-pairing phase $|\mu|>2t$, the energy dispersion is gapped and topologically trivial and remains gapped for $\Delta\to0$.
In the weak-pairing phase $|\mu|<2t$ instead, the energy dispersion is gapped and topologically nontrivial and becomes metallic for $\Delta\to0$.
}
\label{fig:phasespace}
\end{figure}

\Cref{fig:phasespace} shows the phase space of the Kitaev model.
The gapless points $\mu=\pm2t$ separate two qualitatively different regimes, the so-called strong-pairing ($|\mu|>2t$) and weak-pairing ($|\mu|<2t$) phases, following the definition of Read and Green~\cite{read_paired_2000}.
In the strong-pairing phase, the system is insulating for $\Delta=0$ and thus cannot be described by the usual BCS instability scenario (there is no Fermi surface).
In the weak-pairing phase, the system is metallic for $\Delta=0$, and its superconducting phase is BCS-like but with a momentum-dependent order parameter.
The strong-pairing phase can be smoothly connected to the bare electron vacuum (to the state where all particles are occupied) without closing the particle-hole gap and is thus topologically trivial.
On the other hand, the weak-pairing phase $|\mu|<2t$, being qualitatively distinct from the bare electron vacuum~\cite{read_paired_2000}, is topologically nontrivial, as I will discuss in the next Section.

\subsection{Topological invariants\label{sec:topology}}

\subsubsection{Symmetries}

Noninteracting topological insulators and superconductors can be classified in the so-called periodic table of topological invariants~\cite{kitaev_periodic_2009,schnyder_classification_2009,ryu_topological_2010,ludwig_topological_2015,chiu_classification_2016} in terms of the ten Altland–Zirnbauer symmetry classes~\cite{altland_nonstandard_1997} and spatial dimensions.
Gapped phases that can be connected by a continuous transformation without closing the energy gap and without breaking any symmetry are topologically equivalent.
Different equivalence classes within the same symmetry class are characterized by a topological invariant, which is an integer number:
the value of the topological invariant is not affected by small perturbations but can change only in correspondence to a quantum phase transition, e.g., when the particle-hole gap of the superconductor closes and reopens again.
In homogeneous systems with translational invariance, the topological invariant labels the different (inequivalent) homotopy classes partitioning all possible mappings between the $d$-dimensional Brillouin zone and the Hamiltonian $H(\mathbf{k})$.
The relevant symmetries which classify topological phases of matter are the antiunitary symmetries $\mathcal{C}$ and $\mathcal{T}$ defined by antiunitary operators that 
anticommute and commute 
with the Hamiltonian, respectively, and the unitary symmetry defined by their composition $\mathcal{S}=\mathcal{T}\mathcal{C}$.
The $\mathcal C$ symmetry coincides with the particle-hole symmetry of the superconducting condensate, whereas the $\mathcal T$ symmetry usually coincides (but does not need to) with the ordinary time-reversal symmetry.

In a superconductor, each fermionic ``particle'' (with creation operator $c^\dag$) is mirrored by a fermionic ``hole'' (with creation operator $\tilde{c}^\dag=c$), as already mentioned.
The duality between particle and hole is described by the particle-hole symmetry operator $\mathcal{C}$.
The BdG Hamiltonian of a spinless superconductor can be written in matrix form as $\mathcal{H}=\frac12\bm{\Psi}^\dag\cdot H\cdot\bm{\Psi}$ with
\begin{equation}\label{eq:BdGHamiltonianGeneral}
	H=
	\begin{bmatrix}
	\hat{H}_0 & \hat{\Delta}^\dag\\
	\hat{\Delta} & -\hat{H}_0^\intercal & \\
	\end{bmatrix}
	=
	\begin{bmatrix}
	\hat{H}_0 & -\hat{\Delta}^*\\
	\hat{\Delta} & -\hat{H}_0^* & \\
	\end{bmatrix},
\end{equation}
and with $\bm{\Psi}^\dag=[c_1^\dag,\ldots,c_N^\dag,c_1,\ldots,c_N]$ and where $\hat{\Delta}^\dag=-\hat{\Delta}^*$ due to the antisymmetry of the superconducting pairing~\cite{sigrist_phenomenological_1991} and $\hat{H}_0^\intercal=\hat{H}_0^*$ due to hermitianicity.
In the case of the Kitaev model in \cref{eq:H-pwave,eq:H-pwave-BdG}, one has
\begin{equation}\label{eq:H-pwave-BdG-long}
	\hat{H}_0=
	-
	\begin{bmatrix}
	\mu & t & & \\
	t & \mu & & \\[-2mm]
	& & \ddots & \\[-2mm]
	& & &\hspace{-1mm} \mu \\
	\end{bmatrix},
	\quad
	\hat{\Delta}=
	\begin{bmatrix}
	0 & 
	\Delta & & \\
-
	\Delta & 0 & & \\[-2mm]
	& & \ddots & \\[-2mm]
	& & & \hspace{-1mm} 0 \\
	\end{bmatrix}.
\end{equation}
For a generic BdG Hamiltonian in the form \cref{eq:BdGHamiltonianGeneral}, the particle-hole symmetry operator is the antiunitary operator
\begin{equation}\label{eq:PHsymmetry}
	\mathcal{C}=\tau_x\mathcal{K},
\end{equation}
where $\mathcal{K}$ is the complex conjugation operator.
It is easy to verify that $\mathcal{C}^2=1$ and that
\begin{equation}
\mathcal{C}
H
\mathcal{C}^{-1}= - 
H,
\end{equation}
i.e., $\{H,\mathcal{C}\}=0$.
Notice that particle-hole symmetry is a built-in feature of BdG Hamiltonians and is never broken in the superconducting state.

On top of that, the Hamiltonian of the Kitaev model is also symmetric under time-reversal and chiral symmetries.
Indeed, the $p$-wave pairing does not break time-reversal and chiral symmetries in 1D\@.
This is because, contrarily to the case of 2D $p$-wave superconductors, the $p$-wave pairing term in 1D can be made real with a unitary phase rotation~\cite{niu_majorana_2012,tewari_topological_2012a}.
Specifically, the phase $\phi$ of the superconducting pairing can always be absorbed by a unitary rotation in 1D systems, and one can consider $\phi=0$ without loss of generality.
The time-reversal symmetry and chiral operators $\mathcal{T}$ and $\mathcal{S}$ are 
\begin{equation}
	\mathcal{T}= \mathcal{K},\qquad
	\mathcal{S}= \mathcal{T} \mathcal{C}=\tau_x,
\end{equation}
with $\mathcal{T}^2=\mathcal{S}^2=1$.
For the Kitaev model 
in \cref{eq:H-pwave-BdG-long} in the gauge $\phi=0$, the Hamiltonian is real, which yields
\begin{equation}\label{eq:TSsymmetries}
\mathcal{T}
H
\mathcal{T}^{-1}=
H, 
\qquad
\mathcal{S}
H
\mathcal{S}^{-1}=
-
H,
\end{equation}
i.e., $[H,\mathcal{T}]=0$ and $\{H,\mathcal{S}\}=0$.
\Cref{eq:PHsymmetry,eq:TSsymmetries} express the particle-hole, time-reversal, and chiral symmetries.
Unbroken time-reversal symmetry mandates unbroken chiral symmetry due to particle-hole symmetry.
Moreover, unbroken time-reversal symmetry mandates $H=H^*$, i.e., the Hamiltonian is real.

In momentum space, the Hamiltonian density of a single-band spinless superconductor [e.g., the Kitaev model in \cref{eq:H-pwave-kspace-BdG}] can be written as
\begin{align}
	H(k)=&
	\begin{bmatrix}
	\varepsilon_k & \Delta_k^* \\
	\Delta_k & -\varepsilon_k \\
	\end{bmatrix}
	=\nonumber\\=&
	\Re(\Delta_k)\tau_x+
	\Im(\Delta_k)\tau_y+
	\varepsilon_k\tau_z
	,
	\label{eq:BDIHamiltonian}
\end{align}
where $\Delta_{-k}=-\Delta_k$ due to the odd symmetry of the $p$-wave superconducting pairing, and $\varepsilon_{k}=\varepsilon_{k}^*=\varepsilon_{-k}$ due to hermitianicity and 
time-reversal symmetry of the bare-electron dispersion.
Notice that the complex conjugation operator $\mathcal{K}$ acts differently on the momentum basis since one has
\begin{align}
	\mathcal{K}c_k\mathcal{K}=&\frac1{\sqrt{N}}\sum_n\mathcal{K}\ee^{-\ii kna}c_n\mathcal{K}=
	\nonumber\\
	=&\frac1{\sqrt{N}}\sum_n\ee^{\ii kna}c_n=c_{-k}.
\end{align}
Hence, complex conjugation changes the sign of the momentum $k\to-k$, which gives 
$\varepsilon_k\to\varepsilon_{-k}=\varepsilon_{k}$, 
$\Delta_k\to\Delta_{-k}=-\Delta_k$
and $\mathcal{K}H(k)\mathcal{K}=H(-k)^*$.
It 
is easy to verify that \cref{eq:BDIHamiltonian} satisfies
\begin{equation}\label{eq:PHksymmetry}
\mathcal{C}
H(k)
\mathcal{C}^{-1}= 
-
H(k).
\end{equation}
Moreover, 
if $\Re(\Delta_k)=0$ (e.g., the Kitaev model in the gauge $\phi=0$),
one can verify that
\begin{equation}\label{eq:Tksymmetry}
\mathcal{T}
H(k)
\mathcal{T}^{-1}= 
H(k),
\quad
\mathcal{S}
H(k)
\mathcal{S}^{-1}= 
-H(k).
\end{equation}
\Cref{eq:PHksymmetry,eq:Tksymmetry} express the particle-hole, time-reversal, and chiral symmetries in momentum space.

Hamiltonians with unbroken particle-hole $\mathcal{C}$, time-reversal $\mathcal{T}$, and chiral $\mathcal{S}=\mathcal{T}\mathcal{C}$ symmetries with $\mathcal{C}^2=\mathcal{T}^2=1$ belong to the symmetry class BDI\@.
Unbroken time-reversal symmetry mandates that Hamiltonians in class BDI can be written in terms of only two out of three Pauli matrices, 
since \cref{eq:Tksymmetry} mandates $\{H(k),\tau_x\}=0$.
Consequently, these Hamiltonians are real, or can be cast as real Hamiltonians up to a unitary transformation.
In 1D, this class allows topologically nontrivial phases corresponding to a topological invariant that can take in principle all integer values $\nu\in\mathbb{Z}$. 
On the other hand, Hamiltonians with broken time-reversal and chiral symmetries and unbroken particle-hole symmetry with $\mathcal{C}^2=1$ belong to the symmetry class D\@. 
In 1D, this class allows topologically nontrivial phases corresponding to a topological invariant that can take only two values $\nu=0,1\in\mathbb{Z}_2$, corresponding respectively to trivial $\nu=0$ and nontrivial $\nu=1$ phases (see Refs.~\onlinecite{ludwig_topological_2015,chiu_classification_2016} for a thorough overview of the classification of topological phases of matter).
In general, one is interested in the effects of perturbations on the Majorana modes, described by extra terms in the Hamiltonian.
Depending on whether these perturbations preserve or break time-reversal symmetry, the Kitaev model in \cref{eq:H-pwave} can be classified either in class BDI or class D (with topological invariants $\mathbb{Z}$ and $\mathbb{Z}_2$), respectively.

\subsubsection{The winding number}

\begin{figure}[t]
\includegraphics[width=\columnwidth]{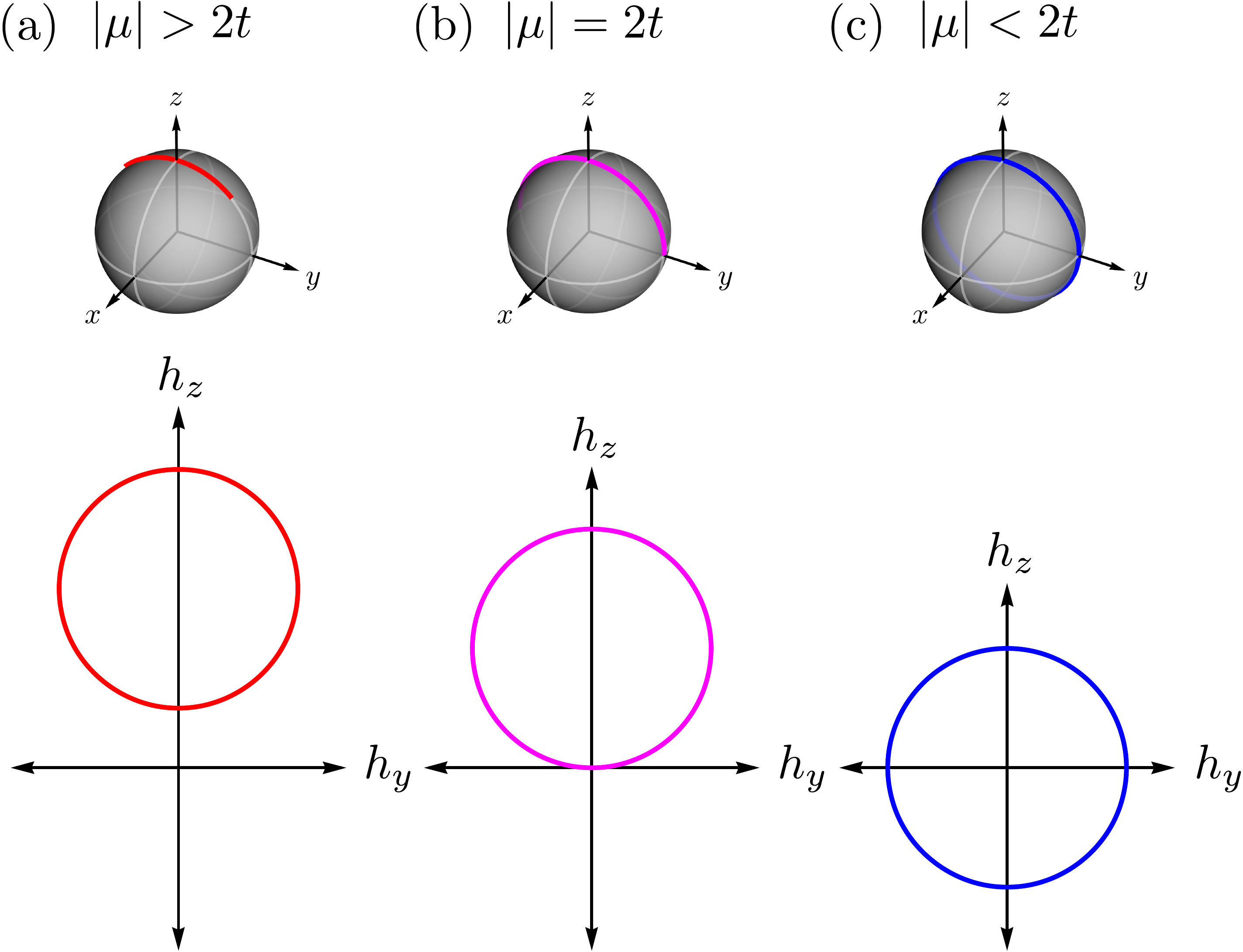}
\caption{
The path followed by the vector $\mathbf{h}(k)$ of the Kitaev model on the $yz$ plane and by the unit vector $\mathbf{u}(k)=\mathbf{h}(k)/|\mathbf{h}(k)|$ on the unit sphere when $k$ runs from 0 to $2\pi$.
The winding number counts the number of turns around the origin.
(a) For $|\mu|>2t$, the unit vector follows a closed loop that does not encircle the origin, and the winding number is $W=0$.
(c) For $|\mu|<2t$ instead, the loop does encircle the origin, following an 
anticlockwise or clockwise 
path for $\Delta>0$ and $\Delta<0$, corresponding to a winding number $W=1$ or $W=-1$, respectively.
(b) The transition between the two distinct topological phases with $W=0$ and $W=\pm1$ coincides with the closing of the particle-hole gap, corresponding to the vector 
on the $yz$ plane
passing through the origin $\mathbf{h}(k)=0$.
}
\label{fig:winding}
\end{figure}

In the case where perturbations do not break time-reversal symmetry, e.g., random disorder or next-nearest neighbor hoppings in \cref{eq:H-pwave}, the topological invariant $\mathbb{Z}$ that characterizes the topological phases in class BDI in 1D is the winding number. 
To define the winding number, let us rewrite the Hamiltonian density in \cref{eq:BDIHamiltonian} in terms of the vector $\mathbf{h}(k)=[\Re(\Delta_k),\Im(\Delta_k),\varepsilon_k]$.
Particle-hole symmetry mandates $h_y(k)=-h_y(-k)$ and $h_z(k)=h_z(-k)$, while chiral symmetry mandates $h_x(k)=\Re(\Delta_k)=0$ [see \cref{eq:PHksymmetry,eq:Tksymmetry}].
Thus, we obtain	$H(k)=\mathbf{h}(k)\cdot\bm\tau$ with
\begin{equation}
	\mathbf{h}(k)=[0,\Im(\Delta_k),\varepsilon_k], 
	\qquad
	|\mathbf{h}(k)|=E_k.
\end{equation}
If the particle-hole gap is open, the unit vector $\mathbf{u}(k)=\mathbf{h}(k)/|\mathbf{h}(k)|$ draws a closed trajectory constrained on the unit circle in the $yz$ plane, as shown in \cref{fig:winding}, with a winding number defined by~\cite{tewari_topological_2012a}
\begin{equation}\label{eq:winding}
	W=\frac1{2\pi}\int_0^{2\pi}\dd\theta(k),
\end{equation}
where $\tan(\theta(k))=h_z(k)/h_y(k)$.
The winding number counts the total number of turns that the vector $\mathbf{h}(k)$ (or equivalently, the unit vector $\mathbf{u}(k)$) makes around the origin in the $yz$ plane as $k$ varies in the Brillouin zone $k\in[0,2\pi]$, with positive or negative signs for 
anticlockwise or clockwise
directions.
This number can only change when the trajectory of the vector $\mathbf{h}(k)$ passes through the origin, i.e., when the particle-hole gap closes.

For the Kitaev model in \cref{eq:H-pwave-kspace-BdG} in the gauge $\phi=0$, the vector $\mathbf{h}(k)=[0
,
2\Delta\sin{(ka)},-\mu-2t\cos{(ka)}]$ and the corresponding unit vector $\mathbf{u}(k)=\mathbf{h}(k)/E_k$ draw a closed loop in the $yz$ plane when $k$ runs from 0 to $2\pi$, as shown in \cref{fig:winding}.
It is easy to see that for $|\mu|>2t$, this closed loop does not encircle the origin, as shown in \cref{fig:winding}(a).
In this case, the winding number is $W=0$.
For $|\mu|<2t$ instead, the closed loop does encircle the origin, as shown in \cref{fig:winding}(c).
The direction of the winding depends on the sign of the superconducting parameter $\Delta$, being 
anticlockwise
for $\Delta>0$, which gives $W=1$, and 
clockwise 
for $\Delta<0$, which gives $W=-1$. 
The transition between the two distinct topological phases with $W=0$ and $W=\pm1$ coincides with the closing of the particle-hole gap at $\mu=\pm2t$, as shown in \cref{fig:winding}(b).
Notice that in class BDI, the two topological phases $W=\pm1$ are not equivalent since they cannot be smoothly connected without breaking time-reversal symmetry.

\subsubsection{The pfaffian invariant}

In the more general case where perturbations do break time-reversal symmetry, e.g., spatial variations of the phase of the superconducting term in \cref{eq:H-pwave}, the $\mathbb{Z}_2$ topological invariant characterizing topological phases in class D in 1D is an integer that admits only two possible values $\nu=0,1$, and such that its value can only change when the particle-hole gap closes.
Notice that if the time-reversal symmetry is broken, the unit vector $\mathbf{u}(k)$ is not constrained anymore on the plane $yz$ since on may have $h_x(k)\neq0$, and thus its winding number is not well-defined.
However, even in this case, the time-reversal and chiral symmetry in \cref{eq:Tksymmetry} are still unbroken for the two time-reversal symmetric momenta $k=0$ and $\pi$, as one can verify since $\Delta_0=\Delta_\pi=0$ due to the particle-hole symmetry.
This mandates
$h_x(k)=h_y(k)=0$ with $\mathbf{h}(k)=[0,0,\varepsilon_k]$ for $k=0,\pi$.
Consequently, the unit vector $\mathbf{u}(k)$ points to the positive $h_z(k)>0$ or negative $h_z(k)<0$ direction on the $z$ axis for $k=0,\pi$.
In this case, one can define the $\mathbb{Z}_2$ topological invariant as
\begin{equation}\label{eq:Z2invariant}
	(-1)^\nu=\sgn{\left(h_z(0)h_z(\pi)\right)}.
\end{equation}
If $\nu=0$, the unit vector $\mathbf{u}(k)$ has the same direction for $k=0,\pi$, and will follow a closed trajectory starting from the north (or south) pole when $k=-\pi$, passing again for the same pole when $k=0$, and returning back again when $k=\pi$.
If $\nu=1$, the unit vector $\mathbf{u}(k)$ has opposite directions for $k=0,\pi$, and will start from the north (or south) pole when $k=-\pi$, passing through the opposite pole when $k=0$, and returning back when $k=\pi$.
Notice that the signs of $h_z(0)$ and $h_z(\pi)$ are both changed by unitary transformations exchanging particle and hole operators.
However, the sign of the product $h_z(0) h_z(\pi)$ can change only by closing the particle-hole gap at either $k=0$ or $\pi$.
Thus, trajectories with $h_z(0)h_z(\pi)>0$ are topologically distinct from trajectories with $h_z(0)h_z(\pi)<0$.

Intuitively, the $\mathbb{Z}_2$ invariant in \cref{eq:Z2invariant} is defined by assigning a parity to the Hamiltonian density $H(k)$ at time-reversal momenta $k=0,\pi$.
This parity coincides with the sign of the component $h_z(k)$ of the 
vector $\mathbf{h}(k)$.
However, there is another, more general way to assign a parity to the Hamiltonian.
First, notice that 
the determinant of the Hamiltonian density $H(k)$ at time-reversal momenta $k=0,\pi$ is the square of
the $h_z(k)$ component up to a minus sign, i.e., $\det(H(k))=-h_z(k)^2$.
Then, let us recall that the determinant of an antisymmetric matrix is the square of its pfaffian (see \cref{sec:pfaffian}).
The matrix $H(k)$ is not antisymmetric, but the matrix 
$H(k)\mathcal{S}$ 
is real and antisymmetric at $k=0,\pi$ due to particle-hole symmetry, and satisfies 
$\det(H(k)\mathcal{S})=-\det(H(k))$.
Indeed, one can verify directly that 
$h_z(k)=\pf(H(k)\mathcal{S})$ 
at the time-reversal momenta $k=0,\pi$.
Thus, the sign of the pfaffian of the matrix $H(k)\mathcal{S}$ coincides with the sign of $h_z(k)$ at the time-reversal momenta $k=0,\pi$, and therefore,
the $\mathbb{Z}_2$ topological invariant can be defined as~\cite{kitaev_unpaired_2001,tewari_topological_2012a,budich_equivalent_2013}
\begin{equation}\label{eq:Z2invariantPfaffian2}
	(-1)^\nu=\sgn\left(\pf\Big(H(0)\mathcal{S}\Big)\pf\Big(H(\pi)\mathcal{S}\Big)\right), 
\end{equation}
with $\mathcal{S}=\tau_x$, where $\pf(A)$ denotes the pfaffian of the antisymmetric matrix $A$ (see \cref{sec:pfaffian}).
The matrix 
$H(k)\tau_x$ 
is real and antisymmetric at the time-reversal symmetry points $k=0,\pi$.
This is a direct consequence of particle-hole symmetry since 
$(H(k)\mathcal{K}\mathcal{C})^\intercal=\mathcal{C}\mathcal{K}H^*(k)=\mathcal{C}H(-k) \mathcal{K}=-H(-k)\mathcal{C}\mathcal{K}$ 
gives $(H(k)\tau_x)^\intercal=-H(k)\tau_x$ for $k=0,\pi$.
The topological invariant defined in \cref{eq:Z2invariant,eq:Z2invariantPfaffian2} is a quantity that admits only two possible values and such that its value changes only if the particle-hole gap closes.
Indeed, due to the symmetry of the $p$-wave pairing term, the particle-hole gap of a spinless superconductor may close only at the time-reversal symmetry points $k=0,\pi$, with the concurrent change of the topological invariant.
The trivial $\nu=0$ and nontrivial $\nu=1$ phases correspond to the cases where the pfaffian 
$\pf(H(k)\mathcal{S})$ 
has the same or opposite signs at the two time-reversal symmetry momenta $k=0,\pi$.
The change of the $\mathbb{Z}_2$ invariant from trivial ($\nu=0$) to nontrivial ($\nu=1$) coincides with the inversion of the particle-hole gap and with the energy level crossing at $k=0,\pi$.

The two definitions of the $\mathbb{Z}_2$ invariant in \cref{eq:Z2invariant,eq:Z2invariantPfaffian2} coincide since 
$h_z(k)=\pf(H(k)\mathcal{S})$.
However, the pfaffian formula in \cref{eq:Z2invariantPfaffian2}
can be straightforwardly applied to more general cases, such as multiband or quasi-1D cases~\cite{potter_multichannel_2010,zhou_crossover_2011}, and generalized to the case of spinful electrons (see \cref{sec:Dinvariant}).
Moreover, the pfaffian $\mathbb{Z}_2$ invariant can be rewritten in real space via a Fourier transform and generalized to cases where the translational symmetry is broken~\cite{budich_equivalent_2013}.
In the case of unbroken time-reversal symmetry (class BDI), the $\mathbb{Z}_2$ invariant corresponds to the parity of the $\mathbb{Z}$ invariant~\cite{tewari_topological_2012a}, i.e., $\nu\equiv W\!\!\mod2$.

Due to the bulk-boundary correspondence~\cite{hatsugai_chern_1993,ryu_topological_2002,teo_topological_2010}, the topological invariant $\nu$ is equal to the number of Majorana end modes localized at each boundary of a finite system or at the boundary with a topologically trivial phase (which has $\nu=0$).
Hence, there is one Majorana mode at each end of a 1D system in the topologically nontrivial phase $\nu=1$, and no Majorana modes in the topologically trivial phase $\nu=0$.
Moreover, since the topological invariant can change only when the gap closes, the existence of Majorana modes is robust against small perturbations which do not close the gap.

For the Kitaev model in \cref{eq:H-pwave-kspace-BdG}, one has that 
$\pf(H(k)\mathcal{S})=h_z(k)=\varepsilon_k=\mu\pm2t$
at $k=0,\pi$ and, consequently, \cref{eq:Z2invariant,eq:Z2invariantPfaffian2} yield
\begin{equation}
(-1)^\nu=\sgn(\mu^2-4t^2),
\end{equation}
which gives $\nu=0,1$ respectively for $|\mu|>2t$ and $|\mu|<2t$, corresponding to the topologically trivial and nontrivial phases.
Notice that the two topological phases with winding numbers $W=\pm1$ (corresponding to $\Delta>0$ and $\Delta<0$) become equivalent in class D since they can be connected by smoothly rotating the phase of the superconducting pairing $\Delta\ee^{\ii\phi}$ and correspond to the same topological phase with $\nu=1$.
Notice also that unitary rotations of the superconducting phase break time-reversal symmetry and are thus not allowed in class BDI.

It is worth mentioning that, by considering systems with unbroken time-reversal symmetry (class BDI) and longer-range couplings~\cite{niu_majorana_2012,vodola_kitaev_2014,viyuela_topological_2016,alecce_extended_2017}, e.g., where both the hopping and superconducting pairing terms extend to the nearest and next nearest neighbors, one can realize topologically nontrivial phases with higher winding numbers $|\nu|>1$.
This corresponds to the existence of a number $|\nu|$ of Majorana end modes at the boundary, which are mutually orthogonal and physically distinguishable~\cite{niu_majorana_2012,alecce_extended_2017}.

\subsection{The Majorana chain regime}

\begin{figure*}[t]
\includegraphics[width=\textwidth]{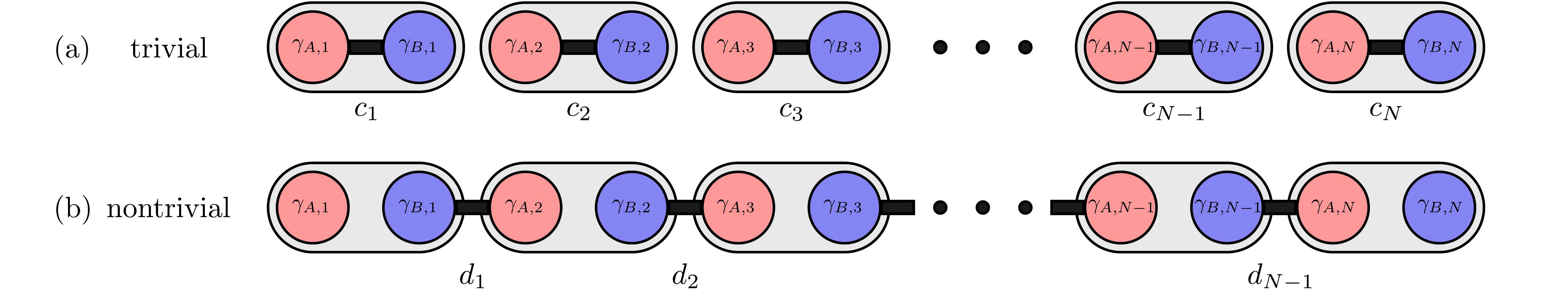}
\caption{
Trivial and nontrivial phases of the Kitaev model in the Majorana chain regime.
(a)
In the trivial phase, the Hamiltonian is diagonalized as a sum of local fermions $c_n$ on each lattice site.
These local operators can be further decomposed into two Majorana operators, which belong to the same site.
(b) In the nontrivial phase, the Hamiltonian is instead diagonalized as a sum of fermions $d_n$, which are the combination of two Majorana operators belonging to two neighboring lattice sites.
Consequently, the two ends of the chain exhibit two dangling Majorana end modes $\gamma_{A,1}$ and $\gamma_{B,N}$.
}
\label{fig:chain}
\end{figure*} 

The presence of Majorana modes in the topologically nontrivial phase $|\mu|<2t$ of the Kitaev model is guaranteed by the bulk-boundary correspondence~\cite{hatsugai_chern_1993,ryu_topological_2002,teo_topological_2010}, as already mentioned.
This result can be verified directly. 
Let us begin to rewrite the Hamiltonian in terms of Majorana operators, defined as
\begin{equation}\label{eq:Majorana}
	\gamma_{
	A
	,n}= c_n^\dag + c_n,
	\qquad
	\gamma_{
	B
	,n}=\ii c_n^\dag - \ii c_n.
\end{equation}
As one can verify directly from the definition above, Majorana operators are self-adjoint and idempotent operators that anticommute 
\begin{equation}\thinmuskip=1mu\medmuskip=1.5mu\thickmuskip=2mu
	\gamma_{\alpha,n}^\dag=\gamma_{\alpha,n},
	\quad
	\gamma_{\alpha,n}^2=1,
	\quad
	\left\{\gamma_{\alpha,n},\gamma_{\beta,m}\right\}=2\delta_{\alpha\beta}\delta_{nm},
\end{equation}
as in \cref{eq:Majorana-Properties-Intro}.
Indeed, it is not possible to define the number operator and the occupation number for an isolated Majorana mode since $\gamma_{\alpha,n}^\dag\gamma_{\alpha,n}=1$.
On the other hand, the fermion operators can be written in terms of the Majorana operators as
\begin{equation}\label{eq:MajoranaReverse}
	c_n=\frac12\left(\gamma_{
	A
	,n}+\ii \gamma_{
	B
,n}\right).
\end{equation}
Consequently, the Hamiltonian in \cref{eq:H-pwave}
in the gauge $\phi=0$
 becomes
\begin{align}
	\mathcal{H}=& 
	-\frac12\sum_{n=1}^N \mu\left(1
	+
	\ii \gamma_{A,n}\gamma_{B,n} \right)+
	\nonumber\\\label{eq:H-pwave-majo}&
	+
	\frac12 \sum_{n=1}^{N-1} \left[ \ii (\Delta+t) \gamma_{B,n}\gamma_{A,n+1} + \ii (\Delta-t) \gamma_{A,n}\gamma_{B,n+1} \right].
\end{align}

In the ``Majorana chain'' regime realized for $\Delta=t$, the Hamiltonian above reduces to 
\begin{align}
	\mathcal{H}= 
	-
	\frac12\sum_{n=1}^N & 
	\mu\left(1
	+
	\ii \gamma_{A,n}\gamma_{B,n} \right)+
	\nonumber\\\label{eq:H-pwave-Majoranachain-regime}
	+
	\frac12 \sum_{n=1}^{N-1} & 2\ii t \gamma_{B,n}\gamma_{A,n+1},
\end{align}
which describes a chain of $2N$ of Majorana modes localized on the two sublattices $A$ and $B$, where each Majorana mode is coupled to its nearest neighbors.
If $|\mu|>2t$, the couplings between Majorana modes on the same lattice sites dominate, while if $|\mu|<2t$, the couplings between Majorana modes on the neighboring lattice sites dominate.
Let us consider the case where only the Majorana modes on the same lattice sites are coupled, which is realized for $\Delta=t=0$ and $\mu\neq0$, which yields
\begin{align}
	\mathcal{H}=-\frac12\sum_{n=1}^N\mu(1
	+
	\ii \gamma_{A,n}\gamma_{B,n} )=
	-\sum_{n=1}^{N}\mu {c}^\dag_n {c}_n.
\end{align}
In this case, the energy spectrum has a bulk gap $\Delta E=|\mu|$ with quasiparticle excitations described by the operators $c^\dag_n$ corresponding to electrons localized on the same lattice site or, alternatively, to the superpositions of two Majorana operators localized on the same lattice site, as illustrated in \cref{fig:chain}(a).
This gapped phase is topologically trivial, having $|\mu|>2t$, as seen in \cref{sec:topology}.

Let us now consider the other, more interesting case, where only Majorana operators on \emph{neighboring} lattice sites are coupled, which is realized for $\Delta=t\neq0$ and $\mu=0$.
In this case, the Hamiltonian can be diagonalized by a set of Bogoliubov operators, which are the superposition of Majorana operators on neighboring sites
\begin{equation}\label{eq:SpecialFermion}
	d_n=\frac12(
\gamma_{B,n}+\ii \gamma_{A,n+1}
),
\end{equation}
for $n=1,\ldots,N-1$,
satisfying the fermion anticommutation relations
$\{d_n,d_m\}=0$ and $\{d_n,d_m^\dag\}=\delta_{nm}$.
Therefore the Hamiltonian becomes
\begin{equation}\label{eq:H-pwave-majo-spatial}
	\mathcal{H}
= 
	\sum_{n=1}^{N-1} \ii t \gamma_{B,n}\gamma_{A,n+1}=
	t \sum_{n=1}^{N-1} \left(2d^\dag_n d_n-1\right).
\end{equation}
In this case, the energy spectrum has a bulk gap $\Delta E=2t$ with quasiparticle excitations described by the operators $d_n$ corresponding to the superpositions of Majorana modes localized on two neighboring lattice sites, as illustrated in \cref{fig:chain}(b).
Indeed, the Majorana operators on the same lattice sites (which correspond to the original fermionic states of the Hamiltonian) are fully decoupled.
This phase is topologically nontrivial, having $|\mu|<2t$, as seen in \cref{sec:topology}.

Notice that we started with a lattice of $N$ fermionic operators $c_n$ and ended up with $N-1$ single-particle states described by the operators $d_n$, i.e., one less than the number of lattice sites.
Where is the missing fermionic mode?
The answer has to be sought in the Majorana operators $\gamma_{A,1}$ and $\gamma_{B,N}$, which are also missing from the Hamiltonian in \cref{eq:H-pwave-majo-spatial}.
These ``dangling'' Majorana modes must commute with the Hamiltonian $[\mathcal{H},\gamma_{A,1}]=[\mathcal{H},\gamma_{B,N}]=0$ and can be combined to form an additional fermionic operator
\begin{equation}\label{eq:MajoranaEdgeStates}
	d_N=\frac12(\gamma_{B,N}
	-
	\ii\gamma_{A,1}),
\end{equation}
which satisfies the usual fermion anticommutation relations $\{d_N,d_N\}=0$ and $\{d_N,d_N^\dag\}=1$.

The state defined in the equation above is the Majorana bound state:
It is a nonlocal single-particle fermion which is the superposition of two Majorana end modes localized at the two opposite ends of the chain, as shown in \cref{fig:chain}(b). 
The Majorana bound state does not appear in the Hamiltonian and thus has zero energy.
However, there is already one state with zero energy, i.e., the groundstate given by the vacuum of the quasiparticle excitations $\ket{0}$, i.e., the state annihilated by any fermionic operator $d_n\ket{0}=0$ for $n=1,\dots,N-1$.
It is now natural to define a number operator for the Majorana bound state as
\begin{equation}\label{eq:Majorananumberoperator}
	n_\mathrm{M}=n_N=d_N^\dag d_N=\frac12(1+
	\ii\gamma_{A,1}\gamma_{B,N}
	).
\end{equation}
The two groundstates are labeled by a different occupation number of the Majorana bound state (i.e., by the eigenvalue of the number operator).
The vacuum state $\ket{0}$ and the state $\ket{1}$ have eigenvalues $n_\mathrm{M}=0$ and $1$, respectively,
\begin{equation}\label{eq:MajorananumberoperatorAndDegenerateGroundstate}
	n_\mathrm{M}\ket{0}=0, \qquad n_\mathrm{M}\ket{1}=\ket{1}.
\end{equation}
Moreover, the two degenerate groundstates have a different fermion number $F=\sum_n c^\dag_n c_n$ and different fermion parity~\cite{kitaev_unpaired_2001}, which is defined by 
\begin{align}
	P=(-1)^F
	=&\prod_n 
	(1-2c^\dag_n c_n)
	=\prod_n 
	(-\ii\gamma_{A,n}\gamma_{B,n})
	=
	\nonumber\\
	=&\prod_n 
	(1-2d^\dag_n d_n),
	\label{eq:Parity}
\end{align}
with $P P^\dag=P^2=1$.
Although the superconducting pairing term in the BCS Hamiltonian does not conserve the total number of fermions, the fermion parity, i.e., the number of fermions modulo 2, is conserved.
Thus, the parity operator commutes with the Hamiltonian, and thus any eigenstate has a well-defined fermion parity given by the eigenvalues $p=\pm1$ of the parity operator.
Since creating or annihilating a fermion correspond to changing the fermion parity, the two degenerate groundstates $\ket{0}$ and $\ket{1}$ have different fermion parity.
These groundstates are the vacuum of the Bogoliubov quasiparticles $\ket{0}$ and the nonlocal single-particle state $\ket{1}$, i.e., the Majorana bound state, which is the superposition of two Majorana end modes localized at the opposite ends of the chain.
The presence of two degenerate groundstates with opposite fermion parity is an example of quantum mechanical supersymmetry, i.e., the symmetry between bosonic and fermionic many-body states corresponding to even and odd fermion parities~\cite{hsieh_all-majorana_2016,huang_supersymmetry_2017,marra_1d-majorana_2022,marra_dispersive_2022}.

\subsection{Majorana bound states\label{sec:MBS}}

One may argue that the appearance of localized Majorana modes is due to the very special choice of the parameters $\mu=0$ and $\Delta=t$.
However, Majorana modes persist in a much larger part of the parameter space.
In other words, the Hamiltonian can be continuously deformed away from $\mu=0$ and $\Delta=t$ without destroying the Majorana modes, as long as the particle-hole gap does not close, i.e., as long as the systems remain in the topologically nontrivial phase.
Hence, as a direct consequence of the bulk-boundary correspondence~\cite{hatsugai_chern_1993,ryu_topological_2002,teo_topological_2010}, the presence of the Majorana modes localized at the opposite ends of the chain is guaranteed as long as $|\mu|<2t$, being indeed a characteristic and distinctive feature of the topologically nontrivial phase in \cref{fig:phasespace}.
More generally, for spatially-dependent chemical potentials $\mu_n$, Majorana modes localize at the boundary between the topologically trivial $|\mu_n|>2t$ and nontrivial phases $|\mu_n|<2t$.

The localized Majorana zero modes can be found by solving the equation of motion and imposing open boundary conditions~\cite{kitaev_unpaired_2001,semenoff_stretching_2006}, analogously to the continuous case described in \cref{sec:pwavecontinuous}, as worked out in detail in \cref{sec:endmodes}.
By doing so, one finds that in the topologically trivial phase $|\mu|>2t$, there are no zero-energy modes that can satisfy the boundary conditions.
In the nontrivial phase $|\mu|<2t$ instead, there exists a pair of zero-energy Majorana end modes satisfying the open boundary conditions.
In the limit of large system sizes $N\to\infty$, their wavefunctions are given by
\begin{equation}\label{eq:MESgeneralNambu}
	\bm\psi_n^A=\varphi_{n}\begin{bmatrix} 1 \\ 1 \\ \end{bmatrix},
	\quad
	\bm\psi_n^B=\varphi_{N+1-n}\begin{bmatrix} \ii \\ -\ii \\ \end{bmatrix},
\end{equation}
in the basis of the particle-hole Nambu spinors $[c_n^\dag,c_n]$, and where
\begin{equation}\label{eq:MESgeneralNambu2}
	\varphi_n=\frac1Z\left(z_+^n - z_-^n\right)
	=\frac1Z\left(\ee^{-\lambda_+ n} - \ee^{-\lambda_- n}\right),
\end{equation}
with $|z_\pm|<1$, $\lambda_\pm=-\log{z_\pm}=-\log|z_\pm|-\ii\arg z_\pm$, and
\begin{subequations}\begin{align}
	\label{eq:zpmmain}
	z_\pm=&\frac{-\mu\pm\sqrt{\mu^2-4(t^2-\Delta^2)}}{2(t+\Delta)},
	\\
	\label{eq:MajoranaNormalizationInfty}
	Z=&\sqrt{\frac{t}{\Delta}\frac{\mu^2-4(t^2-\Delta^2)}{\left(4t^2-\mu^2\right)}}.
\end{align}\end{subequations}
The wavefunctions of the Majorana modes are invariant under particle-hole symmetry, i.e., they are eigenvalues of the particle-hole symmetry operator $\mathcal{C}\bm\psi_n^\alpha=\bm\psi_n^\alpha$ and are always real (up to a constant phase factor).
This is a direct consequence of the fact that the Hamiltonian in \cref{eq:H-pwave-BdG} is purely real for $\phi=0$.

The limiting case $\mu^2=4(t^2-\Delta^2)$ can be obtained as the asymptotic limit of \cref{eq:MESgeneralNambu,eq:MajoranaNormalizationInfty} for $\mu\to\pm\sqrt{4(t^2-\Delta^2)}$.
In particular, for $\mu=0$ and $\Delta=t$, one recovers the case described by the Majorana chain regime in the nontrivial phase, with Majorana nodes perfectly localized at the two opposite lattice sites of the chain, as in \cref{eq:H-pwave-majo-spatial}.
Finally, at the topological phase transition $\mu=\pm2t$, which yields $|z_-|=1$ and $|z_+|=1$, respectively, the zero-energy modes are not localized anymore and become extended along the whole chain.
Physically, this corresponds to the fusion of the two Majorana end modes into a single fermionic mode with a finite energy.

In terms of the Majorana operators, the zero-energy Majorana end modes in \cref{eq:MESgeneralNambu} can be written as 
\begin{equation}\label{eq:MESgeneral} 
	\bm\gamma_A=\sum_n\varphi_{n}\gamma_{A,n},
	\qquad
	\bm\gamma_B=\sum_n\varphi_{N+1-n}\gamma_{B,n},
\end{equation}
One can verify that $\bm\gamma_A$ and $\bm\gamma_B$ are Majorana operators since they are self-adjoint, idempotent operators satisfying the fermionic anticommutation rules, as in \cref{eq:Majorana-Properties-Intro}.
The superposition of these two Majorana modes forms a fermionic single-particle state, the Majorana bound state, given by
\begin{equation}\label{eq:MBSgeneral}
	d_{\mathrm{M}}=\frac12\left(\bm\gamma_B
	-
	\ii \bm\gamma_A\right),
\end{equation}
which again satisfies the usual fermionic anticommutation relations $\{d_{\mathrm{M}},d_{\mathrm{M}}\}=0$ and $\{d_{\mathrm{M}},d_{\mathrm{M}}^\dag\}=1$.
The equation above generalizes \cref{eq:MajoranaEdgeStates} to the case where the Majorana modes are not perfectly localized on the opposite lattice sites of the chain but have a finite localization length.

\begin{figure}[t]
\includegraphics[width=\columnwidth]{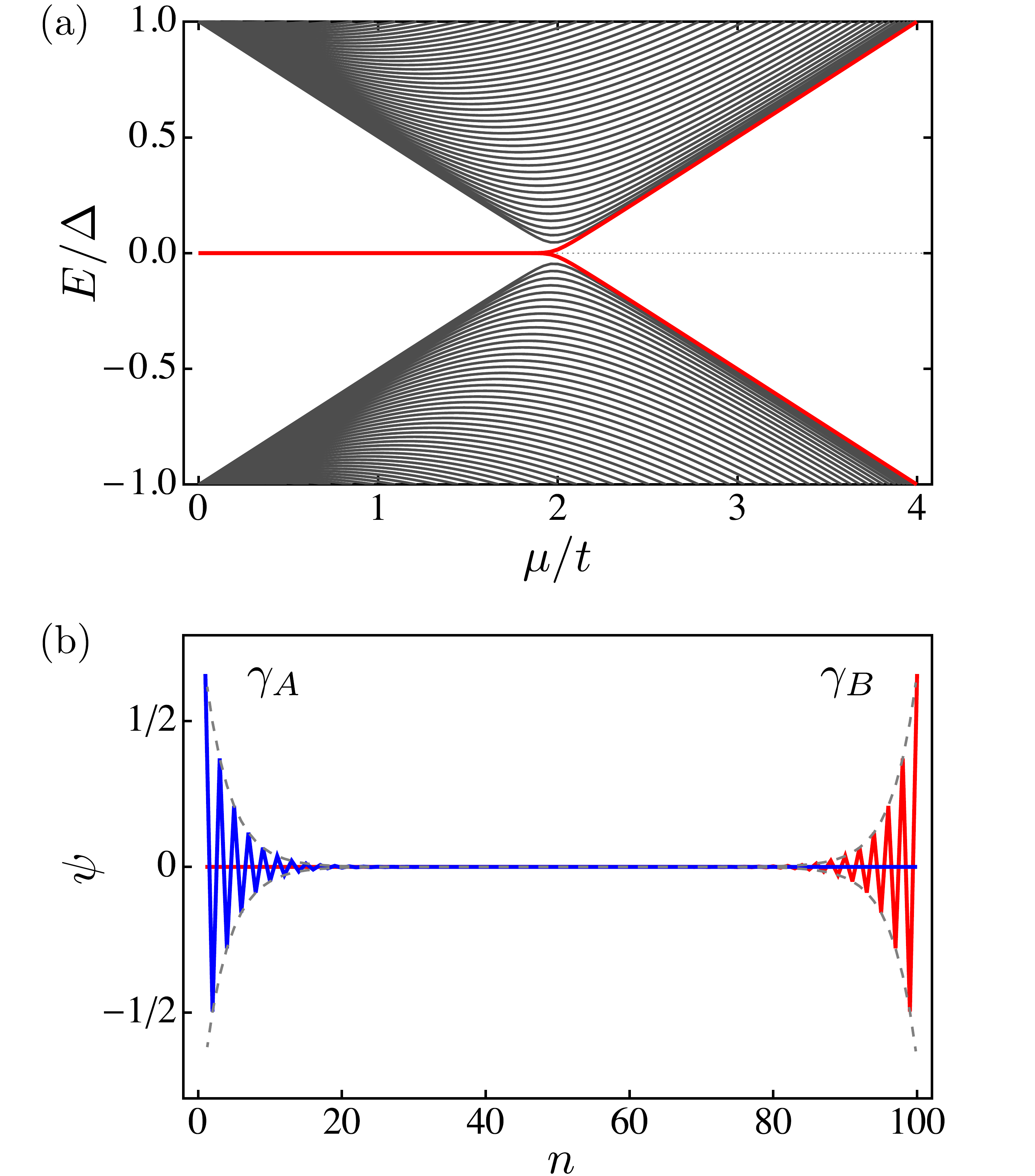}
\caption{
Topological phase transition and wavefunction of the Majorana bound state calculated numerically for $\Delta=t$ for a Kitaev model with $N=100$ lattice sites.
(a) Energy spectra as a function of the chemical potential, showing the topological phase transition with the closing of the particle-hole gap at $\mu=2t$, with the lowest energy level highlighted in color.
(b) Wavefunction of the Majorana bound state in the nontrivial phase for $\mu=1.5t$.
The oscillating curves correspond to the wavefunctions of the two Majorana modes localized at the opposite ends of the chain (notice that the wavefunction is real), while the dashed curves correspond to $\pm|\bm\psi|$.
}
\label{fig:Majo}
\end{figure}

\Cref{fig:Majo} shows the topological phase transition and the Majorana bound state calculated for $|\mu|<2t$.
The topological phase transition at $\mu=2t$ separates the trivial phase from the nontrivial phase with Majorana end modes.
As shown in \cref{fig:Majo}(b), the Majorana modes $\bm\gamma_A$ and $\bm\gamma_B$ localize respectively at the left and right ends of the chain and decay exponentially towards the center as $|\bm\psi_n|\propto\ee^{-n/\xi_\mathrm{M}}$, with the localization length $\xi_\mathrm{M}$ given by
	\begin{equation}\label{eq:MajoranaLocalizationLength}
	\xi_\mathrm{M}=\max\left(\frac1{|\log|z_+||},\frac1{|\log|z_-||}\right).
\end{equation}
In the Majorana chain regime $\Delta=t$, one has $z_+=0$ or $z_-=0$, which correspond to perfectly localized Majorana end modes with $\xi_\mathrm{M}=0$.

Notice that the Majorana bound state in \cref{eq:MBSgeneral} is only an \emph{approximate} eigenstate of the Hamiltonian, which becomes exact only in the infinite-size limit $N\to\infty$.
In fact, the wavefunctions in \cref{eq:MESgeneralNambu} satisfy the boundary conditions $\bm{\psi}_{0}=\bm{\psi}_{N+1}=0$ only up to exponentially small terms for a generic choice of the parameters:
The wavefunctions of $\bm\gamma_A$ and $\bm\gamma_B$ evaluated at $n=N+1$ and $n=0$ do not vanish, being $\propto\ee^{-(N+1)/\xi_\mathrm{M}}$.
Indeed, the energy of the Majorana bound state in a finite chain decays exponentially as $E_\mathrm{M}\propto\ee^{-N/\xi_\mathrm{M}}$ 
with damped oscillations~\cite{hegde_majorana_2016,zeng_analytical_2019,leumer_exact_2020} due to the hybridization of the two Majorana modes at the opposite ends of the chain [see Refs.~\onlinecite{zeng_analytical_2019,leumer_exact_2020}].
The hybridization and energy splitting of the Majorana bound state can be described by an effective Hamiltonian~\cite{kitaev_unpaired_2001}
\begin{equation}\label{eq:HeffW0}
	\mathcal{H}_\mathrm{eff}=\frac\ii2 E_\mathrm{M}
	\bm\gamma_A \bm\gamma_B
	=E_\mathrm{M} \left(n_\mathrm{M}-\frac12\right),
\end{equation}
where $n_\mathrm{M}=d_\mathrm{M}^\dag d_\mathrm{M}$ is the number operator of the Majorana bound state defined in \cref{eq:MBSgeneral}.
The energy splitting $E_\mathrm{M}$ between the Majorana bound state and the vacuum groundstate is zero in the limit of infinite system size.
The crucial point is that the energy splitting scales \emph{exponentially} with the linear dimension.
Consequently, the energy splitting becomes negligible when the linear dimension is larger than the localization length $N\gg\xi_\mathrm{M}$. 
As mentioned in the introduction, this exponential scaling is a distinctive feature of topologically nontrivial states of matter.
Analogously, in the quantum Hall effect~\cite{klitzing_new-method_1980,laughlin_quantized_1981,thouless_quantized_1982}, the conductance is exactly quantized up to exponentially small corrections in the linear dimensions~\cite{niu_quantum_1987,thouless_topological_1998_s73}.
In the presence of 
small perturbations,
e.g., weak disorder, 
the energy splitting remains exponentially small in the system size
and smaller than the particle-hole gap $E_\mathrm{M}\ll\Delta E$,
as long as the particle-hole gap remains open.
The topologically nontrivial phase of the Kitaev model is also robust against weak or moderately strong repulsive and attractive interactions~\cite{hassler_strongly_2012,thomale_tunneling_2013,katsura_exact_2015,gergs_topological_2016}.
See Refs.~\onlinecite{kitaev_anyons_2006,fidkowski_effects_2010,fidkowski_topological_2011} for a more extended discussion of the role and impact of many-body interactions.

\section{Topological quantum computation\label{sec:TQC}}

\subsection{Exchange statistics}

A general statement in quantum mechanics is that the many-body wavefunction describing a collection of identical particles must have a well-defined symmetry under the exchange or permutation of particles.
In particular, bosons are symmetric under exchange, whereas fermions are antisymmetric.
The exchange symmetry determines the particle statistics and is related to the particle spin by the spin-statistics theorem.
Moreover, permutations of fermions or bosons are abelian (i.e., commutative) operations, in the sense that the outcome of a series of successive permutations does not depend on the order in which these permutations are performed.
However, these statements hold true only for \emph{pointlike} particles living in our ordinary 3D world.

In a 2D world, or in a system that is effectively 2D, such as 2D topological superconductors, quasiparticle excitations may have an exchange statistics that differ from those of fermions or bosons~\cite{nayak_non-abelian_2008,stern_anyons_2008,lahtinen_a-short_2017}.
To understand the difference between the 3D and 2D cases, consider two consecutive adiabatic exchanges of two identical pointlike particles.
This double exchange $U^2$ can be realized by moving one particle around the other on a closed path that does not intersect the second particle.
In 3D, this closed path can be continuously deformed to a point, i.e., to the case where the particle does not move at all:
The double exchange is topologically equivalent to no exchange and must coincide with the identity $U^2=1$. 
This leaves only two possibilities, i.e., $U=1$ and $U=-1$ realized respectively by bosons and fermions.
The exchange of identical bosons or fermions is equivalent to multiplying the many-body state by a phase factor equal to $\pm1$, respectively.
In 2D, however, the closed path cannot be continuously deformed to a point without intersecting the second particle.
Therefore, the unitary exchange $U$ is not constrained anymore as in the previous case, and more possibilities arise: 
Pointlike particles in 2D may be bosons, fermions, as well as abelian or nonabelian \emph{anyons}~\cite{leinaas_on-the-theory_1977,wilczek_quantum_1982,moore_nonabelions_1991}.
The exchange of two abelian anyons is equivalent to the multiplication of the many-body state by a phase factor $\ee^{\ii\phi}$, where $\phi$ is \emph{any} rational multiple of $2\pi$.
On the other hand, the exchange of nonabelian anyons corresponds to a multiplication of the many-body state by a unitary matrix $U$.
This necessarily requires the presence of a degenerate many-body groundstate where the unitary matrix $U$ describes unitary evolutions within the groundstate manifold.
Since matrix multiplication is nonabelian (i.e., noncommutative), the outcome of two successive exchange operations depends on the order in which the operations are performed~\cite{moore_nonabelions_1991,nayak_2nquasihole_1996,read_paired_2000,ivanov_non-abelian_2001,stern_geometric_2004}.
Exchange operations of nonabelian anyons are described by braid groups~\cite{nayak_non-abelian_2008}. 
Contrarily to ordinary electronic excitations in condensed matter, Majorana modes bounded to the vortex cores of a topological superconductor exhibit nonabelian exchange statistics~\cite{read_paired_2000,ivanov_non-abelian_2001,kitaev_anyons_2006}.
The nonabelian exchange statistics is a general property of Majorana modes which holds true even in 3D, as shown in Ref.~\onlinecite{teo_majorana_2010}.
Essentially, this is due to the fact that the simple argument discussed before, which rules out the possibility of nonabelian exchange of pointlike particles in 3D, breaks down in the case of Majorana modes due to the intrinsic spatial orientation corresponding to the texture of the superconducting order parameter~\cite{teo_majorana_2010,freedman_projective_2011,freedman_weakly_2011}.

In 1D, however, there is a catch:
there is simply no room in 1D to exchange the position of two identical particles moving one around the other via a continuous transformation.
Hence, it is impossible to meaningfully define exchange operations since there is no way to exchange the positions of two Majorana modes living at the two opposite ends of a quantum wire without overlapping their wavefunctions and thus splitting their energy. 
This is essentially for the same reason why there is no way for two trains to pass each other when moving in opposite directions on the same track or exchange the position of the head and the tail of a train using one single track. 
However, trains avoid collisions by traveling on 1D tracks embedded into 2D (or even 3D) networks.
Therefore, it is possible to assemble a collection of 1D wires into a network that can be regarded as \emph{effectively} higher-dimensional.
In this network, the exchange operations of Majorana modes localized at the ends of the 1D wires are well-defined~\cite{alicea_non-abelian_2011,clarke_majorana_2011,halperin_adiabatic_2012}.

\subsection{Nonabelian statistics of Majorana modes}

Majorana modes bounded to a defect, domain wall, or at the ends of a quantum wire exhibit nonabelian exchange statistics.
The nonabelian nature of the exchange operations of Majorana modes, regardless of spatial dimensions, follows directly from fermion-parity conservation~\cite{clarke_majorana_2011,halperin_adiabatic_2012,sato_majorana_2016}.
Let us consider the adiabatic exchange 
$U$ 
of two zero-energy Majorana modes $\gamma_1$ and $\gamma_2$.
In the presence of a finite energy gap and in the adiabatic limit, these two modes remain Majorana modes during unitary evolutions 
$\gamma_{n}=U\gamma_{m}U^\dag$. 
Since the energy spectra of topological superconductors are gapped and fermion-parity is conserved, the unitary evolution 
$U$ 
of the many-body state of the two Majorana modes is constrained within the groundstate manifold with fixed fermion-parity.
This mandates that the only possible 
exchange process is given by 
\begin{equation}
	U \gamma_1 U^\dag=\chi_2 \gamma_2, 
	\qquad
	U \gamma_2 U^\dag=\chi_1 \gamma_1, 
\end{equation}
where $\chi_1$ and $\chi_2$ are phase factors.
Since Majorana operators are idempotent [see \cref{eq:Majorana-Properties-Intro}], it follows that $(U\gamma_n U^\dag)^2=\chi_n^2=1$, which yields $\chi_1=\pm1$ and $\chi_2=\pm1$.
Now consider the adiabatic evolution of the number operator of the Majorana bound state $n_\mathrm{M}=\frac12(1+\ii
\gamma_1\gamma_2)
$ [cf. \cref{eq:Majorananumberoperator}], which yields
\begin{equation}
	U n_\mathrm{M}U^\dag=
	\frac12(1+\ii U\gamma_1\gamma_2U^\dag)=
	\frac12(1-\ii \chi_1 \chi_2 \gamma_1\gamma_2).
\end{equation}
Now, since fermion-parity is conserved, the occupation number of the Majorana bound state cannot change during the exchange process, i.e., $U n_\mathrm{M}U^\dag=n_\mathrm{M}$, which mandates $\chi_1\chi_2=-1$.
This gives two possibilities, either $\chi=\chi_2=-\chi_1=1$ or $-1$.
The choice between $\chi=1$ and $\chi=-1$ is determined by the Hamiltonian governing the unitary evolution~\cite{clarke_majorana_2011}.

\begin{figure}[t]
\includegraphics[width=\columnwidth]{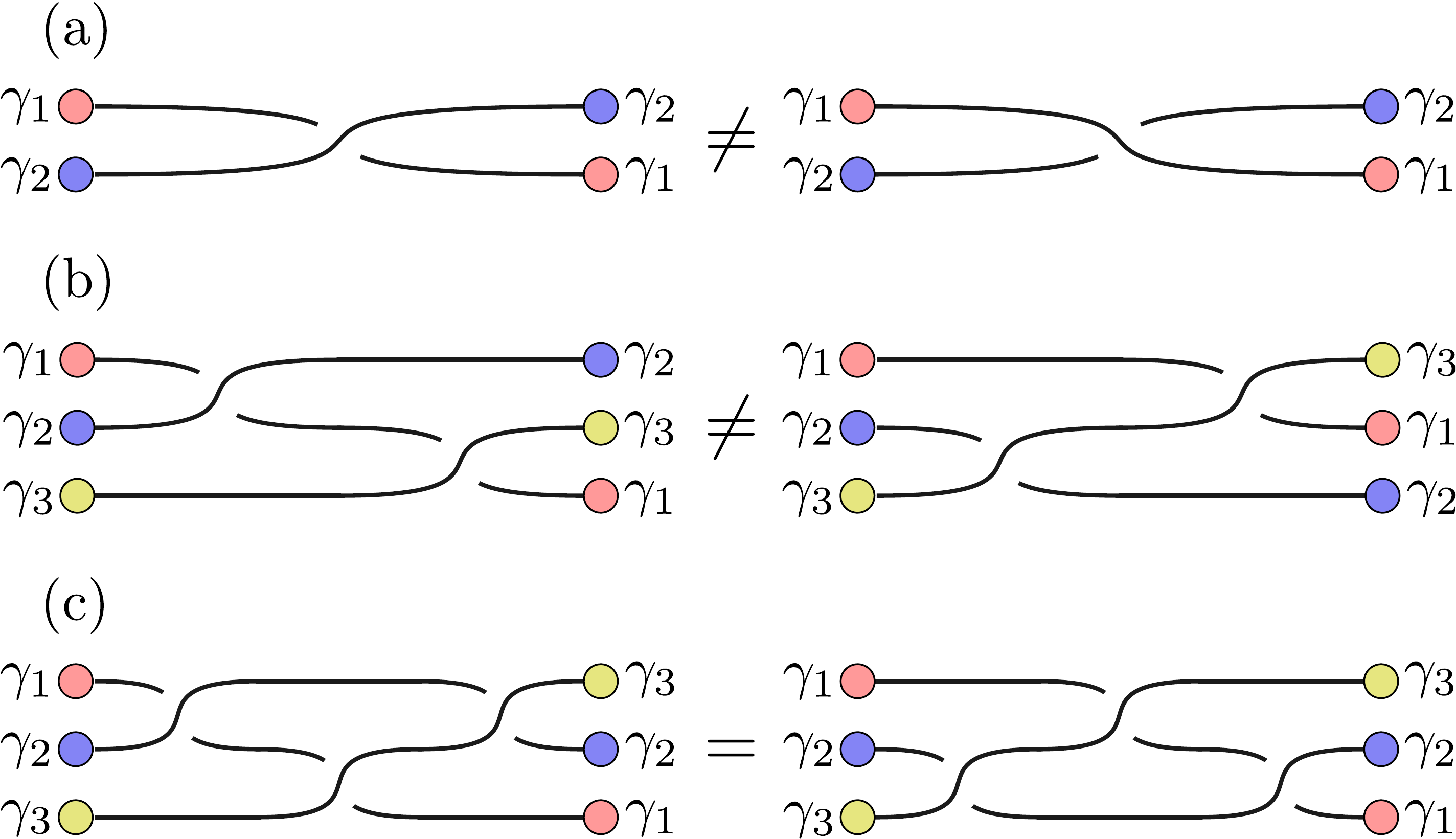}\\
\caption{
Braid operations of Majorana modes described by nonabelian braid operations.
(a) Diagrammatic representations of the braid operation $U_{12}$ and its inverse $U_{21}$.
(b) The composition of braid operations $U_{23}U_{12}$ and $U_{12}U_{23}$. 
The braid operations $U_{12}$ and $U_{23}$ do not commute, cf.~\cref{eq:nonabelian}.
(c) The composition of braid operations $U_{12}$ and $U_{23}$ satisfying the Yang-Baxter equation 
$U_{12}U_{23}U_{12}=U_{23}U_{12}U_{23}$, see \cref{eq:YangBaxter}.
}
\label{fig:braids}
\end{figure}

One can generalize the exchange operation to a system containing several Majorana end modes.
The exchange of two Majorana end modes $\gamma_n$ and $\gamma_m$ is described by 
\begin{subequations}\label{eq:exchangechi}\begin{align}
	\gamma_{n}&\to-\chi\gamma_{m}, 
	\\
	\gamma_{m}&\to+\chi\gamma_{n}, 
\end{align}\end{subequations}
with $\chi=\pm1$ depending on the details of the Hamiltonian.
For $\chi=1$, this exchange corresponds to the unitary transformation $\gamma_p\to U_{nm}\gamma_{p}U_{nm}^\dag$, called the braid operator, defined by
\begin{equation}\label{eq:braiding}
	U_{nm}=
	\frac1{\sqrt2}\left(1+\gamma_n\gamma_m\right)=
	\exp{\left(\frac\pi4\gamma_n\gamma_m\right)},
\end{equation}
up to a complex phase.
The second equality in the equation above is a special case of the identity $\exp{(\alpha\gamma_n\gamma_m)}=\cos{\alpha}+\sin{\alpha}\,\gamma_n\gamma_m$, which can be obtained by expanding the exponentiation as a Taylor series and using the identity $(\gamma_n\gamma_m)^2=-1$.
Braids can be represented in a diagrammatic form, as illustrated in \cref{fig:braids}.
The inverse transformation is the braid operator $U_{nm}^\dag=U_{mn}$ given by
\begin{equation}\label{eq:braidinginv}
	U_{mn}=
	\frac1{\sqrt2}\left(1-\gamma_n\gamma_m\right)=
	\exp{\left(-\frac\pi4 \gamma_n \gamma_m\right)},
\end{equation}
which corresponds to the exchange in \cref{eq:exchangechi} with $\chi=-1$.
Hence, there are two inequivalent ways of braiding the two Majorana modes $U_{mn}\neq U_{nm}$.
Intuitively, braid operations are a generalization of permutations to the case where the exchange of two objects can be performed in several inequivalent ways.
\Cref{fig:braids}(a) represents the braid operator $U_{12}$ and its inverse $U_{21}=U_{12}^\dag$.
Notice that braiding two Majorana modes $\gamma_n$ and $\gamma_m$ is not equivalent to a simple relabeling of their indices $\gamma_n\leftrightarrow\gamma_m$.
In fact, by applying two times the unitary transformation in \cref{eq:braiding}, one does not recover the initial configuration since one can immediately verify that $U_{nm}^2=\gamma_{n}\gamma_{m}$, which gives
\begin{subequations}\begin{align}
	\gamma_{n}\to -\gamma_{m} & \to -\gamma_{n},
	\\
	\gamma_{m}\to +\gamma_{n} & \to -\gamma_{m}.
\end{align}\end{subequations}
For a given set of $2N$ Majorana modes, the braid operators on neighboring Majorana modes $U_{n,n+1}$ generate all braid operators $U_{nm}$, their inverses $U_{mn}$, and all their possible combinations, including the identity operator $\id$.
In other words, the braid operators $U_{n,n+1}$ are the generators of the unitary group describing all unitary evolutions of the $2N$ Majorana modes.

The braiding of Majorana modes is nonabelian, i.e., braid operators do not commute:
The result of the composition of two or more braids may depend on the order they are performed.
Braid operators commute with their inverse, i.e., $[U_{nm},U_{mn}]=0$.
Also, braids of a pair of Majorana modes $\gamma_n$, $\gamma_m$ and another pair and $\gamma_p$, $\gamma_q$ (with $n$, $m$, $p$, and $q$ all distinct) commute $[U_{nm},U_{pq}]=0$.
However, braiding two Majorana modes $\gamma_n$ and $\gamma_m$, and then braiding one of them with a third one $\gamma_p$, is not equivalent to the reverse process.
In particular, one can verify that
\begin{equation}\label{eq:nonabelian}
	[U_{nm},U_{mp}]=\gamma_n\gamma_p,
\end{equation}
with $n$, $m$, and $p$ all distinct.
This equation expresses the fact that the exchange statistics of Majorana modes is nonabelian (noncommutative):
The final outcome depends on the order in which the braiding operations are performed, i.e., $U_{nm}U_{mp}\neq U_{mp}U_{nm}$.
The operations $U_{12}$ and $U_{23}$ and their combinations $U_{23}U_{12}$ and $U_{12}U_{23}$ are shown diagrammatically in \cref{fig:braids}(b).
The noncommutativity of the operators $U_{12}$ and $U_{23}$ is expressed by the fact that the two diagrams are not equivalent.
Moreover, braid operators satisfy
\begin{align}
	U_{nm}U_{mp}U_{nm} 
	&= U_{mp}U_{nm}U_{mp}
	=\nonumber\\
	&=\frac1{\sqrt2}(\gamma_n\gamma_m+\gamma_m\gamma_p),
	\label{eq:YangBaxter}
\end{align}
with $n$, $m$, and $p$ all distinct.
The first line of this equation is known as the Yang-Baxter equation, a ubiquitous and fundamental equation in mathematical physics, which appears in the study of integrable models in statistical mechanics and quantum field theories,
conformal field theories, quantum groups, knot theories, and braided categories~\cite{jimbo_introduction_1989,evans_subfactors_1993,etingof_lectures_1998}.
This equation is represented diagrammatically in \cref{fig:braids}(c).

For a more thorough overview of the nonabelian statistics of Majorana modes and their application to topological quantum computation, see Refs.~\onlinecite{nayak_non-abelian_2008,pachos_introduction_2012,das-sarma_majorana_2015,roy_topological_2017,lahtinen_a-short_2017,stanescu_introduction_2017,beenakker_search_2020,oreg_majorana_2020}.

\subsection{Braiding in trijunctions}

\begin{figure*}[t]
\includegraphics[width=\textwidth]{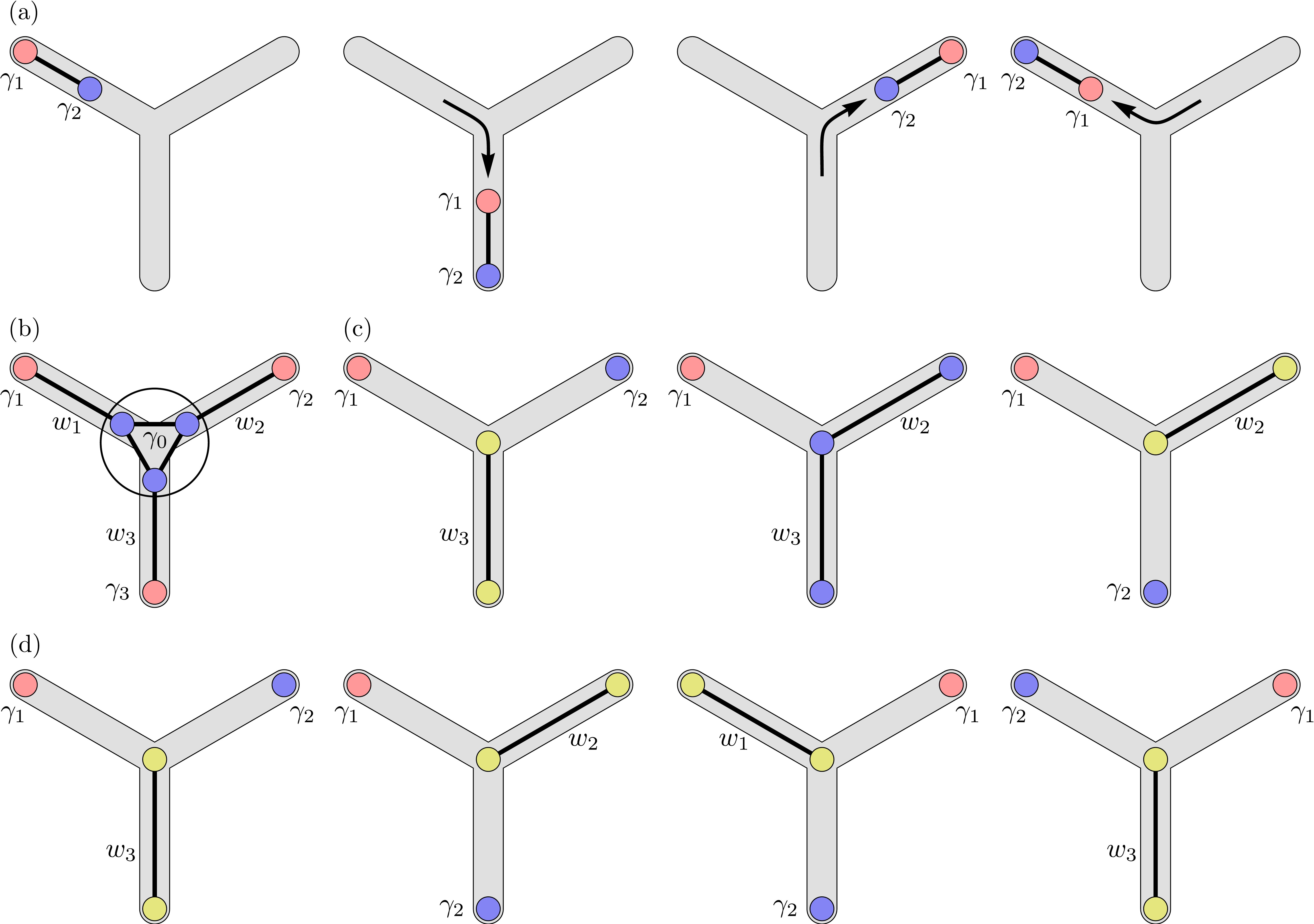}
\caption{
Braiding of Majorana modes in a trijunction.
(a) 
Braid operation of two Majorana modes $\gamma_1$ and $\gamma_2$ (left to right) obtained by physically moving the domain walls between trivial and nontrivial phases.
(b) 
A trijunction formed by bringing together three separated segments in the nontrivial phase, resulting in three Majorana modes $\gamma_n$ with $n=1,2,3$ localized on the outer ends and another Majorana mode $\gamma_0$ at the center, which is the result of the hybridization of three Majorana modes on the inner ends.
(c)
Moving the Majorana mode $\gamma_2$ via manipulating the couplings between the outer ends and the center of the junction (left to right).
(d) 
Braid operation obtained by moving two Majorana modes clockwise around the junction in a three-step process (left to right).
}
\label{fig:braiding}
\end{figure*}

A simple braiding protocol can be engineered by moving around Majorana modes in a 2D network of Majorana nanowires, e.g., in a junction composed of three branches that share a common end~\cite{alicea_non-abelian_2011,clarke_majorana_2011,van-heck_coulomb-assisted_2012,karzig_shortcuts_2015} (trijunction) shown in \cref{fig:braiding}.
The key idea is to manipulate the chemical potential $\mu$ (or another microscopic parameter) in order to slowly shift the boundaries of the trivial and nontrivial regions and, by doing that, adiabatically move the Majorana modes within the junction.
Let us go through the process illustrated in \cref{fig:braiding}(a) from left to right.
We start from a configuration where one segment of the left branch is in the nontrivial phase $|\mu|<2t$, and the rest of the junction is trivial $|\mu|>2t$, with two Majorana modes $\gamma_1$ and $\gamma_2$ localized at the boundaries of the nontrivial segment of the left branch.
We then slowly move the domain walls between the trivial and nontrivial phases, such that the two Majorana modes pass (one after the other) through the center of the junction to reach the lower branch.
Likewise, the two Majorana modes may pass again through the center of the junction to reach the right branch and once again to the left branch.
By doing so, the positions of the two Majorana modes $\gamma_1$ and $\gamma_2$ are exchanged following a clockwise path on the 2D plane of the trijunction.
This corresponds to the adiabatic exchange described by the braid operator $U_{12}$ or $U_{21}$ in \cref{eq:braiding,eq:braidinginv}, depending on the Hamiltonian describing the trijunction but not on the details of the exchange process~\cite{clarke_majorana_2011}.
The reverse process (from right to left), where the Majorana modes are exchanged following an anticlockwise path on the 2D plane, corresponds to the inverse braiding operator, being either $U_{12}^{-1}=U_{21}$ or $U_{21}^{-1}=U_{12}$.
Braiding protocols can be implemented by moving the domain walls between trivial and nontrivial phases (where Majorana modes localize) by tuning the chemical potential, superconducting pairing, or other parameters~\cite{alicea_non-abelian_2011,clarke_majorana_2011,romito_manipulating_2012,halperin_adiabatic_2012,karzig_boosting_2013}, in particular using arrays of electric gates~\cite{alicea_non-abelian_2011,scheurer_nonadiabatic_2013,bauer_dynamics_2018}, nanomagnets~\cite{jardine_integrating_2021,marra_1d-majorana_2022}, tunable spin-valves~\cite{zhou_tunable_2019}, magnetic tunnel junctions~\cite{fatin_wireless_2016,matos-abiague_tunable_2017}, or magnetic stripes~\cite{mohanta_electrical_2019}, or supercurrents~\cite{romito_manipulating_2012}.

Perhaps counterintuitively, braiding can also be performed by exchanging Majorana modes without physically moving them~\cite{flensberg_non-abelian_2011,sau_controlling_2011,van-heck_coulomb-assisted_2012,hyart_flux-controlled_2013,karzig_shortcuts_2015}.
Let us consider a trijunction where all three branches are in the nontrivial phase~\cite{karzig_shortcuts_2015}.
To understand where Majorana modes localize, imagine decoupling the three branches of the trijunction:
This results in three separated segments in the nontrivial phase, arranged as in \cref{fig:braiding}(b), with Majorana modes localized on the inner and outer ends.
When the coupling between these three segments is restored to form the trijunction, the hybridization of the three inner Majorana modes forms a fermion at finite energy and a remaining unpaired Majorana mode at zero energy~\cite{alicea_non-abelian_2011}.
Hence, three Majorana zero-energy modes $\gamma_n$ with $n=1,2,3$ are localized on the outer ends of the trijunction, and another one $\gamma_0$ at the center~\cite{alicea_non-abelian_2011,karzig_shortcuts_2015}.
If the length $L$ of the branches is finite, this system is described by the effective Hamiltonian~\cite{sau_controlling_2011,van-heck_coulomb-assisted_2012,karzig_shortcuts_2015}
\begin{equation}\label{eq:trijunction}
	H=\ii\sum_{n=1}^3 w_n\gamma_0\gamma_n=\ii\gamma_0\sum_{n=1}^3 w_n\gamma_n,
\end{equation}
where $w_n\propto\ee^{-L/\xi_\mathrm{M}}$ are the couplings between the outer Majorana modes $\gamma_n$ and the central mode $\gamma_0$, which depend on the mutual overlap of their wavefunctions.
These couplings can be controlled by changing the lengths of the branches or the Majorana localization lengths by varying, e.g., the chemical potential or other parameters.
An essential requirement is that these coupling can be effectively turned off for $L/\xi_\mathrm{M}\to\infty$ up to exponentially small corrections.
To move the Majorana modes on the trijunction, consider the process illustrated in \cref{fig:braiding}(c).
We start with a configuration where $w_1=w_2=0$ and $w_3>0$, where the Majorana modes $\gamma_1$ and $\gamma_2$ in the upper left and right branches are fully decoupled and have zero energy, while the other two modes are hybridized (with a finite energy splitting).
Next, we increase the coupling $w_2$ to a finite value, leaving the mode $\gamma_1$ unaffected ($w_1=0$).
One can verify that in this configuration, there is a linear combination of the Majorana modes on the right and lower branches which commutes with the Hamiltonian and thus remains pinned at zero energy for all $w_2,w_3>0$.
Therefore the mode $\gamma_2$ becomes a superposition of the two Majorana modes on the right and lower branches.
Then, we reduce the coupling $w_3$ to zero, such that $w_1=w_3=0$ and $w_2>0$.
By doing this, the mode $\gamma_2$ localizes on the lower branch.
In this configuration, the modes $\gamma_1$ and $\gamma_2$ become again fully decoupled zero-energy Majorana modes.
This process amounts to moving the Majorana mode $\gamma_2$ from the right branch to the lower branch.
The exchange of the two Majorana modes $\gamma_1$ and $\gamma_2$ is then realized by three consecutive processes, shown in \cref{fig:braiding}(d). 
First, the mode $\gamma_2$ is moved from the right branch to the lower branch, then $\gamma_1$ from the left to the right branch, and finally, $\gamma_2$ from the lower branch to the left branch.
The whole process amounts to exchanging the two Majorana modes $\gamma_1$ and $\gamma_2$ following a clockwise path.
In the adiabatic limit, one can show that this process is described by the unitary operator $U_{12}=\exp(\frac\pi4\gamma_1\gamma_2)$~\cite{van-heck_coulomb-assisted_2012}.
This braiding protocol exchanges Majorana modes in parameter space without physically moving them, and can be realized by directly controlling the tunnel couplings via electric gates~\cite{sau_controlling_2011,karzig_shortcuts_2015} or the Coulomb couplings between Majorana modes via magnetic fluxes~\cite{hassler_the-top-transmon:_2011,van-heck_coulomb-assisted_2012,hyart_flux-controlled_2013} or electric gates~\cite{aasen_milestones_2016}, or by removing or adding single electrons in Coulomb blockade regimes~\cite{flensberg_non-abelian_2011}.
Another approach is to emulate a braiding process without moving the Majorana modes via measurement-only protocols using projective measurements of the fermion parity~\cite{bonderson_measurement-only_2008,bonderson_measurement-only_2009,bonderson_measurement-only_2013,vijay_teleportation-based_2016,plugge_majorana_2017,karzig_scalable_2017}.

More elaborate 2D or 3D networks can be engineered using multiple trijunctions as building blocks~\cite{sau_universal_2010,alicea_non-abelian_2011,sau_controlling_2011,clarke_majorana_2011,romito_manipulating_2012,halperin_adiabatic_2012,van-heck_coulomb-assisted_2012,hyart_flux-controlled_2013,karzig_shortcuts_2015,aasen_milestones_2016,karzig_scalable_2017,plugge_majorana_2017}.
Contiguous Majorana modes on the same junction can be exchanged using the braiding protocols described above, while Majorana modes on different junctions can be exchanged via a composition of successive exchanges of Majorana modes on the same junction.
Braiding protocols realized in a 2D network of 1D topological superconductors are analogous to the braiding of Majorana modes localized at the vortices of a 2D topological superconductor~\cite{ivanov_non-abelian_2001}. 

\subsection{Topological qubits}

The nonabelian braid operations of Majorana modes can be used to realize the qubits of topological quantum computers~\cite{kitaev_fault-tolerant_2003,nayak_non-abelian_2008}.
A Majorana qubit is realized by the superposition of degenerate groundstates with a fixed fermion parity.
The quantum gates are implemented by unitary transformations on qubits composed of different combinations of braid operators.
Consider a network of $2N$ localized Majorana modes $\gamma_i$ which commute with the Hamiltonian.
These Majorana modes can be combined into pairs corresponding to a set of $N$ nonlocal fermionic operators
\begin{equation}
	d_i=\frac12(\gamma_{2i-1}+\ii\gamma_{2i}),
\end{equation}
where $\gamma_{2i-1}$ and $\gamma_{2i}$ are pairs of neighboring Majorana modes, separated either by a trivial or a nontrivial region.
Notice that the way one combines Majorana modes into pairs is an arbitrary choice, with different choices being unitary equivalent and corresponding to unitary rotations of the basis of fermionic operators.
The state of the system is fully described by the set of number operators $n_i=d^\dag_i d_i$ of each fermionic mode.
The creation or annihilation of such fermionic modes does not change the total energy since the fermionic operators commute with the Hamiltonian, but it changes the fermion parity
\begin{equation}
	P=(-1)^F=
	\prod_{i=1}^N 
	(1-2d^\dag_i d_i)=
	\prod_{i=1}^{N} (-\ii \gamma_{2i-1}\gamma_{2i}),
\end{equation}
where $F=\sum_i n_i$ is the fermion number [cf.~\cref{eq:Parity}].
Thus, every fermionic mode can be either occupied $n_i=1$ or empty $n_i=0$, giving two degenerate states for each pair of Majorana modes.
Consequently, the groundstate manifold has degeneracy $2^N$:
The collection of these degenerate groundstates is spanned by the linear combination of all possible eigenstates $\ket{n_1,n_2,\ldots,n_N}$ labeled by the occupation numbers $n_i$ of the fermionic modes.
Notice that the groundstate manifold includes states with both even and odd fermion parities.
Since braid operations do not break parity conservation, they do not mix the even and odd parity sectors.
Hence, for a set of $2N$ Majorana modes, the dimension of the groundstate manifold with fixed fermion parity is $2^{N-1}$.
Quantum gates correspond to braid operations $U_{nm}$ acting on this manifold, which is separated from higher-energy quasiparticle excitations by the particle-hole gap.
Hence, if performed adiabatically, braid operations drive the system from one groundstate to another: 
The unitary evolution of the qubit is constrained within the groundstate manifold with fixed fermion parity.

A simple Majorana qubit can be implemented by $4$ Majorana modes $\gamma_i$, corresponding to $2$ fermionic operators $d_1=(\gamma_1+\ii\gamma_2)/2$ and $d_2=(\gamma_3+\ii\gamma_4)/2$.
The occupancy of the fermionic states is described by the number operators $n_1=d_1^\dag d_1=(1+\ii\gamma_1\gamma_2)/2$ and $n_2=d_2^\dag d_2=(1+\ii\gamma_3\gamma_4)/2$ [cf.~\cref{eq:Majorananumberoperator}].
All possible braid operations in this manifold are generated by the braiding of neighboring Majorana modes $U_{12}$, $U_{23}$, $U_{34}$ given by
\begin{subequations}\begin{align}
	U_{12}&=
	\exp\left(\frac\pi4\gamma_1\gamma_2\right)=
	\exp\left(\ii\frac\pi4\left(1 - 2 d_1^\dag d_1 \right)\right),
	\\
	U_{23}&=\exp\left(\frac\pi4\gamma_2\gamma_3\right)=
	\frac1{\sqrt2}\left(1+\gamma_2\gamma_3\right)=
	\nonumber\\&=
	\frac1{\sqrt2}\left(1+ \ii \left( d_1^\dag-d_1 \right)\left( d_2^\dag + d_2 \right) \right),\\
	U_{34}&=
	\exp\left(\frac\pi4\gamma_3\gamma_4\right)=
	\exp\left(\ii\frac\pi4\left(1 - 2 d_2^\dag d_2 \right)\right).
\end{align}\end{subequations}
The groundstate is $4$-fold degenerate and spanned by the eigenstates $\ket{00}$, $\ket{11}$, $\ket{10}$, and $\ket{01}$ of the number operators.
In this basis, the braid operators can be written in matrix form as~\cite{ivanov_non-abelian_2001}
\begin{subequations}\label{eq:ivanovbraids}\begin{align}
	U_{12}&=
	\begin{bmatrix}
	\ee^{\ii\pi/4}&&&\\
	&\ee^{-\ii\pi/4}&&\\
	&&\ee^{-\ii\pi/4}&\\
	&&&\ee^{\ii\pi/4}\\
	\end{bmatrix},
	\\
	U_{23}&=
	\frac1{\sqrt2}
	\begin{bmatrix}
	1&\ii&&\\
	\ii&1&&\\
	&&1&\ii\\
	&&\ii&1\\
	\end{bmatrix},
	\\
	U_{34}&=
	\begin{bmatrix}
	\ee^{\ii\pi/4}&&&\\
	&\ee^{-\ii\pi/4}&&\\
	&&\ee^{\ii\pi/4}&\\
	&&&\ee^{-\ii\pi/4}\\
	\end{bmatrix}.
\end{align}\end{subequations}
One can easily verify (algebraically or diagrammatically) that the other braid operators can be written as $U_{13}=U_{12}U_{23}U_{21}$, $U_{24}=U_{23}U_{34}U_{32}$, and $U_{14}=U_{12}U_{23}U_{34}U_{32}U_{21}$.

Braid operators mix states with different number of fermions but preserve fermion parity, e.g., $U_{23}\ket{00}\propto\ket{00}+\ii\ket{11}$. 
Indeed, the braid operators are block diagonal with respect to the fermion parity.
If the initial state of the qubit has, e.g., even fermion parity, the unitary evolution of the qubit is constrained on the manifold spanned by the many-body states $\ket{00}$ and $\ket{11}$.
Projecting on this manifold, the braid operators in \cref{eq:ivanovbraids} become
\begin{equation}\label{eq:braidfixedparity}
	U_{12}'=U_{34}'=
	\begin{bmatrix}
	\ee^{\ii\pi/4}&\\
	&\ee^{-\ii\pi/4}\\
	\end{bmatrix},
	\quad
	U_{23}'=
	\frac1{\sqrt2}
	\begin{bmatrix}
	1&\ii\\
	\ii&1\\
	\end{bmatrix}.
\end{equation}
The braid operators act on a manifold with dimension $2^{N-1}=2$, i.e., a two-level quantum system corresponding to a single qubit.
The fact that fermion parity is conserved but not the fermion number is an intrinsic property of the superconducting condensate: 
In this sense, topological quantum computation is protected by fermion parity conservation.

\subsection{Quantum gates}

\begin{figure}[t]
\includegraphics[width=\columnwidth]{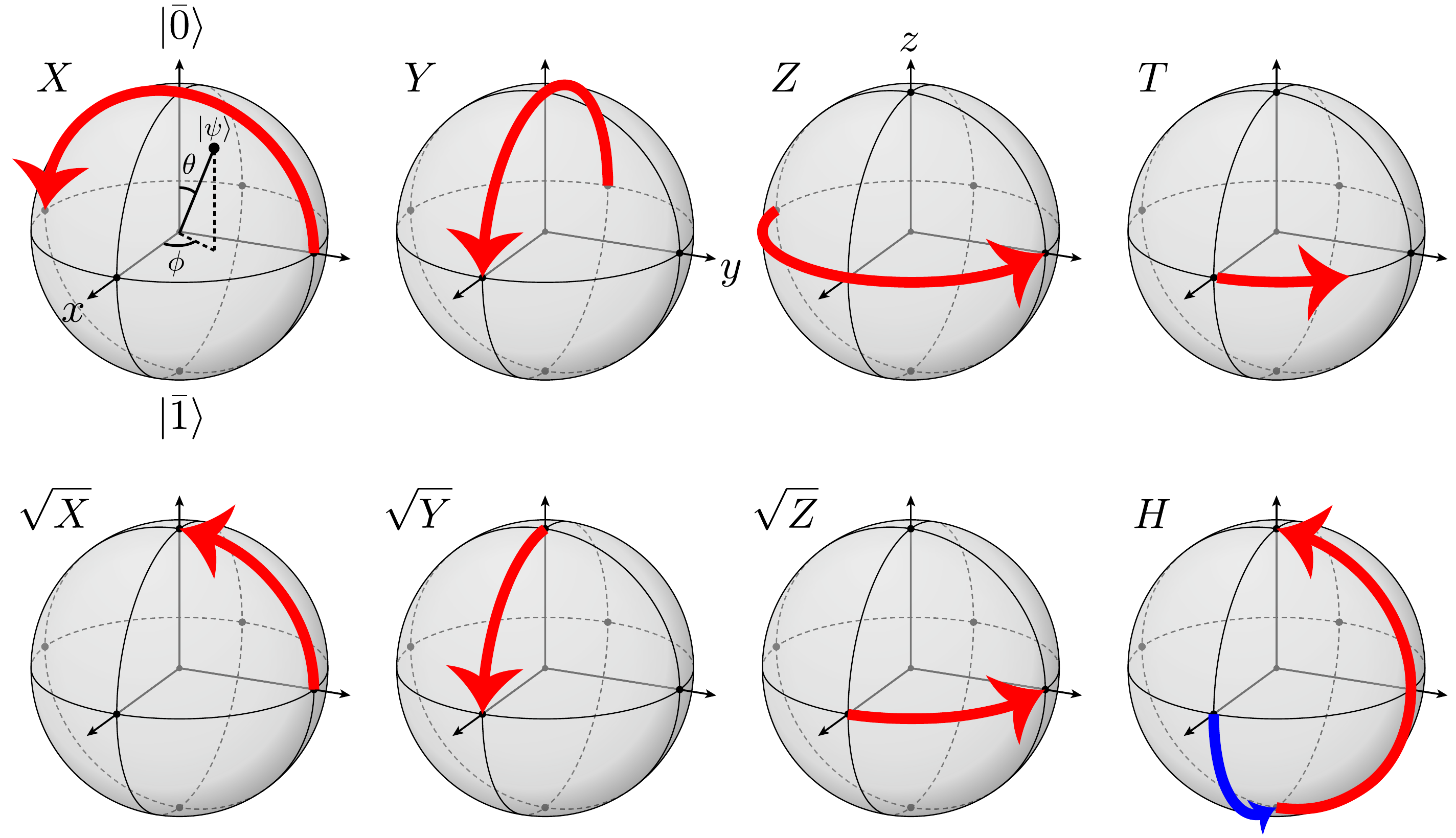}
\caption{
Single-qubit quantum gates as rotations on the Bloch sphere.
The Pauli gates $X$, $Y$, and $Z$ correspond to rotations by $\pi$, and their square roots $\sqrt{X}$, $\sqrt{Y}$, and $\sqrt{Z}$ to rotations by $\pi/2$ around the $x$, $y$, and $z$ axes.
The $T$ gate corresponds to a rotation by $\pi/4$ around the $z$ axis, and the Hadamard gate to a $\pi/2$ rotation around the $y$ axis followed by a $\pi$ rotation around the $x$ axis.
}
\label{fig:Bloch}
\end{figure}

Quantum gates are unitary operations acting on a set of qubits.
For a set of $2N$ Majorana modes, the groundstate manifold with fixed fermion parity has dimension $2^{N-1}$ and corresponds to $N-1$ qubits.
The set of all quantum gates acting on these $N-1$ qubits coincides with the unitary group $\mathrm{U}(2^{N-1})$.
A single qubit ($2N=4$ Majorana modes) corresponds to a two-level quantum system, and quantum gates acting on it correspond to the unitary group $\mathrm{U}(2)$.
Each single-qubit quantum gate $U\in\mathrm{U}(2)$ can be represented as rotations on the Bloch sphere, which is the unit sphere spanned by the eigenstates 
$\ket{\bar0}$ and $\ket{\bar1}$ of the qubit, which corresponds to the many-body states with fixed fermion parity, e.g., $\ket{\bar0}=\ket{00}$ and $\ket{\bar1}=\ket{11}$.
In this representation, any pure state of the qubit $\ket{\psi}=\cos(\frac\theta2)\ket{\bar0}+\ee^{\ii\phi}\sin(\frac\theta2)\ket{\bar1}$ can be represented as a point on the unit sphere with polar angle $\theta$ and azimuthal angle $\phi$, as in \cref{fig:Bloch}.
Some notable examples of single-qubit gates are the Pauli gates, given by the three Pauli matrices acting on a single qubit, i.e.,
\begin{equation}
	X=
	\begin{bmatrix}
	&1\\1&\\
	\end{bmatrix},
	\quad
	Y=
	\begin{bmatrix}
	&-\ii\\\ii&\\
	\end{bmatrix},
	\quad
	Z=
	\begin{bmatrix}
	1&\\&-1\\
	\end{bmatrix}.
\end{equation}
The Pauli-$X$ gate is also called the NOT gate since it is the quantum version of the logic NOT gate in classical computers.
The Pauli gates $X$, $Y$, and $Z$ perform rotations by $\pi$ around the $x$, $y$, and $z$ axes of the Bloch sphere.
Moreover, the square roots of the Pauli gates $\sqrt{X}$, $\sqrt{Y}$, and $\sqrt{Z}$ perform rotations by $\pi/2$ around the $x$, $y$, and $z$ axes of the Bloch sphere, as shown in \cref{fig:Bloch}.
Other examples of single-qubit gates are the so-called phase-shift gates
\begin{equation}
	P(\phi)=
	\begin{bmatrix}
	1&\\&\ee^{\ii\phi}\\
	\end{bmatrix},
\end{equation}
which include the Pauli-$Z$ gate as a special case, as well as the $S$ and $T$ gates 
\begin{subequations}\begin{align}
	Z&=P(\pi)=
	\begin{bmatrix}
	1&\\&-1\\
	\end{bmatrix},
	\\
	S&=P\left(\tfrac\pi2\right)=\sqrt{Z}=
	\begin{bmatrix}
	1&\\&\ii\\
	\end{bmatrix},
\\
	T&=P\left(\tfrac\pi4\right)=\sqrt{S}=\sqrt[4]{Z}=
	\begin{bmatrix}
	1&\\&\ee^{\ii\pi/4}\\
	\end{bmatrix},
\end{align}\end{subequations}
where $T^2=S$ and $S^2=Z$.
The $Z$, $S$, and $T$ gates above correspond respectively to rotations by $\pi$, $\pi/2$, and $\pi/4$ around the $z$ axis of the Bloch sphere, as shown in \cref{fig:Bloch}.
Another single-qubit gate is the Hadamard gate
\begin{equation}
	H=\frac1{\sqrt\ii} X {\sqrt{Y}}=\frac1{\sqrt2}\left(X+Z\right)=
	\frac1{\sqrt2}
	\begin{bmatrix}
	1&1\\1&-1\\
	\end{bmatrix},
\end{equation}
with $H^2=\id$.
The Hadamard gate corresponds to a $\pi/2$ rotation around the $y$ axis followed by a $\pi$ rotation around the $x$ axis, as shown in \cref{fig:Bloch}.
Combinations of the $S$ gate and the Hadamard $H$ gate generate the Pauli gates $X$, $Y$, and $Z$ up to a phase, as well as their square roots $\sqrt{X}=HSH$, $\sqrt{Y}=\sqrt{\ii}XH$, and $\sqrt{Z}=S$.
Hence, the $S$ gate and the Hadamard $H$ gate generate all rotations around the three axes of the Bloch sphere by an angle multiple of $\pi/2$.
A remarkable result is that only by adding rotations of $\pi/4$ around a single axis (e.g., using the $T$ gate) one can generate all possible rotations (by arbitrary angles around arbitrary axes) up to arbitrary precision.
Indeed, all unitary transformations on a single qubit can be performed using only the Hadamard $H$ gate and the $T$ gate:
For any unitary transformation $U\in\mathrm{U}(2)$, there exists a finite product of $H$ and $T$ that approximates $U$ with arbitrary precision, up to a complex phase~\cite{boykin_on-universal_1999}.
In other words, arbitrary rotations in the Bloch sphere can be efficiently approximated by a combination of the Hadamard $H$ and $T$ gates.

Examples of two-qubit gates $U\in\mathrm{U}(4)$ are the controlled gates, where the state of the \emph{control} qubit determines the quantum operation performed on the \emph{target} qubit.
A controlled-$U$ gate performs a unitary operation $U$ on a target qubit if the control qubit is in the state $\ket{\bar1}$ and leaves the target qubit unchanged if the control qubit is in the state $\ket{\bar0}$ and is represented by
\begin{equation}
	C(U)=\begin{bmatrix} \id & \\ & U \\ \end{bmatrix}.
\end{equation}
Notable examples are the controlled Pauli gates, i.e., the controlled-$X$ (CX), also called the controlled-NOT gate (CNOT), the controlled-$Y$ (CY), and the controlled-$Z$ (CZ) gates respectively with $U=X,Y,Z$.

A set of quantum gates is called \emph{universal} if any unitary evolution of $N$ qubits can be approximated to arbitrary accuracy by a finite sequence of gates in the set.
A universal quantum computer can thus implement all possible unitary operations on its qubits with arbitrary accuracy.
Moreover, the Solovay–Kitaev theorem~\cite{kitaev_quantum_1997,nielsen_quantum_2010} guarantees that this can be done \emph{efficiently}, i.e., the number of unitary operations required to approximate an arbitrary unitary operation to a given accuracy $\varepsilon$ grows logarithmically with $1/\varepsilon$. 
A common choice (but not unique) for a universal set of quantum gates is given by the $T$ gate, Hadamard $H$ gate, and the CNOT gate:
The unitary evolution of an arbitrary number of qubits can be efficiently approximated by a finite sequence of these gates.
A smaller subset of unitary evolutions in $\mathrm{U}(2^{N-1})$ is represented by the Clifford group $\mathrm{C}_{N-1}$, which is generated by the Clifford gates, i.e., the $S$ gate, Hadamard $H$ gate, and the CNOT gate.
The Clifford gates alone are not universal but become universal with the addition of the $T$ gate~\cite{boykin_on-universal_1999}.
This is essentially because the Clifford gates alone can only generate rotations of $\pi/2$ around the three axes of the Bloch sphere:
however, the addition of $\pi/4$ rotations via the $T$ gate allows rotations by arbitrary angles around arbitrary axes.
A quantum computer based only on Clifford gates can be efficiently simulated by a classical computer, according to the Gottesman-Knill theorem~\cite{aaronson_improved_2004} (see also Ref.~\onlinecite{nielsen_quantum_2010}):
Hence, only universal quantum computers can provide a fundamental advantage over classical computers.

The Pauli gates, the Hadamard $H$ gate, and the $S$ gate can be realized via Majorana qubits as a composition of braid operators acting on a qubit with fixed fermion parity as in \cref{eq:braidfixedparity}, which gives
\begin{subequations}\begin{align}
	X& = -\ii 
	(U_{23})^2
	=-\ii\gamma_2\gamma_3,\\
	Y&= -\ii
	(U_{31})^2
	=-\ii\gamma_3\gamma_1,\\
	Z&= -\ii 
	(U_{12})^2
	=-\ii\gamma_1\gamma_2,\\
	H& 
	=-\ii 
	U_{12}U_{23}U_{12}=
	-\ii
	U_{23}U_{12}U_{23},
	\label{eq:HUUU}
	\\
	S&
	=\ee^{\ii\pi/4} U_{21}.
\end{align}\end{subequations}
Notice that the last identity of \cref{eq:HUUU} is given by the Yang-Baxter equation in \cref{eq:YangBaxter}. 
Furthermore, the two-qubit CNOT gate can be implemented in a topologically-protected way by combining fermion-parity projective measurements and single-qubit Clifford gates implemented by braiding~\cite{bravyi_fermionic_2002,beenakker_charge_2004,bravyi_universal_2006,zilberberg_controlled-not_2008}.
Hence, Majorana modes can implement the full set of topologically-protected Clifford gates.

An intrinsic limitation of Majorana qubits is that the $T$ gate cannot be realized by braiding.
Nevertheless, the $T$ gate can still be implemented in a topologically-unprotected way by bringing two Majorana modes close to each other and allowing them to dynamically dephase or by manipulating the geometric phase~\cite{karzig_universal_2016,karzig_robust_2019} of the trijunction Hamiltonian in \cref{eq:trijunction}.
The accuracy of topologically-unprotected $T$ gates can be significantly improved by using the so-called ``magic-state distillation''~\cite{bravyi_universal_2005,bravyi_universal_2006,sau_universal_2010,bonderson_implementing_2010}.
Another route to provide the missing phase gate is by coupling topological qubits with conventional (nontopological) qubits~\cite{hassler_anyonic_2010,bonderson_topological_2011,leijnse_quantum_2011,jiang_interface_2011,hoffman_universal_2016,rancic_entangling_2019}.
However, all these approaches necessarily break topological protection and cannot provide universal quantum computation in a fully topologically-protected way.

\subsection{Fusion rules}

A quantum computer consists of a set of quantum gates combined into a quantum circuit.
The quantum information is encoded in the initial state (the input) and then processed via the unitary evolution of the quantum gates.
The final step of the computation is the readout of the quantum information stored in the final state of the qubits (the output).
This step requires measuring the fermion parity of the Majorana modes of the qubit, which can be obtained by fusing the Majorana modes into a single fermionic mode, i.e., by adiabatically moving two Majorana modes close to each other~\cite{kitaev_fault-tolerant_2003,stone_fusion_2006,souto_fusion_2022}.
Depending on the fermionic parity of the initial groundstate, the occupation number of the Majorana bound state is either $n=0$ or $1$ with opposite fermion parity.
Consequently, the fusion of two Majorana modes results either in the fermionic vacuum or in a single fermionic mode with finite energy~\cite{fu_superconducting_2008}.
This process can be formally described by the fusion rule
\begin{equation}
	\gamma\times\gamma=1+\psi,
\end{equation}
where $\gamma$, $\psi$, and $1$ denote respectively Majorana fermions, conventional Dirac fermions, and the vacuum.
In principle, the two outcomes can be distinguished experimentally.
For instance, the readout of the fermion parity can be achieved by parity-to-charge conversion~\cite{aasen_milestones_2016,plugge_majorana_2017,steiner_readout_2020,seoane-souto_timescales_2020,zhou_fusion_2022,krojer_demonstrating_2022}.
In this approach, the Majorana qubit is coupled via tunneling with a quantum dot, such that the two become entangled.
In this case, the occupation number and the parity of the Majorana qubit become correlated to the charge state of the quantum dot.
Hence, measuring the charge state of the quantum dot amounts to measuring the parity of the qubit~\cite{steiner_readout_2020}.

\subsection{Topological protection}

Braid operations are topological in the sense that their outcome does not depend on the details of the exchange process of the Majorana modes and their precise trajectory in the parameter space.
Also, since the Majorana qubits encode quantum information nonlocally and are topologically protected, they are extremely robust against local perturbations.
Besides, the finite gap in the quasiparticle spectra in the superconducting phase suppresses thermal fluctuations and confines the unitary evolution within the degenerate groundstate. 
As a result, topological protection significantly reduces computational errors, providing a considerable advantage over nontopological quantum computation schemes.
However, braiding errors cannot be completely avoided due to the occurrence of nonadiabatic processes~\cite{cheng_nonadiabatic_2011,karzig_boosting_2013,scheurer_nonadiabatic_2013,karzig_optimal_2015,pedrocchi_majorana_2015,amorim_majorana_2015,knapp_the-nature_2016,sekania_braiding_2017,bauer_dynamics_2018} or processes that break fermion parity conservation, such as quasiparticle poisoning, i.e., the tunneling of superconducting quasiparticles from the external environment that breaks fermion parity conservation~\cite{leijnse_scheme_2011,rainis_majorana_2012,budich_failure_2012}.
Indeed, braid operations need to be performed slow enough to be adiabatic but fast enough to avoid decoherence.
If the adiabaticity condition is not met, the braiding process will produce a superposition of many-body states with higher-energy excitations above the particle-hole gap and away from the groundstate manifold.
Therefore, the exchange process must be performed on time scales longer than the time scale determined by the particle-hole gap $\tau_\mathrm{ph}=\hbar/\Delta E_\mathrm{ph}$ to avoid nonadiabatic effects, where $\Delta E_\mathrm{ph}$ is the energy gap separating the excited states from the groundstate.
This time scale can be, in principle, reduced by implementing the so-called shortcuts to adiabaticity, i.e., by suppressing nonadiabatic effects via counterdiabatic terms in the Hamiltonian introduced via additional couplings between noncontiguous Majorana modes~\cite{karzig_shortcuts_2015,ritland_optimal_2018}. 
On the other hand, the braiding process must be performed on time scales shorter than the coherence time of the Majorana qubit. 
The coherence time is determined by the inverse of the tunneling rate between Majorana modes at finite distance $\Gamma_L\propto\ee^{-L/\xi_\mathrm{M}}$~\cite{bonderson_quasi-topological_2013,das-sarma_majorana_2015,beenakker_search_2020}, the rate of thermal excitations $\Gamma_T\propto\ee^{-\Delta/k_\mathrm{B}T}$~\cite{cheng_topological_2012,bonderson_quasi-topological_2013,das-sarma_majorana_2015,beenakker_search_2020}, and the quasiparticle poisoning rate $\Gamma_\mathrm{QP}$.
Moreover, as mentioned before, Majorana modes cannot realize universal quantum computation in a fully topologically-protected way, since the $T$ gate cannot be implemented only by braiding Majorana modes.
From these considerations, it is clear that topological quantum computation still needs to be supplemented by quantum error correction schemes~\cite{kitaev_quantum_1997,kitaev_fault-tolerant_2003}, which can be provided by the so-called Majorana toric code~\cite{xu_fractionalization_2010,terhal_from_2012,roy_quantum_2017,roy_charge_2018,ziesen_topological_2019}, surface codes~\cite{vijay_majorana_2015,vijay_physical_2016,landau_towards_2016,plugge_roadmap_2016,manousakis_majorana_2017,litinski_quantum_2018}, or color codes~\cite{litinski_combining_2017,litinski_quantum_2018}.

In this Section, I only discussed some specific aspects of topological quantum computation with Majorana modes.
For a more thorough overview, see Refs.~\onlinecite{nayak_non-abelian_2008,stern_non-abelian_2010,pachos_introduction_2012,das-sarma_majorana_2015,roy_topological_2017,lahtinen_a-short_2017,stanescu_introduction_2017,beenakker_search_2020,oreg_majorana_2020}.
For a general discussion on quantum gates, qubits, and quantum computation, see Refs.~\onlinecite{steane_quantum_1998,kitaev_classical_2002,nielsen_quantum_2010}.

\section{Majorana nanowires\label{sec:swave}}

\subsection{The three main ingredients\label{sec:ingredients}}

\subsubsection{Time-reversal symmetry breaking}

At first glance, engineering a spinless topological superconductor in a real system seems an impossible task.
Spinless electrons do not exist, and superconductivity is not stable against fluctuations in 1D~\cite{giamarchi_quantum_2003}.
Naively, one may think to address the first issue by simply generalizing \cref{eq:H-pwave} to good old-fashioned spin-1/2 electrons by considering two copies of the Hamiltonian $\mathcal{H}_\up$ and $\mathcal{H}_\down$, one for each of the two spin channels.
Hence, in an open chain $\mathcal{H}_\up+\mathcal{H}_\down$ with $|\mu|<2t$, one will obtain two Majorana end modes per boundary, having opposite spins $\gamma_\up$ and $\gamma_\down$.
This system realizes the so-called Majorana Kramers pairs, which appear in time-reversal-invariant topological superconductors in 1D~\cite{wong_majorana_2012,zhang_time-reversal-invariant_2013,nakosai_majorana_2013,keselman_inducing_2013,gaidamauskas_majorana_2014,haim_time-reversal-invariant_2014,schrade_proximity-induced_2015,reeg_diii_2017,schrade_parity-controlled_2018} (see Ref.~\onlinecite{haim_time-reversal-invariant_2019} for a review on the topic).
However, since Majorana Kramers pairs are protected by time-reversal symmetry, they are not robust against local perturbations which break this symmetry, e.g., magnetic impurities.
Indeed, without any additional symmetry constraints, local perturbations may hybridize the two degenerate modes at each end.
Consequently, their energies become finite, lifting the groundstate degeneracy.
Hence, the Hamiltonian can be continuously deformed without closing the particle-hole gap into a trivial Hamiltonian without end modes if time-reversal symmetry is not enforced.

Consequently, to obtain unpaired Majorana modes localized at opposite ends and robust against local perturbations, one needs to lift the Kramers degeneracy between the two spin channels:
The only way to escape Kramers theorem is to break time-reversal symmetry.
This can be done, for example, by including a ferromagnetic insulator layer~\cite{sau_generic_2010,alicea_majorana_2010} or a half-metal~\cite{chung_topological_2011,duckheim_andreev_2011} in the heterostructure or by applying an external magnetic field~\cite{sato_non-abelian_2009,sato_non-abelian_2010,alicea_majorana_2010,lutchyn_majorana_2010,oreg_helical_2010}.
The resulting spin-polarized Hamiltonian will describe two distinct fermionic sectors which can be \emph{effectively} regarded as spinless.
If the lowest energy sector becomes topologically nontrivial, the system will exhibit \emph{unpaired} and nondegenerate Majorana modes localized at its ends.
An alternative way to break time-reversal symmetry is to drive a supercurrent through a superconductor (the current density is odd under time-reversal)~\cite{romito_manipulating_2012,kotetes_topological_2015,lesser_majorana_2022}.

\subsubsection{Proximity-induced pairing}

In 1D, quantum fluctuations prevent superconducting phase transitions at finite temperatures~\cite{giamarchi_quantum_2003}.
To overcome this issue, one can consider a 1D system, e.g., a quantum nanowire, placed in contact with a bulk 3D superconductor through an interface.
In this heterostructure, electrons can tunnel between the superconducting bulk and the nanowire, such that electrons in the wire acquire finite superconducting correlations, which can be described as an effective long-range superconducting pairing~\cite{stanescu_proximity_2010,sau_robustness_2010,potter_engineering_2011,sau_experimental_2012,cole_effects_2015,stanescu_proximity-induced_2017}.
The magnitude and symmetry of this proximity-induced superconducting pairing are largely determined by the bulk order parameter and interface transparency.

The superconducting pairing is described in momentum space by
\begin{equation}\label{eq:pairing}
{\mathcal{H}}_\mathrm{SC}=
\frac12
\int \dd k\,
\left[ \Psi_{\up}(k), \Psi_{\down}(k) \right]
\cdot
\hat\Delta(k)
\cdot
\begin{bmatrix} \Psi_{\up}(k) \\ \Psi_{\down}(k) \end{bmatrix}
+\text{h.c.},
\end{equation}
where the superconducting pairing satisfies $\hat\Delta(k)=-\hat\Delta(-k)^\intercal$ due to the antisymmetry of the fermionic wavefunction of the superconducting electrons~\cite{sigrist_phenomenological_1991}.
If no other degrees of freedom are involved, the pairing symmetry can be classified in terms of the spin and angular momentum of the coupled electrons (Cooper pairs).
The total spin of the Cooper pairs can be either $S=0$ (spin-singlet) or $S=1$ (spin-triplet),
which correspond to antisymmetric and symmetric spin states, respectively.
Since the superconducting pairing needs to be antisymmetric, spin-singlet pairing requires even-parity angular momentum (e.g., $L=0$, $s$-wave, or $L=2$, $d$-wave) while spin-triplet pairing requires odd-parity angular momentum (e.g., $L=1$, $p$-wave).
Ideally, one would like to employ odd-parity, spin-triplet bulk superconductors~\cite{kallin_chiral_2016,sato_majorana_2016,sato_topological_2017} to couple electrons with parallel spins and achieve an effective ``spinless'' superconductivity separately in the two spin channels.
However, these unconventional superconductors have low critical temperatures $<\text{\SI{1}{\kelvin}}$ and a rather small superconducting gap.
On the other hand, conventional $s$-wave singlet-paring superconductors have a relatively higher critical temperature (e.g., $T_\mathrm{c}=\text{\SI{9.25}{\kelvin}}$ for niobium) and larger superconducting gap.
Moreover, they are abundant in nature, and their physics is well understood in terms of the microscopic BCS theory~\cite{bardeen_microscopic_1957,bardeen_theory_1957}.

In a nanowire proximitized with a conventional superconductor, one expects a proximity-induced pairing comparable to or smaller than the bulk superconducting pairing.
Proximity effects are effectively described by modeling the superconducting correlations of the semiconductor-superconductor interface at a mean-field level and integrating out the degrees of freedom of the bulk~\cite{stanescu_proximity_2010,sau_robustness_2010,potter_engineering_2011,sau_experimental_2012,cole_effects_2015,stanescu_proximity-induced_2017}. 
At low energies, the resulting proximity-induced pairing can be written as $\Delta=\Gamma\Delta_0/(\Gamma+\Delta_0)$, where $\Delta_0$ is the bulk superconducting pairing and $\Gamma$ is the coupling between the bulk superconductor and the nanowire, which depends on the density of states and the transparency of the interface~\cite{stanescu_proximity_2010,sau_robustness_2010,sau_non-abelian_2010,potter_engineering_2011,stanescu_majorana_2011,sau_experimental_2012,cole_effects_2015,stanescu_proximity-induced_2017}.
The proximity effect will also induce a renormalization of the other energy scales of the nanowire~\cite{potter_engineering_2011,stanescu_majorana_2011,sau_experimental_2012,cole_effects_2015,stanescu_proximity-induced_2017}.
For a more extended discussion on the proximity effect in Majorana wires, see Ref.~\onlinecite{stanescu_majorana_2013}.

\subsubsection{Spin-orbit coupling}

The last ingredient is spin-orbit coupling, which is necessary to avoid the closing of the particle-hole gap in the topologically nontrivial phase.
In a single-band 1D without spin-orbit coupling (neglecting orbital degrees of freedom) and in a magnetic field, electron spin is a good quantum number and a conserved quantity, and consequently, electrons with spin parallel and antiparallel to the field direction form two distinct bands with a finite energy splitting.
In particular, electrons in the lowest energy band crossing the Fermi level have the same spin.
In this case, the proximity-induced superconducting pairing cannot couple electrons within the same band, since spin-singlet pairing only couples electrons with opposite spins and momenta.
Consequently, the particle-hole gap closes even in the presence of a finite superconducting pairing.
However, to realize topologically protected end modes, one needs to keep the particle-hole gap open. 
This can be done by tilting the directions of the spin of electrons having opposite momenta.
A finite spin-orbit coupling induces an effective magnetic field in the direction of the electron momentum, which competes with the bare magnetic field in fixing the direction of the electron spin.
Consequently, electrons with the same energy and opposite momenta have their spins tilted in different directions.
This tilting allows electrons within the same band and opposite momenta to couple via the conventional superconducting pairing, thus avoiding the closing of the particle-hole gap at finite magnetic fields.
As I will clarify in \cref{sec:band-inversion}, the combined effect of spin-orbit coupling, time-reversal symmetry breaking, and proximitized conventional (spin-singlet) superconductivity is equivalent to an effective spin-triplet superconducting pairing between effectively spinless electrons.

Spin-orbit coupling in heterostructures can emerge as the direct effect of the presence of an electric field $\mathbf{E}$.
For a relativistic electron in an electric field, the Dirac equation can be approximated at the second-order perturbation theory in powers of $1/c$ by an effective low-energy Hamiltonian density~\cite{berestetskii_quantum_1982_s33,sakurai_advanced_1967_s33} that reads 
\begin{align}
H&=
\frac{\mathbf {p}^2}{2m}+eV-\frac{\mathbf{p}^4}{8m^3c^2}+
\nonumber\\
&-\frac{e\hbar}{4m^2c^2}
\bm{\sigma}\cdot\left(\mathbf{E}\times\mathbf{p}\right)
-\frac{e\hbar^2}{8m^2c^2}\nabla\mathbf{E},
\end{align}
with $\bm{\sigma}=[\sigma_x,\sigma_y,\sigma_z]$ the vector of Pauli matrices.
Here, the fourth term $\propto\bm{\sigma}\cdot(\mathbf E\times\mathbf{p})$ is the spin-orbit coupling induced by the electric field. 
Notice that this term does not break time-reversal symmetry since $\mathbf{E}$ is even, whereas $\mathbf{p}$ and $\bm\sigma$ are both odd under time-reversal.
In nanowire heterostructures, the dominant contribution to the spin-orbit coupling is due to the broken space-reflection symmetry at the interface between the wire and its substrate.
This is the case of the so-called Rashba spin-orbit coupling, which arises due to a net nonzero electric field $\mathbf{E}=-\nabla V$ induced by the sudden variation of the electrostatic potential at the interface~\cite{rashba_properties_1960} (structural inversion asymmetry). 
Another contribution is given by the Dresselhaus spin-orbit coupling, induced by the net nonzero electric field produced by the broken space inversion symmetry in noncentrosymmetric lattices~\cite{dresselhaus_spin-orbit_1955} (bulk inversion asymmetry). 
Strong spin-orbit couplings are observed in nanowire-superconductor heterostructures made with InAs, InSb (or other III-V semiconductors) wires~\cite{manchon_new-perspectives_2015,van-weperen_spin-orbit_2015,shabani_two-dimensional_2016}.

Another route to obtain an \emph{effective} spin-orbit coupling is to consider magnetic fields with a direction that rotates along the nanowire.
It can be shown that such a rotating field is unitarily equivalent to a system with a uniform field along a fixed direction and an effective spin-orbit coupling with a strength proportional to the spatial derivative of the field angle~\cite{braunecker_spin-selective_2010,kjaergaard_majorana_2012,klinovaja_topological_2013} (see also Ref.~\onlinecite{laubscher_majorana_2021}).
In nanowires, rotating magnetic fields can be achieved using substrates with magnetic textures~\cite{desjardins_synthetic_2019,mohanta_electrical_2019}, helical-magnetic superconducting compounds~\cite{martin_majorana_2012}, or periodic arrays of nanomagnets~\cite{kjaergaard_majorana_2012,klinovaja_transition_2012}.

\subsection{Minimal continuous model}

\subsubsection{The Oreg-Lutchyn Hamiltonian\label{sec:OLmodel}}

\begin{figure*}[t]
\includegraphics[width=\textwidth]{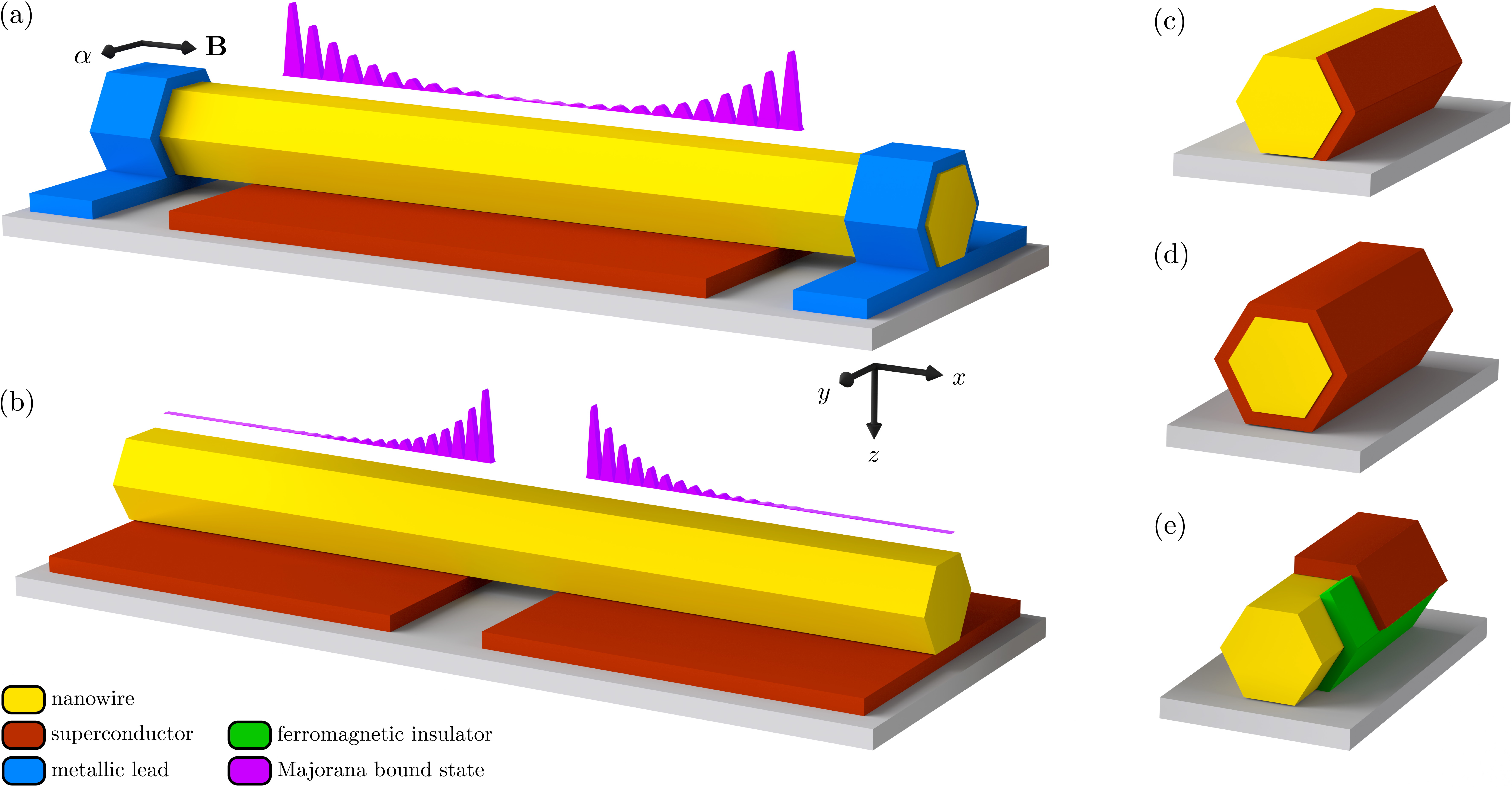}
\caption{
Majorana nanowires, i.e., semiconductor nanowires (e.g., InAs, InSb) with strong spin-orbit coupling $\alpha$ proximitized by a conventional superconductor substrate (e.g., Al, NbTiN) in an external magnetic field $\mathbf{B}$.
The magnetic field (in the $zx$ plane) is applied perpendicularly to the spin-orbit coupling field (along the $y$ axis).
(a) In typical setups, the nanowire is connected to metallic leads through tunneling barriers, and one measures the current through the barriers as a function of the applied voltage.
(b) In another possible setup, the nanowire is proximitized by two separate superconducting leads at opposite sides, realizing a Josephson junction.
(c)
In more advanced devices, the superconductor substrate is replaced by a superconducting thin film epitaxially grown on the facets of the nanowire.
Other setups are possible:
(d)
The superconducting film encloses the nanowire completely, covering all facets (full-shell nanowire);
(e)
The nanowire is partially covered by a ferromagnetic insulator thin film (e.g., EuS) and by a superconductor.
This last setup does not necessarily require the presence of an applied magnetic field.
}
\label{fig:nanowire}
\end{figure*}

As discussed before, only three ingredients are required to obtain a topologically nontrivial superconducting state with Majorana end modes in a realistic 1D condensed matter system: broken time-reversal symmetry (i.e., finite magnetic fields), proximity-induced superconducting pairing, and finite spin-orbit coupling~\cite{sato_non-abelian_2009,sato_non-abelian_2010,lutchyn_majorana_2010,oreg_helical_2010,alicea_majorana_2010,sau_generic_2010,sau_non-abelian_2010,sau_robustness_2010,akhmerov_quantized_2011,stanescu_majorana_2011}.
Ideally, one would like to have a large superconducting pairing induced by proximity with a superconductor with a relatively high critical temperature and critical magnetic fields for the superconducting phase to be stable against disorder and thermal fluctuations.
In addition, since the magnetic field inevitably suppresses the superconducting pairing, one would like to have a large $g$-factor to maximize the resulting Zeeman fields with relatively small externally-applied magnetic fields.
Finally, one would like to have large spin-orbit couplings to keep the particle-hole gap open in the nontrivial phase, even in the presence of disorder.

So far, the most promising physical systems are the so-called Majorana nanowires, i.e., semiconductor nanowires with strong spin-orbit coupling (e.g., InAs, InSb) proximitized by a conventional superconductor in an external magnetic field~\cite{lutchyn_majorana_2010,oreg_helical_2010}, as in \cref{fig:nanowire}(a).
The applied magnetic field $\mathbf{B}$ breaks the time-reversal symmetry, whereas the superconducting substrate induces a finite superconducting pairing in the nanowire by proximity effect.
Moreover, a finite spin-orbit coupling is provided by the broken space-reflection symmetry at the interface between the nanowire and the substrate and by the intrinsic broken space inversion symmetry of the semiconductor crystal lattice~\cite{manchon_new-perspectives_2015}.
An ideal Majorana nanowire, with a single electronic band with parabolic dispersion at the Fermi level and in an external magnetic field, proximity-induced pairing, and finite spin-orbit coupling, is described by an effective continuum Hamiltonian, known as the Oreg-Lutchyn minimal model~\cite{oreg_helical_2010,lutchyn_majorana_2010}, which reads 
\begin{align}
{\mathcal{H}}=\int \dd x\Big[&
\Psi^\dag(x)\left(\frac{p_x^2}{2m}+\frac{\alpha}{\hbar}\sigma_y p_x -\mu+\mathbf{b}\cdot\bm{\sigma} \right)\Psi(x)
+\nonumber\\\label{eq:H-swaveC}+&
\Psi(x)\left(\frac12\Delta\ee^{\ii\phi}\ii\sigma_y\right)\Psi(x)+\text{h.c.}
\Big],
\end{align}
where $\Psi(x)=[\Psi_\up(x),\Psi_\down(x)]$ is the real-space electronic field, $p_x=-\ii\hbar\partial_x$ the momentum operator, $m$ the effective mass, $\alpha$ the spin-orbit coupling strength, $\mu\ge0$ the chemical potential measured from the bottom of the electronic band, $\mathbf{b}=(g\mu_\mathrm{B}/2)\mathbf{B}$ the Zeeman field, with $g$ the effective Landé $g$-factor, $\mu_\mathrm{B}$ the Bohr magneton, $\mathbf{B}$ the applied magnetic field, and $\Delta\ee^{\ii\phi}$ the proximity-induced superconducting pairing with $\Delta>0$.
Again, since the phase of the superconducting pairing can always be absorbed by a unitary transformation, let us assume $\phi=0$ for simplicity.
The cartesian coordinates are chosen such that the wire lays along the $x$ axis, the surface of the superconductor is perpendicular to the $z$ axis [see \cref{fig:nanowire}(a)], and the spin-orbit coupling is 
$\propto\bm{\sigma}\cdot(\mathbf{E}\times\mathbf{p})=E \sigma_y p_x$.
The values of the parameters entering the Hamiltonian depend on the precise details of the specific experimental setup.
Majorana nanowires have been realized with InAs or InSb semiconducting nanowires proximitized by thin Al or NbTiN superconducting layers.
InAs and InSb provide a strong spin-orbit coupling and large $g$-factor, allowing large Zeeman fields with relatively small magnetic fields. 
See \cref{tab:materials} for the relevant material parameters of Majorana nanowires.

\begin{table}\footnotesize
\newcommand\Tstrut{\rule{0pt}{2.6ex}} 
\begin{tabular}{l|l|l}
\hline\hline\\[-2.5mm]
	Semiconductors 					& InAs						& InSb						\\[1mm]
	effective electron mass $m$ 	& 0.023 $m_\mathrm{e}$		& 0.014 $m_\mathrm{e}$ 	\\
									& \SI{12000}{\eV\per c^2}	& \SI{7200}{\eV\per c^2}	\\[1mm]
	effective $g$-factor 			& \SIrange{5}{20}{} 						& \SIrange{35}{60}{} 					\\
	$b/B=g \mu_\mathrm{B}/2$		& \SIrange{0.1}{0.6}{\milli\eV\per\tesla} 	& \SIrange{1}{1.7}{\milli\eV\per\tesla} \\[1mm]
	spin-orbit energy 	$E_\mathrm{SO}\!=\!\frac{m\alpha^2}{2\hbar^2}$ 							& \SIrange{0.05}{1}{\milli\eV} 		& \SIrange{0.05}{1}{\milli\eV} 		\\
	spin-orbit coupling $\alpha$																& \SIrange{0.2}{0.8}{\eV\angstrom} 	& \SIrange{0.2}{1}{\eV\angstrom} 	\\
	spin-orbit length 	$l_\mathrm{SO}\!=\! k_\mathrm{SO}^{-1}\!=\!\frac{\hbar^2}{m\alpha}$ 	& \SIrange{40}{180}{\nano\meter} 	& \SIrange{50}{230}{\nano\meter} 	\\[1mm]
\hline\hline
\multicolumn{1}{c}{}\\
\hline\hline\\[-2.5mm]
	Superconductors 									& Al						& NbTiN					\\[1mm]
	critical temperature	$T_\mathrm{c}$				& \SI{1.2}{\kelvin} 		& \SI{16}{\kelvin} 		\\
	superconducting gap		$\Delta_0$ 					& \SI{0.2}{\milli\eV} 	 	& \SI{2.5}{\milli\eV} 	\\
	critical field			$B_\mathrm{c}$ (at $T=0$)	& \SI{10}{\milli\tesla} 	& \SI{10}{\tesla} 		\\[1mm]
\hline\hline
\multicolumn{1}{c}{}\\
\hline\hline\\[-2.5mm]
	Heterostructures 							& InAs/Al, InSb/Al 				& InSb/NbTiN 				\\[1mm]
	maximum proximity-induced gap	$\Delta$ 	& \SI{0.2}{\milli\eV} 	& \SI{1}{\milli\eV} 	\\[1mm]
\hline\hline
\end{tabular}
\caption{
Material parameters of semiconductors and superconductors typically used in Majorana nanowire experiments.
Effective electron masses are in units of the bare electron mass $m_\mathrm{e}$.
The $g$-factor and spin-orbit coupling parameters depend on the heterostructure details (e.g., nanowire diameter, interface geometry) and can be partially controlled by an external back gate voltage~\cite{manchon_new-perspectives_2015,shabani_two-dimensional_2016}.
The ranges given in the table correspond to the $g$-factors measured via tunneling spectroscopy~\cite{weperen_quantized_2013,kammhuber_conductance_2016,kammhuber_conductance_2017,vaitiekenas_effective_2018} and to the spin-orbit coupling strengths measured via magnetotransport experiments~\cite{dhara_magnetotransport_2009,estevez-hernandez_spin-orbit_2010,roulleau_suppression_2010,liang_strong_2012,van-weperen_spin-orbit_2015}.
The proximity-induced gaps have been measured via tunneling experiments in InAs/Al~\cite{chang_hard_2015,deng_majorana_2016}, InSb/Al~\cite{zhang_large_2021}, and InSb/NbTiN~\cite{zhang_ballistic_2017,gul_hard_2017,gul_ballistic_2018} heterostructures.
Bulk properties in InAs and InSb semiconductors~\cite{gildenblat_handbook_1996} and Al~\cite{cochran_superconducting_1958,court_energy_2007} and NbTiN~\cite{yen_superconducting_1967,hong_terahertz_2013} superconductors do not necessarily coincide with the properties measured in heterostructures.
Table adapted with permission from Ref.~\onlinecite{lutchyn_majorana_2018}, copyright 2018 Springer Nature.
}
\label{tab:materials}
\end{table}

\subsubsection{Bare energy dispersion}

\begin{figure}[t]
\includegraphics[width=\columnwidth]{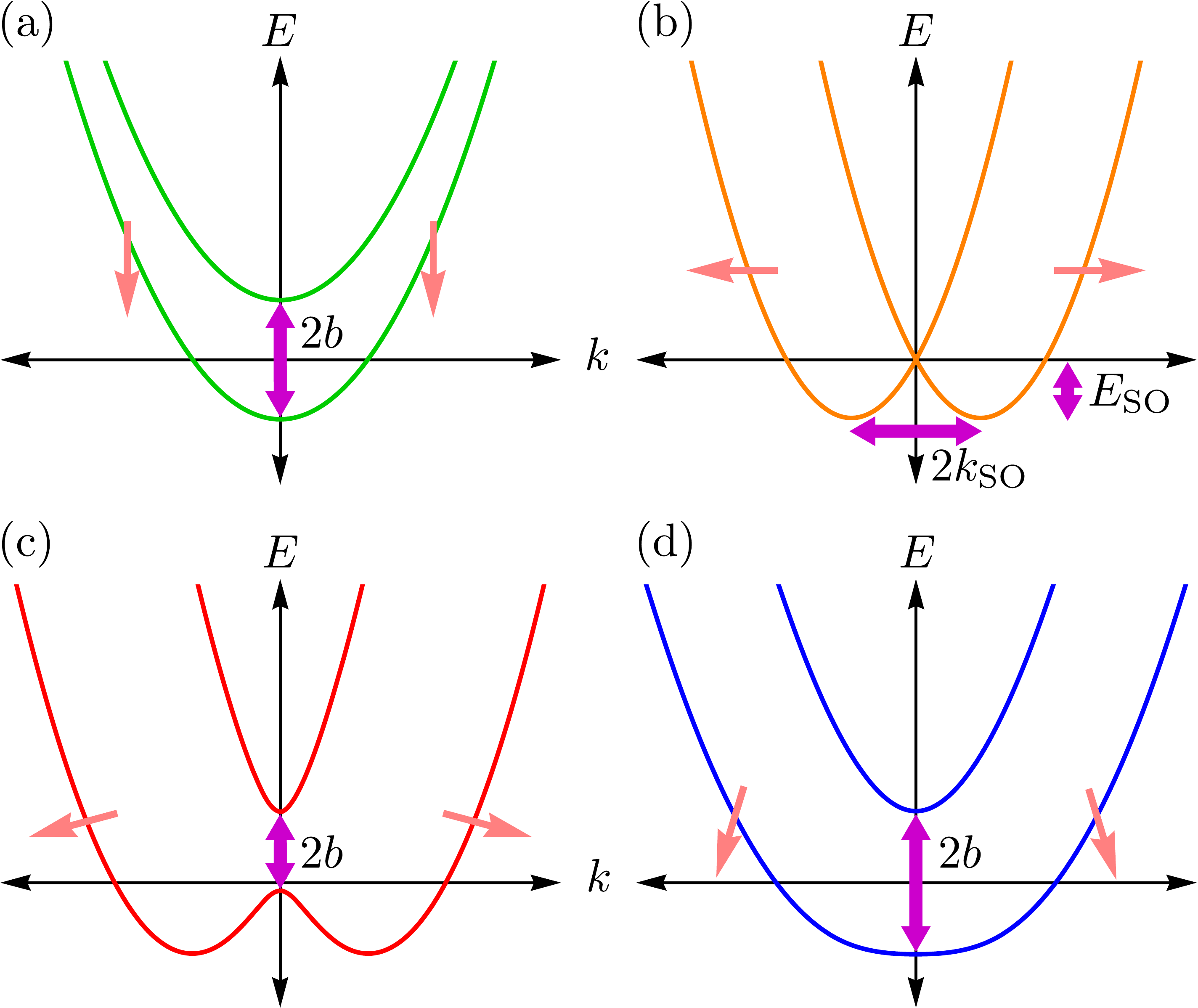}
\caption{
Energy dispersion and spin directions of the bare electrons in a Majorana nanowire in different regimes. 
(a)
Zero spin-orbit coupling and finite magnetic field:
The dispersion exhibits two parabolic bands shifted in energy and with opposite spins aligned with the field direction. 
Electrons with the same energy and opposite momenta have the same spin.
(b)
Finite spin-orbit coupling and zero magnetic field:
The dispersion exhibits two parabolic bands shifted on the momentum axis by $\pm k_\mathrm{SO}$, where electrons with the same energy and opposite momenta have opposite spins.
The energy bands are now lowered by a finite amount $E_\mathrm{SO}$, with minima at $\pm k_\mathrm{SO}$.
[(c) and (d)] Finite spin-orbit coupling and magnetic field perpendicular to each other:
The dispersion exhibits two bands separated by a finite gap $2b$ at zero momentum.
(c)
If the spin-orbit coupling dominates, i.e., for $2k_\mathrm{SO}^2>k_\mathrm{Z}^2$, the lower band has a double-well shape 
(minima at $k=\pm\sqrt{k_\mathrm{SO}^2-k_\mathrm{Z}^4/4k_\mathrm{SO}^2}$), 
and electrons with opposite directions have approximately opposite spins at large momenta.
(d)
If the magnetic field dominates, i.e., for $2k_\mathrm{SO}^2<k_\mathrm{Z}^2$, the double-well flattens out, and the spins progressively align with the magnetic field direction.
}
\label{fig:swavebarebulk}
\end{figure}

Let us first understand the physics of the Oreg-Lutchyn Hamiltonian in \cref{eq:H-swaveC} when the superconducting pairing is switched off.
For $\Delta=0$, the Fourier transform $\Psi(x)=(1/2\pi)\int\dd{k}\,\ee^{\ii kx}\Psi(k)$ yields in momentum space
\begin{equation}
\thinmuskip=1mu\medmuskip=1.5mu\thickmuskip=2mu
\mathcal{H}=\!\int\!\dd k\,
\Psi^\dag(k)\left(\frac{\hbar^2k^2}{2m} + {\alpha k} \sigma_y - \mu+ \mathbf{b}\cdot\bm{\sigma} \right)\Psi(k),
\end{equation}
where $\Psi^\dag(k)=[\Psi^\dag_{\up}(k),\Psi^\dag_{\down}(k)]$.
The Hamiltonian density of the bare electron can be written in a compact form as
\begin{equation}\label{eq:hkswave}
h(k)=\varepsilon_k+\mathbf{r}_k\cdot\bm{\sigma},
\end{equation}
with
\begin{equation}
\varepsilon_k=\dfrac{\hbar^2k^2}{2m}-\mu,
\qquad
\mathbf{r}_k=\mathbf{b}+ {\alpha k}\mathbf{\hat y},
\end{equation}
where $\varepsilon_k$ is the free electron dispersion and $\mathbf{r}_k$ an effective magnetic field given by the sum of the bare magnetic field and spin-orbit coupling, and which depends on the momentum $k$. 
The bare electron dispersion is easily obtained via the characteristic equation $\det(h(k)-E(k))=0$, which gives
\begin{align}
	E_\pm(k)&=\varepsilon_k\pm r_k=
	\nonumber\\
	&=\frac{\hbar^2k^2}{2m}-\mu\pm\sqrt{b_x^2+(b_y+\alpha k)^2+b_z^2},
	\label{eq:OLbaredispersion}
\end{align}
where $r_k=|\mathbf{r}_k|$. 
The two energy bands are nondegenerate at all momenta with $E_+(k)>E_-(k)$ for $b>0$, while they become degenerate at $k=0$ for $b=0$.
In both cases, the two bands are characterized by the direction of the electron spin.
Intuitively, one can expect that the electron spin aligns with the effective magnetic field $\mathbf{r}_k$.
Thus, let us consider the spin of the electron in the direction parallel to the unit vector $\mathbf{u}(k)=[u_x(k),u_y(k),u_z(k)]=\mathbf{r}_k/r_k$, described by the operator
\begin{equation}\label{eq:swavespindirection}
\mathbf{S}(k)=
\frac12 \ \frac{\mathbf{r}_k\cdot\bm\sigma}{r_k}=
\frac12 \ \mathbf{u}(k)\cdot\bm\sigma,
\end{equation}
which is the spin-1/2 operator along the direction of the effective field $\mathbf{r}_k$.
The spin operator $\mathbf{S}(k)$ commutes with the bare electron Hamiltonian density, which indeed can be written as 
$h(k)=\varepsilon_k+2r_k \mathbf{S}(k)=\varepsilon_k+r_k \mathbf{u}(k)$.
Hence, the two bands are eigenfunctions of the spin operator with eigenvalues $\pm1/2$, with wavefunctions given by
\begin{equation}
	\psi_\pm(k)=\frac{1}{\sqrt{2\pm 2u_z(k)}}\begin{bmatrix} 1 \pm u_z(k) \\ \pm u_x(k) \pm \ii u_y(k)\end{bmatrix}.
\end{equation}
The lower and higher energy bands $E_-(k)$ and $E_+(k)$ correspond to spin antiparallel and parallel to the effective magnetic field $\mathbf{r}_k=\mathbf{b}+\alpha k\hat{\mathbf{y}}$, given by the sum of the magnetic and spin-orbit fields.
Within each band, electrons can be effectively regarded as spinless since their spins are locked to their momenta via the effective magnetic field $\mathbf{r}_k$.
The presence of the spin-orbit coupling competing with the magnetic field tilts the spin direction in a way that depends on the momentum $k$:
at small momenta, electron spins are almost aligned with the magnetic field direction, while at larger momenta, they increasingly tilt in the direction of the spin-orbit coupling.
As a result, electrons with the same energy and opposite momenta have a spin direction tilted in different directions.
As already mentioned, this tilting is crucial to allow the pairing of electrons with opposite momenta by proximity effect.

The energy scales set by the chemical potential, Zeeman field, and spin-orbit coupling determine the shape of the bare electron dispersion and the Fermi momenta, i.e., the momenta where the dispersion crosses the Fermi energy $E(\pm k)=0$.
It is easy to verify that these Fermi momenta are 
$k_\mu=\sqrt{2m\mu}/\hbar$ for $b=\alpha=0$, 
$k_\mathrm{Z}=\sqrt{2mb}/\hbar$ for $\mu=\alpha=0$, and 
$2k_\mathrm{SO}=2m\alpha/\hbar^2$ for $b=\mu=0$.
More generally, the Fermi momenta are obtained as the solutions of the equation $\varepsilon_k^2=r_k^2$ [see \cref{eq:OLbaredispersion}], which simplifies to $\varepsilon_k^2= b^2+\alpha^2k^2$ if the magnetic field is in the $zx$ plane perpendicular to the spin-orbit coupling.
In this case, the bare electron dispersion crosses zero energy at the Fermi momenta $k_\mathrm{F_\pm}$ and $-k_\mathrm{F_\pm}$ with
\begin{equation}\label{eq:FermiMomenta}\thinmuskip=1mu\medmuskip=1.5mu\thickmuskip=2mu
k_\mathrm{F_\pm}=\sqrt{2 k_\mathrm{SO}^2 + k_\mu^2 \pm \sqrt{(2 k_\mathrm{SO}^2 + k_\mu^2)^2 - k_\mu^4 + k_\mathrm{Z}^4}},
\end{equation}
which gives a total of two $\pm k_\mathrm{F_+}$, three $\pm k_\mathrm{F_+}$, 0, or four Fermi momenta $k_\mathrm{F_\pm}$, $-k_\mathrm{F_\pm}$, respectively, if $|k_\mu|<|k_\mathrm{Z}|$, $|k_\mu|=|k_\mathrm{Z}|$, or $|k_\mu|>|k_\mathrm{Z}|$ (equivalently, if $\mu<b$, $\mu= b$, or $\mu>b$).
The two momenta $k_\mathrm{F_+}>k_\mathrm{F_-}$ are referred to as the exterior and interior Fermi momenta, corresponding respectively to the lower and higher energy bands $E_-(k)$ and $E_+(k)$ crossing the Fermi level at zero energy.

\Cref{fig:swavebarebulk}(a) shows the bare electron dispersion in the case of zero spin-orbit coupling and finite magnetic field.
The dispersion is given by two parabolic bands $E_\pm(k)=\hbar^2k^2/2m-\mu\pm b$ shifted in energy, corresponding to states with opposite spins aligned with the magnetic field direction.
Time-reversal symmetry is broken, Kramers degeneracy is lifted, and the spin direction does not depend on the momentum: 
Electrons with the same energy and opposite momenta have the same spin.
The gap between the two bands is equal to $2b$.
\Cref{fig:swavebarebulk}(b) shows the case of finite spin-orbit coupling and zero magnetic field.
The dispersion is given by two parabolic bands $E_\pm(k)=\hbar^2k^2/2m-\mu\pm\alpha k$ shifted by $\pm k_\mathrm{SO}$ on the momentum axis, corresponding to states with spin locked into opposite directions $\pm\hat{\mathbf{y}}$, respectively, and parallel to the spin-orbit coupling field.
The spin-orbit coupling does not break time-reversal symmetry, but the spin direction is now coupled to the momentum:
Electrons with the same energy and opposite momenta have opposite spin. 
The spin-orbit coupling shifts the energy bands by a finite amount $E_\mathrm{SO}=m\alpha^2/2\hbar^2$, with the minima of the energy dispersion $-(\mu+E_\mathrm{SO})$ at $\pm k_\mathrm{SO}$.
Figures~\ref{fig:swavebarebulk}(c) and~\ref{fig:swavebarebulk}(d) show the case where both the spin-orbit coupling and the magnetic field are nonzero, with the field perpendicular to the spin-orbit coupling, i.e., with $b_y=0$.
The dispersion is given by two bands $E_\pm(k)=\hbar^2k^2/2m-\mu\pm\sqrt{\alpha^2k^2+b^2}$ separated by a finite gap $2b$ at zero momentum.
The time-reversal symmetry is broken, and spin-degeneracy is completely lifted: 
Electrons with the same energy and opposite momenta have different spins, aligned as in \cref{eq:swavespindirection}.
If the spin-orbit coupling dominates, i.e., $2k_\mathrm{SO}^2>k_\mathrm{Z}^2$, the lower energy band has a double-well shape, 
with minima at $k=\pm\sqrt{k_\mathrm{SO}^2-k_\mathrm{Z}^4/4k_\mathrm{SO}^2}$.
In this case, electrons with opposite momenta and same energy have approximately opposite spin, as in \cref{fig:swavebarebulk}(c).
If the magnetic field dominates instead, $2k_\mathrm{SO}^2<k_\mathrm{Z}^2$, the double-well flattens out, and the spins tend to align with the magnetic field direction.
In this case, electrons with opposite momenta and same energy have approximately parallel spin as in \cref{fig:swavebarebulk}(d).

\subsubsection{Quasiparticle dispersion}

In the presence of a finite superconducting pairing $\Delta>0$
and in the gauge $\phi=0$, 
the Oreg-Lutchyn Hamiltonian in \cref{eq:H-swaveC} becomes in momentum space
\begin{align}
\mathcal{H}=\int \dd k\Big[&
\Psi^\dag(k)\left(\frac{\hbar^2k^2}{2m}+{\alpha k}\sigma_y-\mu+\mathbf{b}\cdot\bm{\sigma}\right)\Psi(k)
+\nonumber\\\label{eq:H-swaveCk}+&
\Psi(k)\left(\frac12\Delta
\ii\sigma_y\right)\Psi(
-k
)+\text{h.c.}
\Big].
\end{align}
By formally substituting $\Psi_s^\dag(k)\Psi_{s'}(k)\to(\Psi_s^\dag(k)\Psi_{s'}(k)-\Psi_{s'}(k)\Psi_s^\dag(k))/2$, the Hamiltonian above
can be written in the BdG form as $\mathcal{H}=\frac12\int\dd{k}\,\bm{\Psi}^\dag(k)\cdot H(k)\cdot\bm{\Psi}(k)$
with
\begin{equation}\label{eq:H-swave-kspaceC-BdG}\thinmuskip=1mu\medmuskip=1.5mu\thickmuskip=2mu
H(k)=
\begin{bmatrix}
h(k) & (\Delta\ii\sigma_y)^\dag \\
\Delta\ii\sigma_y & -h(-k)^\intercal \\
\end{bmatrix}
=
\begin{bmatrix}
h(k) & -\Delta\ii\sigma_y \\
\Delta \ii\sigma_y & -h(-k)^* \\
\end{bmatrix},
\end{equation}
where $\bm\Psi^\dag(k)=[\Psi^\dag_{\up}(k),\Psi^\dag_{\down}(k),\Psi_{\up}(-k),\Psi_{\down}(-k)]$ is the Nambu spinor in momentum space. 
It can be convenient to redefine the Nambu spinor as $\bm\Psi^\dag(k)=[\Psi^\dag_{\up}(k),\Psi^\dag_{\down}(k),\Psi_{\down}(-k),-\Psi_{\up}(-k)]$ via a unitary transformation (see \cref{sec:nambu}), such that the Hamiltonian density $H(k)$ becomes
\begin{align}
H(k)=&
\begin{bmatrix}
h(k) & -\Delta \\
-\Delta & -\mathcal{T}h(k)\mathcal{T}^{-1} \\
\end{bmatrix}=
\nonumber\\
=&
\left(
\varepsilon_k + \alpha k \sigma_y
\right)
\tau_z + 
\mathbf{b}\cdot\bm{\sigma} 
-\Delta\tau_x,
\label{eq:H-swave-kspaceC-BdG-newnambu}
\end{align} 
where $\mathcal{T}=\ii\sigma_y\mathcal{K}$ is the time-reversal symmetry operator, and $\sigma_{xyz}$ and $\tau_{xyz}$ the Pauli matrices respectively in spin and particle-hole space.
In this basis, the hole sector of the Hamiltonian is easily obtained from the particle sector via antiunitary time-reversal symmetry, i.e., by changing the sign to all time-reversal invariant terms (e.g., kinetic energy, chemical potential, spin-orbit coupling, electric fields) but not to the terms which break time-reversal symmetry (e.g., magnetic field).
Be aware that, although slightly confusing, the two conventions in the choice of the Nambu basis are both used in the literature.

The quasiparticle dispersion for $\Delta>0$ is obtained as usual by solving the characteristic equation $\det(H(k)-E(k))=0$.
If the magnetic field is perpendicular to the spin-orbit coupling, such that $r_k^2=b^2+\alpha^2k^2$, the four dispersion branches $E_\pm(k)$ and $-E_\pm(k)$ are given by
\begin{equation}\label{eq:OLgappeddispersion}
	E_\pm(k)=\sqrt{\varepsilon_k^2+r_k^2+\Delta^2\pm2\sqrt{\varepsilon_k^2 r_k^2 + \Delta^2 b^2}}.
\end{equation}
The effect of the paring is to open a gap in the correspondence of the Fermi momenta in \cref{eq:FermiMomenta}.
In the case of a conventional superconductor, i.e., with no spin-orbit coupling $\alpha=0$ and in zero field $b=0$, the pairing opens a gap $2\Delta$ at the Fermi momenta $k=\pm k_\mu$, as expected.
This gap is renormalized for finite spin-orbit couplings and magnetic fields $\alpha,b>0$.
For small superconducting pairing $\Delta\ll\alpha k_\mathrm{F_+},b$, the width of the gap opening near the exterior Fermi momentum can be calculated by expanding the lowest energy band of the equation above, which yields
\begin{equation}
	\Delta E_\mathrm{SC}
	=2E_-(k_\mathrm{F_+})
	\approx\frac{2\alpha\Delta|k_\mathrm{F_+}|}{\sqrt{b^2+\alpha^2k_\mathrm{F_+}^2}},
\end{equation}
up to the second order in $\Delta$. 
The gap $\Delta E_\mathrm{SC}$ is always open for finite spin-orbit coupling $\alpha>0$ and pairing $\Delta>0$.
On the other hand, at zero momentum the quasiparticle dispersion becomes $E_\pm(0)=|b\pm\sqrt{\mu^2+\Delta^2}|$ and the gap is given by 
\begin{equation}\label{eq:EZ_gap}
	\Delta E_\mathrm{Z}=2E_-(0)=2|b-\sqrt{\mu^2+\Delta^2}|.
\end{equation}
Consequently, the particle-hole gap closes at $k=0$ if $b=\sqrt{\mu^2+\Delta^2}$.
The particle-hole gap $\Delta E_\mathrm{Z}$ is often referred to as the minigap, to distinguish it from the superconducting pairing gap 
$2\Delta$ opening at the Fermi momenta in a conventional superconductor. 
The closing of the gap corresponds to a topological quantum phase transition between trivial $b<\sqrt{\mu^2+\Delta^2}$ and nontrivial $b>\sqrt{\mu^2+\Delta^2}$ gapped phases~\cite{oreg_helical_2010,lutchyn_majorana_2010}, topologically equivalent to the trivial and nontrivial phases of a spinless topological superconductor, respectively, as I will show in the next Section.

\subsubsection{Band inversion and topological transition\label{sec:band-inversion}}	

\begin{figure*}[t]
\includegraphics[width=\textwidth]{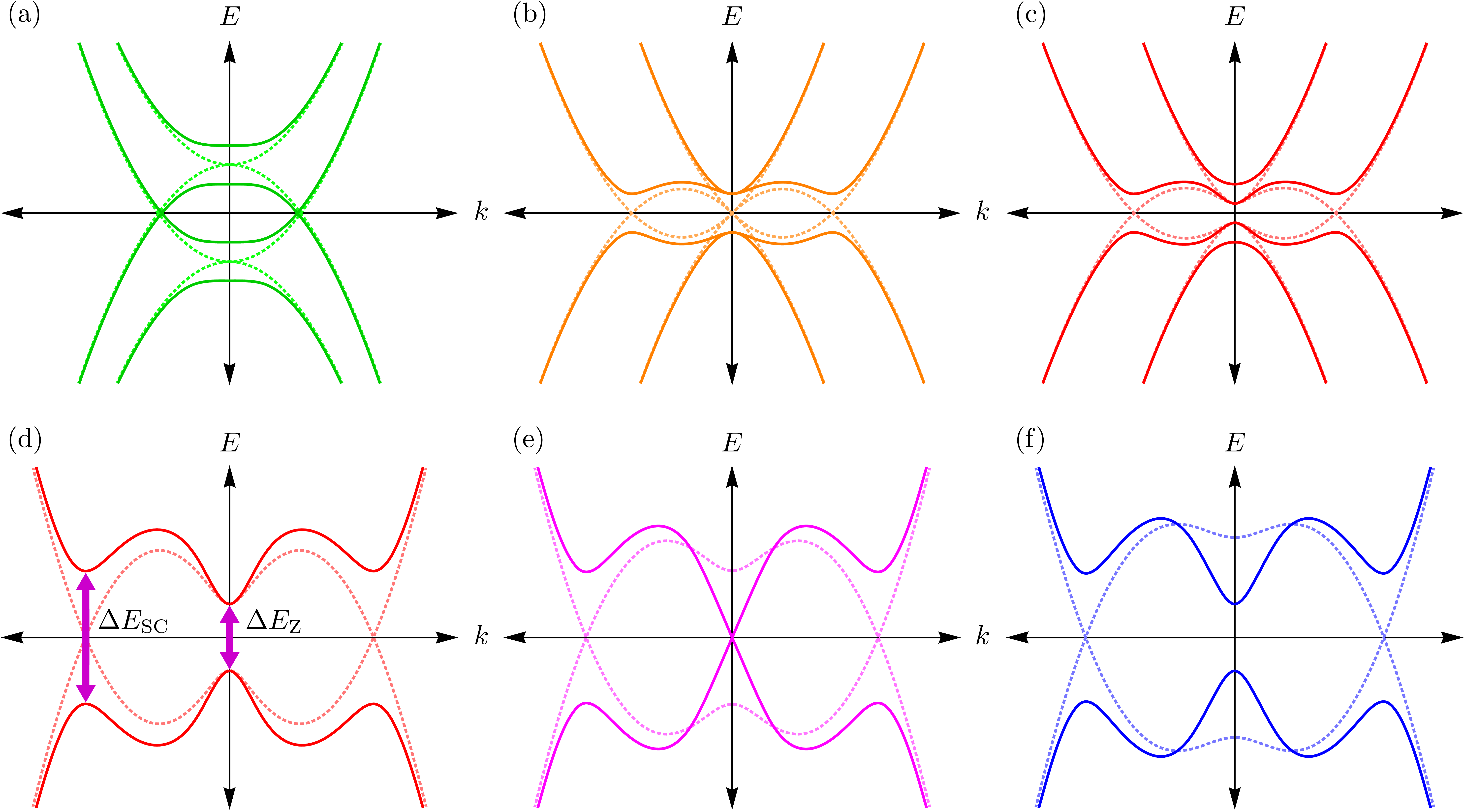}
\caption{
Energy dispersion $E_\pm(k)$ (continuous lines) of the superconducting quasiparticles compared with the bare electron dispersion $\varepsilon_k$ (dotted lines) in a Majorana nanowire in different regimes (both particle and hole levels with opposite energies are shown).
(a)
Zero spin-orbit coupling and finite magnetic field:
time-reversal symmetry is broken, but the gap is closed. 
(b)
Finite spin-orbit coupling and zero magnetic field:
the closing of the gap is avoided by the spin-orbit coupling, but time-reversal symmetry is unbroken. 
(c)
Finite spin-orbit coupling and magnetic field perpendicular to each other:
the gapped phase now breaks time-reversal symmetry. 
[(d)-(f)]
Transition between topologically trivial and nontrivial phases with finite spin-orbit coupling and magnetic fields.
The gap at finite momenta $\Delta E_\mathrm{SC}$ remains open for finite spin-orbit couplings and finite superconducting pairings.
The central gap $\Delta E_\mathrm{Z}$ is open at low magnetic fields $b<\sqrt{\mu^2+\Delta^2}$ (d), closes at the critical value $b=\sqrt{\mu^2+\Delta^2}$ (e), and reopens again for $b>\sqrt{\mu^2+\Delta^2}$ (f).
The two regimes correspond to topologically inequivalent phases, respectively trivial (d) and nontrivial phases (f) for $b^2\lessgtr\mu^2+\Delta^2$.
}
\label{fig:swavegappedbulk}
\end{figure*}

The Oreg-Lutchyn model~\cite{oreg_helical_2010,lutchyn_majorana_2010} contains all the necessary ingredients to engineer a 1D topological superconductor: magnetic field, proximity-induced superconducting pairing, and finite spin-orbit coupling~\cite{alicea_majorana_2010,sau_generic_2010,sau_robustness_2010,sau_non-abelian_2010}. 
This system can realize a nontrivial phase that is topologically equivalent to the nontrivial phase of the Kitaev model.
Precisely, one can show that the Oreg-Lutchyn Hamiltonian is equivalent to two decoupled spinless $p$-wave superconductors corresponding to two energy bands with opposite helicity~\cite{alicea_majorana_2010,shen_topological_2017_s104}.
Following Ref.~\onlinecite{shen_topological_2017_s104}, after a unitary rotation of the spin basis into the direction $\mathbf{u}_k=\mathbf{r}_k/r_k$ and a unitary Bogoliubov transformation, the Oreg-Lutchyn Hamiltonian in \cref{eq:H-swaveCk} decouples into two orthogonal sectors $\mathcal{H}=\mathcal{H}_++\mathcal{H}_-$ given by
\begin{equation}
\mathcal{H}_\pm=
\frac12
\int \dd k\,
\bm{\Psi}_\pm^\dag(k)
H_\pm(k)
\bm{\Psi}_\pm(k),
\end{equation}
where the Nambu spinors $\bm\Psi_\pm^\dag(k)=[\Psi^\dag_\pm(k),\Psi_\pm(-k)]$ represents states with opposite helicity, and with
\begin{equation}\label{eq:H-swave-kspaceC-ShenDirac}
	H_\pm(k)=\left( \sqrt{\varepsilon_k^2+\frac{\Delta^2 b^2}{r_k^2}} \pm r_k \right) \tau_z \pm \frac{\alpha k\Delta}{r_k}\tau_y.
\end{equation}
The two sectors $H_\pm(k)$ describe two decoupled spinless superconductors with an effective spin-triplet $p$-wave superconducting pairing equal to $\alpha k\Delta/r_k\approx\alpha k\Delta/b$ at small momenta.
The fact that the effective pairing depends explicitly on the magnetic field, proximitized pairing, and spin-orbit coupling, confirms that all these ingredients are necessary to realize topological superconductivity.
This Hamiltonian is topologically equivalent to the Hamiltonian of a 1D $p$-wave superconductor in \cref{eq:H-pwaveC}.
Indeed, at large momenta $k\to\infty$, the quadratic term dominates $H_\pm(k)\approx(\hbar^2/2m)k^2\tau_x$, and the energy levels are $\propto k^2$, as in \cref{eq:H-pwaveC}.
At small momenta instead, one has $\varepsilon_k\approx-\mu$ and $r_k\approx b$ at the first order in $k$, which yields
\begin{equation}
	H_\pm(k)\approx
	\left( \sqrt{\mu^2+\Delta^2}\pm b \right) \tau_z \pm \frac{\alpha k \Delta}{b}\tau_y,
\end{equation}
which is formally equivalent to the Dirac Hamiltonian of a Majorana fermion with a mass gap $\propto\sqrt{\mu^2+\Delta^2}\pm b$ and analogous to the Hamiltonian of a 1D $p$-wave superconductor in \cref{eq:H-pwaveC} at small momenta $p\approx0$.
The Hamiltonian sector $H_+(k)$ is always gapped and can be continuously deformed into the Hamiltonian of a free electron, and it is thus topologically trivial.
The Hamiltonian sector $H_-(k)$ instead has two gapped phases with a positive and negative mass gap, respectively, for $b^2\lessgtr\mu^2+\Delta^2$, separated by a topological quantum phase transition at $b=\sqrt{\mu^2+\Delta^2}$, where the particle-hole gap closes at $k=0$.
The gapped phase with a positive mass gap is topologically equivalent to the vacuum, and it is thus topologically trivial.
On the other hand, the gapped phase with a negative mass gap at large magnetic fields $b>\sqrt{\mu^2+\Delta^2}$ is topologically nontrivial.
The fact that the particle-hole gap closes exactly at $k=0$ is again not coincidental: it is a direct consequence of the $p$-wave symmetry of the effective superconducting pairing $\propto\alpha k\Delta/b$.
Since the $p$-wave pairing is an odd function of the momentum, it can only vanish at $k=0$.
Thus, the particle-hole gap can close only at $k=0$ if the mass gap $\sqrt{\mu^2+\Delta^2}-b$ vanishes altogether.

\begin{figure}[t]
\includegraphics[width=\columnwidth]{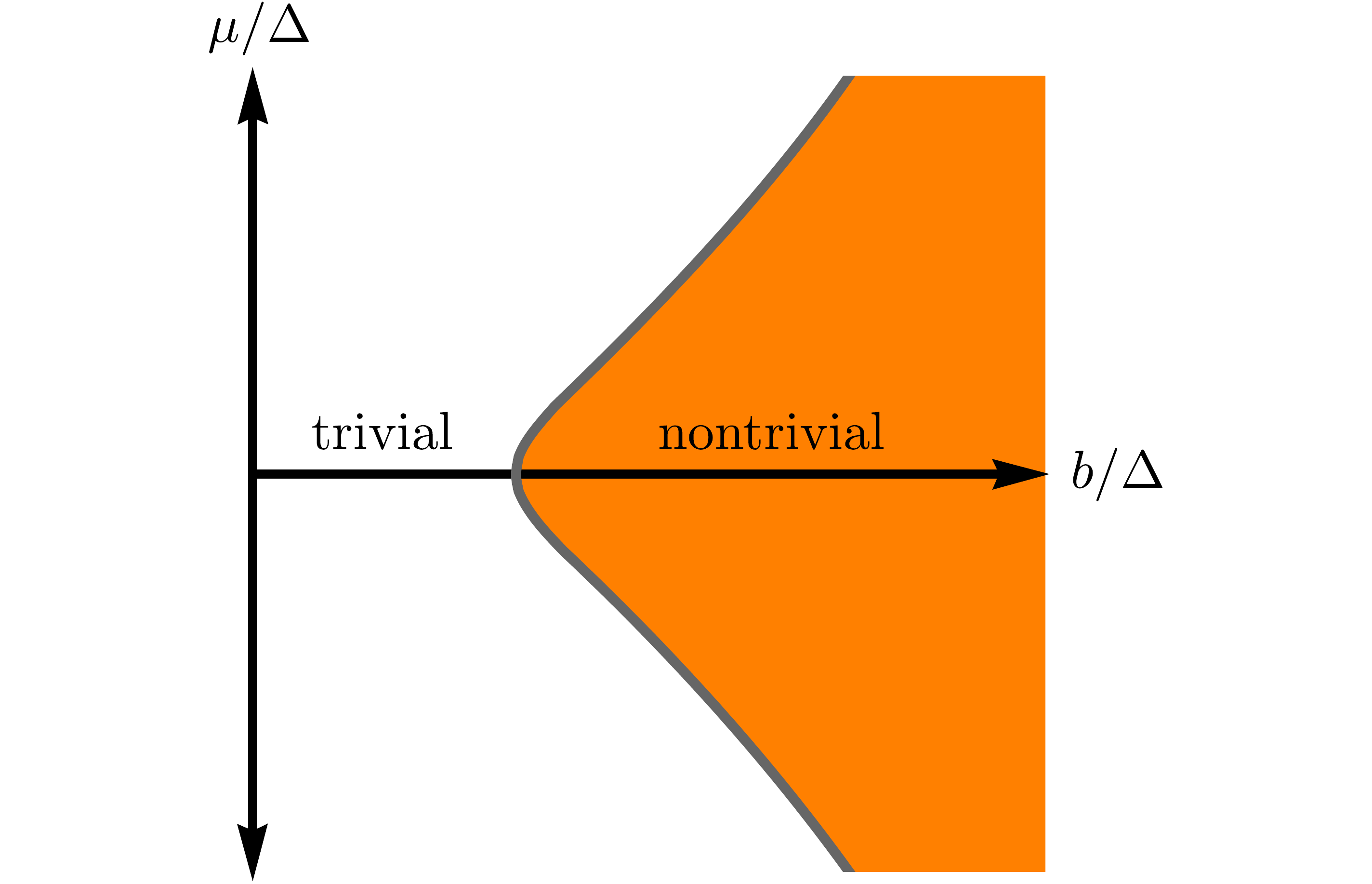}
\caption{
Phase space of a Majorana nanowire as a function of the magnetic field $b$ and chemical potential $\mu$.
The topologically trivial and nontrivial phases are realized respectively at low $b<\sqrt{\mu^2+\Delta^2}$ and high magnetic fields $b>\sqrt{\mu^2+\Delta^2}$.
}
\label{fig:phasespaceNW}
\end{figure}

\Cref{fig:swavegappedbulk} shows the energy dispersion $E_\pm(k)$ of the superconducting quasiparticles compared with the bare electron dispersion $\varepsilon_k$ for different regimes.
\Cref{fig:swavegappedbulk}(a) shows the quasiparticle dispersion in the case of zero spin-orbit coupling and finite magnetic field.
In this case, time-reversal symmetry is broken, the Kramers degeneracy is lifted, but the gap is closed at finite momenta. 
This is because electrons at zero energy have parallel spin, and thus their energy cannot be gapped by the spin-singlet superconducting pairing, which only couples antiparallel spin components.
\Cref{fig:swavegappedbulk}(b) shows the case of finite spin-orbit coupling and zero magnetic field.
In this case, the particle-hole gap remains open, but the gapped phase does not break time-reversal symmetry and thus cannot realize nontrivial superconductivity. 
\Cref{fig:swavegappedbulk}(c) shows the case where both the spin-orbit coupling and the magnetic field are nonzero, with the field perpendicular to the spin-orbit coupling.
In this case, time-reversal symmetry is broken, and spin-orbit coupling allows a finite superconducting pairing between electrons with opposite momenta.
Hence, the particle-hole gap remains open at finite fields, closing only at $b=\sqrt{\mu^2+\Delta^2}$, which results in a topological phase transition between topologically inequivalent phases.
Figures~\ref{fig:swavegappedbulk}(d)-\ref{fig:swavegappedbulk}(f) show the transition between the topologically trivial phase at low magnetic fields $b<\sqrt{\mu^2+\Delta^2}$ (d) and the topologically nontrivial phase at high fields $b>\sqrt{\mu^2+\Delta^2}$ (f), separated by a topological quantum phase transition $b=\sqrt{\mu^2+\Delta^2}$ (e) corresponding to the closing of the particle-hole gap $\Delta E_\mathrm{Z}$ at zero momentum.
The phase space of the Majorana nanowire is shown in \cref{fig:phasespaceNW}.

In the above discussion, the magnetic field direction has been chosen to be parallel to the wire direction ($x$ axis), perpendicular to the surface of the superconductor ($z$ axis), or lying in the plane $zx$ [see \cref{fig:nanowire}(a)].
Indeed, if the magnetic field is instead applied along the transverse direction $y$, i.e., parallel to the spin-orbit coupling field, the particle-hole gap cannot remain open at finite magnetic fields.
In this case, electrons have a well-defined spin along $y$ (it is a good quantum number) and have a finite energy splitting between states with spin $s_y=\pm\hbar/2$.
Consequently, electrons at zero energy have parallel spin, the superconducting pairing cannot lift their energy, and the particle-hole gap remains closed.
For arbitrary directions of the magnetic field $b_y=b\cos{\theta}$, the particle-hole gap decreases and finally closes for $|b_y|>\Delta$, preventing the existence of the topologically nontrivial phase~\cite{osca_effects_2014,rex_tilting_2014,perez-daroca_phase_2021}.

\subsection{Beyond the Oreg-Lutchyn model}

The Oreg-Lutchyn Hamiltonian is a minimal low-energy effective model that can be adapted and generalized to obtain a more realistic description of Majorana nanowires and similar systems.
For instance, the chemical potential, magnetic field, superconducting pairing, and spin-orbit coupling may become spatial-dependent, i.e., $\mu\to\mu(x)$, $\mathbf{b}\to\mathbf{b}(x)$, $\alpha\to\alpha(x)$, and $\Delta\to\Delta(x)$, in order to describe, e.g., smooth variations of the chemical potential or partial proximization of the wire~\cite{kells_near-zero-energy_2012,prada_transport_2012,chevallier_mutation_2012,stanescu_disentangling_2013,stanescu_nonlocality_2014}, spatially-dependent magnetic fields~\cite{martin_majorana_2012,kjaergaard_majorana_2012,mohanta_electrical_2019}, or the effect of disorder~\cite{brouwer_probability_2011,brouwer_topological_2011,pientka_enhanced_2012}.
Another straightforward generalization is to consider arbitrary directions of the spin-orbit coupling, for example, by rewriting the spin-orbit coupling term in \cref{eq:H-swaveC} as $\bm{\sigma}\cdot
(\bm{\alpha}(x)\times\mathbf{p})=(\alpha_z(x)\sigma_y - \alpha_y(x)\sigma_z) p_x$.
Another generalization
is obtained by allowing spatial variations of the superconducting phase $\Delta\to\Delta\ee^{\ii\phi(x)}$, which is unitarily equivalent to shifting the momentum $p_x\to p_x-\tau_z\partial_x\phi/2$, with the gradient of the phase proportional to the current density, in order to describe the effect of supercurrents driven through the superconducting substrate~\cite{romito_manipulating_2012,lesser_majorana_2022}.
Moreover, nanowires typically exhibit several occupied bands near the Fermi level.
Hence, a more realistic model must include the effect of the multiband dispersion, i.e., considering 2D or 3D Hamiltonian with strong confinement in the transverse direction and projecting into a low-energy sector near the Fermi energy~\cite{potter_majorana_2011,lutchyn_interacting_2011,lutchyn_search_2011,stanescu_majorana_2011,lim_magnetic-field_2012,tewari_topological_2012a,tewari_topological_2012b,san-jose_mapping_2014}.
Furthermore, the proximity-induced pairing acquires an explicit energy dependence at energies comparable with the bulk superconducting gap:
A more realistic description of proximity effects is obtained by including a self-energy term describing the coupling between the bulk superconductor and the wire~\cite{stanescu_proximity_2010,sau_robustness_2010,sau_non-abelian_2010,potter_engineering_2011,stanescu_majorana_2011,sau_experimental_2012,cole_effects_2015,stanescu_proximity-induced_2017} and solving the resulting Hamiltonian self-consistently.
Furthermore, the model may be generalized to include many-body interactions~\cite{gangadharaiah_majorana_2011,stoudenmire_interaction_2011,lutchyn_interacting_2011,lobos_interplay_2012,manolescu_coulomb_2014}.
The Oreg-Lutchyn Hamiltonian is also the starting point to study more realistic nanowire geometries, e.g., hexagonal nanowires with epitaxially grown superconducting coatings, shown in \cref{fig:nanowire}(c), by taking into account the 3D geometry of the heterostructure~\cite{nijholt_orbital_2016,vuik_effects_2016,woods_effective_2018,antipov_effects_2018,winkler_unified_2019,escribano_effects_2019,escribano_improved_2020}, also including orbital effects~\cite{nijholt_orbital_2016,winkler_unified_2019}, and the effect of the electrostatic environment via self-consistent Schrödinger-Poisson approaches~\cite{vuik_effects_2016,dominguez_zero-energy_2017,mikkelsen_hybridization_2018,antipov_effects_2018,woods_effective_2018,escribano_interaction-induced_2018,wojcik_tuning_2018,winkler_unified_2019,escribano_effects_2019,escribano_improved_2020}.

Moreover, the Oreg-Lutchyn Hamiltonian can be modified to effectively describe other experimental setups.
These setups include the so-called full-shell nanowires, i.e., semiconducting nanowires fully coated by a superconducting shell~\cite{vaitiekenas_flux-induced_2020,penaranda_even-odd_2020}, shown in \cref{fig:nanowire}(d).
In these devices, a magnetic field parallel to the wire induces a finite magnetic flux through the nanowire section and the consequent winding of the superconducting phase around the wire.
Moreover, the breaking of the local inversion symmetry at the interface between the wire and the superconductor provides a radial electric field that corresponds to a finite spin-orbit coupling.
By approximating the hexagonal wire as a cylinder, the zero angular momentum sector of the Bogoliubov-de~Gennes Hamiltonian becomes equivalent to the Oreg-Lutchyn Hamiltonian, with an effective Zeeman field determined by the magnetic flux~\cite{vaitiekenas_flux-induced_2020}.
Another setup is realized by hexagonal nanowires where two side facets are covered by a ferromagnetic insulator (e.g., EuS), and two facets covered by a superconductor, with one facet covered by both~\cite{vaitiekenas_zero-bias_2021}, shown in \cref{fig:nanowire}(e).
In these hybrid nanowires, the role of the Zeeman field is played by the exchange coupling induced by the proximity effect from the ferromagnetic insulator~\cite{liu_electronic_2021,maiani_topological_2021,langbehn_topological_2021,escribano_tunable_2021,woods_electrostatic_2021}.
Finally, the Oreg-Lutchyn model can be employed to describe other 1D platforms, such as metallic nanowires~\cite{hell_two-dimensional_2017,pientka_topological_2017} or 1D stripes in epitaxial semiconductor-superconductor heterostructures (e.g., planar Josephson junctions)~\cite{shabani_two-dimensional_2016,hell_two-dimensional_2017,hell_coupling_2017,pientka_topological_2017,haim_benefits_2019,setiawan_topological_2019,melo_supercurrent-induced_2019,stern_fractional_2019,zhou_phase_2020,laeven_enhanced_2020,mohanta_skyrmion_2021,paudel_enhanced_2021}.

\subsection{Discrete lattice (tight-binding) model}

In order to use numerical methods to diagonalize the Hamiltonian and calculate relevant physical observables, it is useful to discretize the continuum Hamiltonian in \cref{eq:H-swaveC} into a lattice (tight-binding) model (see \cref{sec:discretization}).
By doing so, one obtains a discrete Hamiltonian that reads
\begin{align}\thinmuskip=1mu\medmuskip=1.5mu\thickmuskip=2mu
\mathcal{H}=&
\sum_{n=1}^{N-1}
\begin{bmatrix} c^\dag_{n\up}, & c^\dag_{n\down} \\ \end{bmatrix}
\!\cdot\! \left( -t -\ii\widetilde{\alpha}\sigma_y \right) \!\cdot\! 
\begin{bmatrix} c_{n+1\up} \\ c_{n+1\down} \\ \end{bmatrix}
+\text{h.c.}
+\nonumber\\&
+\sum_{n=1}^N 
\begin{bmatrix} c^\dag_{n\up}, & c^\dag_{n\down} \\ \end{bmatrix}
\!\cdot\! \left( \mathbf{b} \!\cdot\! \bm{\sigma}+2t-\mu \right) \!\cdot\! 
\begin{bmatrix} c_{n\up} \\ c_{n\down} \\ \end{bmatrix}
+\nonumber\\\label{eq:H-swave}
&+\sum_{n=1}^N 
\begin{bmatrix} c_{n\up}, & c_{n\down} \\ \end{bmatrix}
\!\cdot\! \left(\frac12\Delta\ee^{\ii\phi}\ii\sigma_y\right) \!\cdot\! 
\begin{bmatrix} c_{n\up} \\ c_{n\down} \\ \end{bmatrix}
+\text{h.c.},
\end{align}
where $[c^\dag_{n\up},c^\dag_{n\down}]$ is the spinor whose components are the fermionic operators with up and down spin on each lattice site, $t=\hbar^2/2ma^2$ the hopping parameter, 
$\widetilde{\alpha}=\alpha/2a=\sqrt{tE_\mathrm{SO}}$ 
the spin-flip hopping parameter, and $N$ the number of sites of the discrete lattice with lattice parameter $a$.
For simplicity, one can assume $\phi=0$ as usual.
Note that the lattice parameter of the discrete Hamiltonian does not need to coincide with the microscopic atomic lattice parameter, and it is usually chosen to be much larger.
This makes numerical computations much more efficient, and it is justified by the fact that the typical length scales (e.g., nanowire length, Majorana localization length $\xi_\mathrm{M}$, spin-orbit length $l_\mathrm{SO}=k_\mathrm{SO}^{-1}$) as well as possible inhomogeneities of the magnetic field, chemical potential, and superconducting pairing, are much larger than the atomic length scales.

One can rewrite the Hamiltonian in the BdG form analogously to the case of the spinless topological superconductor. 
In the gauge $\phi=0$, using the fermion anticommutation relations, one obtains the identity $c_{ns}^\dag c_{n's'}=(\delta_{nn'}\delta_{ss'}+c_{ns}^\dag c_{n's'}-c_{n's'}c_{ns}^\dag)/2$, which yields
\begin{align}
\mathcal{H}=
\frac12\!\sum_{n=1}^N\!
\bm\Psi_n^\dag
\cdot
&
{
\begin{bmatrix}
\mathbf{b}\!\cdot\!\bm{\sigma}+2t-\mu	&	-\Delta\ii\sigma_y \\
\Delta\ii\sigma_y						&	-\left(\mathbf{b} \!\cdot\! \bm{\sigma}+2t-\mu\right)^\intercal
\end{bmatrix}}
\cdot
\bm\Psi_n
+\nonumber\\\label{eq:H-swave-BdG}
+\frac12\!\sum_{n=1}^{N-1}\!
\bm\Psi_n^\dag
\cdot
&
{
\begin{bmatrix} 
-t-\ii\widetilde\alpha\sigma_y & 0 \\
0 & t+\ii\widetilde\alpha\sigma_y \\
\end{bmatrix}}
\cdot
\bm\Psi_{n+1}
+\text{h.c.},
\end{align}
up to a constant term, and where $\bm\Psi_n^\dag=[c^\dag_{n\up},c^\dag_{n\down},c_{n\up},c_{n\down}]$ are the Nambu spinors.
Note that the Hamiltonian in \cref{eq:H-swave-BdG} is identical to the Hamiltonian in \cref{eq:H-swave} (up to a constant term).
The BdG Hamiltonian can be written in matrix form as
\begin{equation}
	\mathcal{H}=\frac12
	\bm{\Psi}^\dag
	\cdot
	\begin{bmatrix}
	\hat{H}_0 & \hat{\Delta}^\dag\\
	\hat{\Delta} & -\hat{H}_0^\intercal & \\
	\end{bmatrix}
	\cdot
	\bm{\Psi},
\end{equation}
where $\bm{\Psi}^\dag=[c_{1\up}^\dag,c_{1\down}^\dag,\ldots,c_{N\up}^\dag,c_{N\down}^\dag,c_{1\up},c_{1\down},\ldots,c_{N\up},c_{N\down}]$ and
\begin{equation}
\hat{H}_0=
{\arraycolsep=0.8\arraycolsep
\begin{bmatrix}
V & T & & \\
T^\dag & V & & \\[-2mm]
& & \hspace{-2mm}\ddots\hspace{-2mm} & \\[-1mm]
& & &\hspace{-1mm} V \\
\end{bmatrix}},
\quad
\hat{\Delta}=
\Delta\!
{\arraycolsep=0.8\arraycolsep
\begin{bmatrix}
\ii\sigma_y & 0 & & \\
0 & \ii\sigma_y & & \\[-2mm]
& & \hspace{-2mm}\ddots\hspace{-2mm} & \\[-1mm]
& & &\hspace{-1mm} \ii\sigma_y \\
\end{bmatrix}},
\end{equation}
where $V=\mathbf{b}\cdot\bm{\sigma}+2t-\mu$ and $T=-t-\ii\widetilde\alpha\sigma_y$.
The tight-binding model can naturally incorporate the effect of random disorder in the wire and in the superconducting substrate~\cite{akhmerov_quantized_2011,liu_zero-bias_2012,pikulin_a-zero-voltage_2012,rainis_towards_2013,cole_proximity_2016,pan_physical_2020}.
Furthermore, the discrete model can describe several other physical setups, e.g., arrays of superconducting quantum dots~\cite{sau_realizing_2012,fulga_adaptive_2013} or arrays of magnetic atoms deposited on a conventional superconductor substrate~\cite{choy_majorana_2011,nadj-perge_proposal_2013,vazifeh_self-organized_2013,pientka_unconventional_2014,heimes_majorana_2014}.

After imposing periodic boundary conditions, via a Fourier transform $c_{ns}=(1/\sqrt{N})\sum_k\ee^{\ii kna}c_{ks}$, and again using the fermion anticommutation relations which yield $c_{ks}^\dag c_{k's'}=(\delta_{kk'}\delta_{ss'}+c_{ks}^\dag c_{k's'}-c_{k's'}c_{ks}^\dag)/2$, one recovers the BdG Hamiltonian in \cref{eq:H-swave-kspaceC-BdG} or \cref{eq:H-swave-kspaceC-BdG-newnambu}, where $h(k)$ is now given by
\begin{equation}\label{eq:H-swave-kspace-BdG0}
	h(k)=\varepsilon_k+\mathbf{r}_k\cdot\bm{\sigma},
\end{equation}
with
\begin{equation}\label{eq:H-swave-kspace-BdG}
	\varepsilon_k=2t-\mu-2t\cos{(ka)},
	\quad
	\mathbf{r}_k=\mathbf{b}+2\widetilde\alpha\sin{(ka)}\ \mathbf{\hat y},
\end{equation}
where the momenta are quantized as $ka=2\pi m/N$ with $m=1,\ldots,N$ integer.
Analogously to the continuous case, the Hamiltonian has dispersion given by \cref{eq:OLgappeddispersion} and can be rewritten as in \cref{eq:H-swave-kspaceC-BdG-newnambu} in terms of the Pauli matrices.
At small momenta, the discrete Hamiltonian approximates the continuous Hamiltonian in \cref{eq:hkswave}, as one can directly verify by substituting $\sin{(ka)}\approx ka$, $\cos{(ka)}\approx1-(ka)^2/2$ in the equation above (see \cref{sec:discretization}).
Physically, the only difference compared with the continuous case is the finite bandwidth in the bare electron dispersion and the presence of a well-defined Brillouin zone.

\subsection{Topological invariant\label{sec:Dinvariant}}

The first step to understanding the topological properties of the Oreg-Lutchyn model and any of its generalizations is to determine its symmetry classification.
For a superconductor made of ordinary spin-1/2 electrons, the particle-hole, time-reversal, and chiral symmetries are defined by 
\begin{equation}\label{eq:CTS}
	\mathcal{C}= \tau_x \mathcal{K}, 	\qquad
	\mathcal{T}= \ii\sigma_y \mathcal{K}, \qquad
	\mathcal{S}= \mathcal{T} \mathcal{C} = \ii\sigma_y \tau_x, 
\end{equation}
in the basis where the Nambu spinor is given by $\bm\Psi^\dag(k)=[\Psi^\dag_{\up}(k),\Psi^\dag_{\down}(k),\Psi_{\up}(-k),\Psi_{\down}(-k)]$ or, alternatively,
\begin{equation}\label{eq:CTSnewnambu}
	\mathcal{C}= \sigma_y\tau_y \mathcal{K}, \qquad
	\mathcal{T}= \ii\sigma_y \mathcal{K}, \qquad
	\mathcal{S}= \mathcal{T} \mathcal{C} = \ii\tau_y,
\end{equation}
in the unitary equivalent basis where the Nambu spinor becomes $\bm\Psi^\dag(k)=[\Psi^\dag_{\up}(k),\Psi^\dag_{\down}(k),\Psi_{\down}(-k),-\Psi_{\up}(-k)]$ (see \cref{sec:nambu}).
One has $\mathcal{C}^2=1$, $\mathcal{T}^2=-1$, and $\mathcal{S}^2=-1$.
In the superconducting state, particle-hole symmetry is always unbroken, i.e., $\mathcal{C}\mathcal{H}\mathcal{C}^{-1}=-\mathcal{H}$.
Time-reversal symmetry is unbroken by spin-orbit coupling and by the superconducting pairing term, as long as the superconducting phase $\phi$ is uniform.
Indeed, a uniform superconducting phase can always be gauged away $\phi\to0$ such that the superconducting pairing term becomes manifestly invariant under the antiunitary transformation $\mathcal{T}$.
However, time-reversal symmetry is broken by the magnetic field since one has
\begin{equation}
\mathcal{T} (\mathbf{b}\cdot\bm\sigma)\mathcal{T}^{-1}=-\mathbf{b}\cdot\bm\sigma.
\end{equation}
and in this case, chiral symmetry $\mathcal{S}=\mathcal{T}\mathcal{C}$ is broken as well.
Unbroken particle-hole and broken time-reversal symmetries correspond to the symmetry class D, where the topologically inequivalent phases in 1D are labeled by the $\mathbb{Z}_2$ invariant.

Generally, the symmetry operator $\mathcal{T}$ coincides with the physical time-reversal symmetry, which transforms the magnetic field as $\mathbf{B}\to-\mathbf{B}$.
However, concerning symmetry classes, the operator $\mathcal{T}$ does not need to coincide with the ordinary time-reversal symmetry, as already mentioned.
Indeed, one may consider a second set of artificial ``time-reversal'' and ``chiral'' symmetries~\cite{tewari_topological_2012a,tewari_topological_2012b} defined by
\begin{subequations}\label{eq:fakeTS}\begin{align}
	\label{eq:fakeTS-T}
	\mathcal{T}'&= \mathcal{K},
	\\
	\label{eq:fakeTS-S} 
	\mathcal{S}'&= \mathcal{T}' \mathcal{C}
\end{align}\end{subequations}
where $\mathcal{T}'^2=1$ and $\mathcal{S}'^2=1$ with $\mathcal{S}'=\tau_x$ or $\mathcal{S}'=\sigma_y\tau_y$ depending on the choice of the Nambu basis.
The artificial ``time-reversal'' $\mathcal{T}'$ and chiral $\mathcal{S}'$ symmetries are still unbroken by spin-orbit coupling and by superconducting pairings with a uniform phase.
However, regarding the magnetic field, one has $b_{xz}\to b_{xz}$ and $b_y\to -b_y$ under $\mathcal{T}'$:
Hence, the artificial ``time-reversal'' symmetry $\mathcal{T}'$ is broken by any finite component $b_y>0$ of the magnetic field in the $y$-direction, i.e., the direction of the spin-orbit coupling field.
Therefore, if the magnetic field is perpendicular to the direction of the spin-orbit coupling $y$, the ``time-reversal'' symmetry $\mathcal{T}'$ is unbroken.
Unbroken particle-hole and time-reversal symmetries with $\mathcal{T}'^2=\mathcal{S}'^2=1$ correspond to the symmetry class BDI, where the topologically inequivalent phases in 1D are labeled by the $\mathbb{Z}$ invariant.
Analogously to the spinless case in \cref{eq:winding}, the $\mathbb{Z}$ invariant can be calculated as the winding number of a vector 
constructed from the chiral-symmetric representation of the Hamiltonian~\cite{tewari_topological_2012a,tewari_topological_2012b,budich_equivalent_2013} (see also Ref.~\onlinecite{stanescu_majorana_2013}).

In a realistic experimental setup, the artificial ``time-reversal'' $\mathcal{T}'$ symmetry is inevitably broken by magnetic disorder, misalignment of the magnetic field, or interband spin-orbit couplings in multiband nanowires~\cite{lim_magnetic-field_2012}.
In all these cases, the topological invariant describing the Oreg-Lutchyn model in symmetry class D is the pfaffian $\mathbb{Z}_2$ invariant, which can be defined as~\cite{tewari_topological_2012a,budich_equivalent_2013}
\begin{equation}\label{eq:Z2invariantPfaffianSpinfull}
	(-1)^\nu=\sgn\left(\pf\left(H(0) \mathcal{S}' \right) \pf\left(H(\pi)\mathcal{S}' \right)\right),
\end{equation}
in momentum space, with $\mathcal{S}'$ depending on the choice of the Nambu basis.
The antisymmetry of the matrix $H(k)\mathcal{S}'$ at $k=0,\pi$ is guaranteed by particle-hole symmetry, as in the spinless case [see \cref{eq:Z2invariantPfaffian2}].

For the Oreg-Lutchyn model on a discrete lattice with $\phi=0$, using \cref{eq:H-swave-kspaceC-BdG}, one gets
\begin{equation}
(-1)^\nu=
\sgn\!\left(
\!\pf 
\begin{bmatrix}
-\ii\Delta\sigma_y & h(0) \\
- h(0)^\intercal &\hspace{-1mm} \ii\Delta\sigma_y \\
\end{bmatrix}
\!\pf 
\begin{bmatrix}
-\ii\Delta\sigma_y & h(\pi) \\
- h(\pi)^\intercal &\hspace{-1mm} \ii\Delta\sigma_y \\
\end{bmatrix}
\right),
\end{equation}
where $h(0)=\mathbf{b}\cdot\bm{\sigma}-\mu$ and $h(\pi)=\mathbf{b}\cdot\bm{\sigma}+4t-\mu$ from \cref{eq:H-swave-kspace-BdG0,eq:H-swave-kspace-BdG}, which yields
\begin{equation}\thinmuskip=1mu\medmuskip=1.5mu\thickmuskip=2mu
(-1)^\nu=\sgn\left((\mu^2+\Delta^2-b^2)[(4t-\mu)^2+\Delta^2-b^2]\right).
\end{equation}
In the physical regime, where the kinetic term dominates $t\gg\mu,b,\Delta$, the second factor in the equation above is always positive, and therefore one has
\begin{equation}
(-1)^\nu=
\begin{cases}
+1 & \text{ for } b^2<\mu^2+\Delta^2, \\
-1 & \text{ for } b^2>\mu^2+\Delta^2. \\
\end{cases}
\end{equation}
The topologically trivial ($\nu=0$) and nontrivial ($\nu=1$) phases are therefore realized for $b^2\lessgtr\mu^2+\Delta^2$, respectively.

\subsection{Properties of Majorana bound states\label{sec:MajoranaProperties}}

A 1D topological superconductor in the nontrivial phase exhibits Majorana modes at its boundaries due to the bulk-boundary correspondence~\cite{hatsugai_chern_1993,ryu_topological_2002,teo_topological_2010}.
A wire with open boundary conditions can be thought of as the same wire with the left and right ends connected to the vacuum, which is topologically trivial.
Therefore, the wire ends separate two topologically inequivalent phases, i.e., the nontrivial phase of the topological superconductor and the trivial vacuum.
However, topologically inequivalent phases cannot be connected by a smooth transformation without closing the particle-hole gap.
Consequently, the gap must vanish at the boundary between trivial and nontrivial phases.
This mandates the existence of a zero-energy mode localized at the boundary.
In the case of a Majorana nanowire, these zero-energy modes are the Majorana end modes $\gamma_\mathrm{L}$ and $\gamma_\mathrm{R}$ localized at the left and right ends of the wire.
The same reasoning applies when the topological nanowire is connected through a barrier to a metallic lead, a topologically trivial insulator, or a trivial superconductor:
Moreover, if the chemical potential, magnetic field, or superconducting pairing magnitude are nonuniform such that, e.g., $\sqrt{\mu(x)+\Delta(x)}\lessgtr b(x)$ for $x\lessgtr x_0$, only a segment of the Majorana nanowire becomes nontrivial, and consequently a Majorana mode localizes at the boundary between the trivial and nontrivial segments.

As a general statement, end modes in a topological superconductor are zero-energy and self-adjoint Majorana modes if their spatial separation is much larger than their localization length $L\gg\xi_\mathrm{M}$.
To see this, let us consider the wavefunction of an end mode in a 1D superconductor, which can be written as
\begin{align}
	\gamma&=
	\int \dd x\, \bm\psi(x)^\intercal
	\cdot\bm\Psi(x)
	=\nonumber\\\label{eq:MajoranaModeGeneral}
	&=\sum_{s=\updown}
	\int \dd x\, \left( u_{s}(x)\Psi_{s}(x) + v_{s}(x)\Psi^\dag_{s}(x) \right),
\end{align}
where $\bm\Psi^\dag(x)=[\Psi^\dag_{\up}(x),\Psi^\dag_{\down}(x),\Psi_{\up}(x),\Psi_{\down}(x)]$ is the Nambu spinor in real space, and where the wavefunction can be expressed as 
\begin{equation}
	\bm\psi(x)=\begin{bmatrix} u_{\up}(x) \\ u_{\down}(x) \\ v_{\up}(x) \\ v_{\down}(x) \\ \end{bmatrix}.
\end{equation}
Due to particle-hole symmetry $\mathcal{C}\mathcal{H}\mathcal{C}^{-1}=-\mathcal{H}$ the mode described by $\mathcal{C}\bm\psi(x)$ must have opposite energy with respect to the mode $\bm\psi(x)$.
Since we assumed the existence of only one mode on each side of the wire, and the hybridization between modes localized at the opposite ends can be neglected (assuming $L\gg\xi_\mathrm{M}$), the mode $\bm\psi(x)$ has zero energy and must coincide with its particle-hole conjugate $\mathcal{C}\bm\psi(x)=\bm\psi(x)$, i.e., $\bm\psi(x)$ is an eigenstate of the particle-hole symmetry operator $\mathcal{C}$ with zero energy.
This mandates $v_s^*(x)=u_s(x)$, which yields
\begin{equation}\label{eq:MajoranaCondition}
	\begin{bmatrix}	v^*_{\up}(x) \\ v^*_{\down}(x) \\ u^*_{\up}(x) \\ u^*_{\down}(x) \end{bmatrix}
	=
	\begin{bmatrix} u^\nostar_{\up}(x)\\ u^\nostar_{\down}(x) \\ v^\nostar_{\up}(x) \\ v^\nostar_{\down}(x)\end{bmatrix},
\end{equation}
which gives $\gamma^\dag=\gamma$, i.e., the localized zero-energy mode is described by self-adjoint Majorana operator.
Having zero energy, this mode commutes with the Hamiltonian.
Hence, Majorana zero modes are described by 
\begin{equation}
	\bm\psi(x)=
	\begin{bmatrix} u^\nostar_{\up}(x)\\ u^\nostar _{\down}(x) \\ u^*_{\up}(x) \\ u^*_{\down}(x) \end{bmatrix},
\end{equation}
which is an eigenstate of the particle-hole symmetry operator $\mathcal{C}$ and the Hamiltonian $\mathcal{H}$.

The above argument is strictly valid only in the infinite size limit, i.e., when the spatial separation of the Majorana modes is infinitely larger than their spatial separation.
Unfortunately, nanowires have a finite length in the real world.
Consequently, Majorana modes localized at the opposite ends become partially overlapped and hybridized when the wire length $L$ becomes comparable with their localization length $L\gtrsim\xi_\mathrm{M}$.
Indeed, the Majorana wavefunction decays exponentially with fast oscillations~\cite{klinovaja_composite_2012,das-sarma_splitting_2012,stanescu_dimensional_2013} as $\bm\psi\propto\ee^{-x/\xi_\mathrm{M}}\ee^{\pm\ii k_\mathrm{M}x}$ where $\xi_\mathrm{M}$ is the localization length and $k_\mathrm{M}\sim k_\mathrm{F}$ the wavevector associated with the Majorana modes~\cite{das-sarma_splitting_2012}.
The resulting energy splitting becomes nonzero, being exponentially small in the system size~\cite{prada_transport_2012,das-sarma_splitting_2012,rainis_towards_2013,stanescu_dimensional_2013,fleckenstein_decaying_2018} and given by~\cite{das-sarma_splitting_2012}
\begin{equation}\label{eq:MajoranaOscillations}
	E_\mathrm{M}\approx\frac{\hbar^2 k_\mathrm{F}}{m\xi_\mathrm{M}} \cos{\left(k_\mathrm{M} L\right)}
	\ee^{-2L/\xi_\mathrm{M}}.
\end{equation}
Hence, the energy of a Majorana bound state approaches zero (up to exponentially small corrections) only for long wires $L\gg\xi_\mathrm{M}$, being strictly zero only in the limit $L/\xi_\mathrm{M}\to\infty$. 
Notice that, at the topological transition, the localization length diverges $\xi_\mathrm{M}\to\infty$, and the Majorana modes become fully delocalized on the whole wire.
Away from the topological transition instead, one has $\xi_\mathrm{M}\sim \alpha/\Delta$ and $k_\mathrm{M}\approx2k_\mathrm{SO}$ in the strong spin-orbit regime $E_\mathrm{SO}\gg b,\Delta$, and 
$\xi_\mathrm{M}\sim(\alpha/\Delta)b_z/E_\mathrm{SO}=2l_\mathrm{SO}b_z/\Delta$
and $k_\mathrm{M}\approx k_\mathrm{Z}$ in the weak spin-orbit regime $E_\mathrm{SO}\ll b,\Delta$~\cite{klinovaja_composite_2012,mishmash_approaching_2016} (see also Refs.~\onlinecite{aguado_majorana_2017,laubscher_majorana_2021}).
Since the wavevector $k_\mathrm{M}$ depends on the magnetic field $b$ and chemical potential $\mu$, the oscillating factor $\propto\cos{(k_\mathrm{M}L)}$ in the equation above leads to oscillations of the Majorana energy splitting as a function of the magnetic field and chemical potential.
These oscillations can be experimentally detected by measuring the variations in the differential conductance in tunneling experiments in the Coulomb blockade regime (see \cref{sec:DifferentialConductanceSplitting,sec:CoulombBlockade}), which would provide a strong signature of the presence of Majorana modes in short nanowires~\cite{das-sarma_splitting_2012,rainis_towards_2013}.
However, these oscillations may be suppressed by Coulomb interactions~\cite{das-sarma_splitting_2012,escribano_interaction-induced_2018} or the interaction with the dielectric surrounding~\cite{dominguez_zero-energy_2017}.

\begin{figure*}[t]
\includegraphics[width=\textwidth]{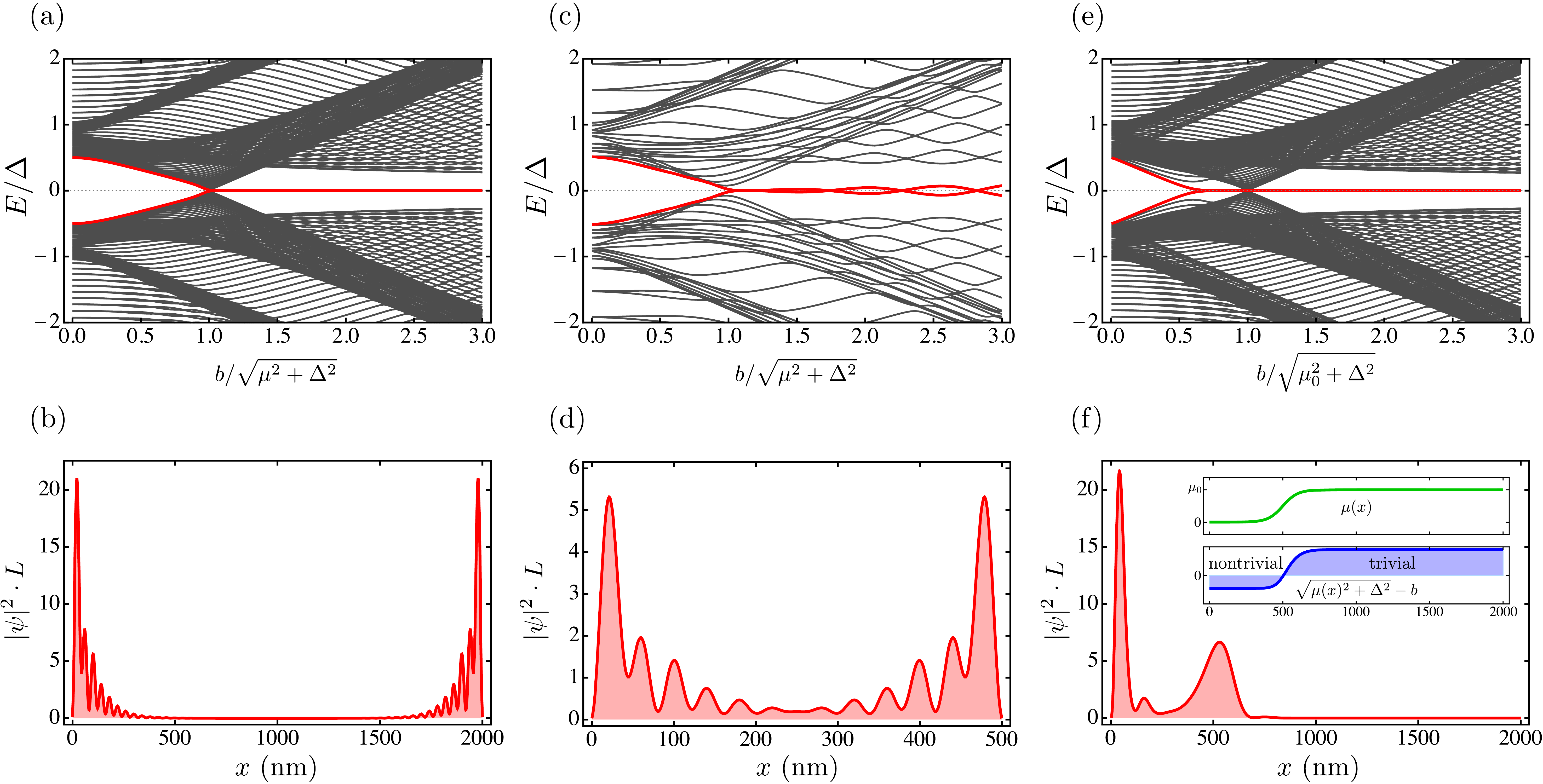}
\caption{
Topological phase transition and Majorana bound states in Majorana nanowires.
(a) Energy levels as a function of the magnetic field, showing the topological phase transition with the closing of the particle-hole gap at $b=\sqrt{\mu^2+\Delta^2}$, and (b) probability density $|\bm\psi|^2$ of the Majorana bound state in the nontrivial phase $b=1.5\sqrt{\mu^2+\Delta^2}$, calculated numerically using the tight-binding model in \cref{eq:H-swave-BdG} for an InSb nanowire with $\mu=1.5\Delta$, $\Delta=\text{\SI{3}{\milli\eV}}$, and $\alpha=\text{\SI{1}{\eV\angstrom}}$ (cf.~\cref{tab:materials}) for a long wire of length $L=\text{\SI{2000}{\nano\meter}}$.
The lowest energy level is highlighted in color.
[(c) and (d)] 
Same as before, but for a short wire of length $L=\text{\SI{500}{\nano\meter}}$.
The Majorana bound state is a superposition of two Majorana modes exponentially localized at the opposite ends of the wire.
The hybridization between Majorana modes increases at shorter nanowire lengths, i.e., when the length $L$ becomes comparable with the Majorana localization length $\xi_\mathrm{M}$.
Consequently, the Majorana bound state acquires a finite energy, which oscillates as a function of the magnetic field.
(e) Energy levels as a function of the magnetic field 
and 
(f) probability density of the quasi-Majorana bound state for $b=0.7\sqrt{\mu_0^2+\Delta^2}$, right before the topological phase transition,
calculated for a spatially-dependent chemical potential $\mu(x)$ (shown in the inset).
The quasi-Majorana bound state (lowest energy level, highlighted in color) before the topological phase transition coalesces into a nontrivial Majorana bound state after the closing and reopening of the bulk gap with the onset of the topological phase transition to the nontrivial phase.
This state is formed by two overlapping Majorana modes localized respectively at the left end $x=0$ and close to $x=\text{\SI{500}{\nano\meter}}$, where the quantity $\sqrt{\mu(x)^2+\Delta^2}-b$ changes sign, as shown in the inset.
}
\label{fig:MajoDensity}
\end{figure*}

\Cref{fig:MajoDensity} shows the topological phase transition and the onset of the Majorana bound state in the nontrivial phase $b>\sqrt{\mu^2+\Delta^2}$ for different choices of the wire length. 
The topological phase transition at $b=\sqrt{\mu^2+\Delta^2}$ [\cref{fig:MajoDensity}(a) and~\ref{fig:MajoDensity}(c)] separates the trivial phase at low fields from the nontrivial phase at higher magnetic fields. 
In the nontrivial phase, Majorana end modes localize at the ends of the wire and decay exponentially towards the center [\cref{fig:MajoDensity}(b) and~\ref{fig:MajoDensity}(d)].
For long wires, the Majorana modes are well separated in space, and their energy is zero up to exponentially small corrections, as one can see in \cref{fig:MajoDensity}(a) and~\ref{fig:MajoDensity}(b).
The hybridization between Majorana modes in short nanowires, due to the overlap between their wavefunctions, induces a finite energy splitting $E_\mathrm{M}$ which oscillates as a function of the magnetic field, as one can see in \cref{fig:MajoDensity}(c) and~\ref{fig:MajoDensity}(d).
For short nanowires, the transition between the topologically trivial and nontrivial phase may occur without an apparent closing and reopening of the particle-hole gap~\cite{prada_transport_2012,stanescu_to-close_2012,stanescu_disentangling_2013,mishmash_approaching_2016,huang_metamorphosis_2018} with a subgap state at finite energy detaching from the bulk energy spectrum and gradually approaching zero energy~\cite{prada_transport_2012,penaranda_quantifying_2018,moore_two-terminal_2018,vuik_reproducing_2019}, as one can also see in \cref{fig:MajoDensity}(c).

As a direct consequence of their topological origin, Majorana modes are robust:
Their existence and their properties are not affected by perturbations, as long as these perturbations are local and do not close the particle-hole gap. 
Local perturbations having a length scale shorter than the Majorana localization length cannot couple Majorana modes localized at the opposite ends of the wire.
For instance, a Majorana nanowire exhibits 
spatially-separated 
Majorana modes in its nontrivial phase, even in the presence of 
disorder or other local perturbations, 
as long as the perturbations do not close the particle-hole gap.
Majorana modes are also expected to be free from quantum decoherence, which is a fundamental advantage over other physical implementations of quantum computing~\cite{kitaev_fault-tolerant_2003,nayak_non-abelian_2008}.
This is the direct consequence of fermion-parity conservation, spatial separation, and the presence of a finite particle-hole gap.
Indeed, transitions between occupied $n_\mathrm{M}=1$ to unoccupied Majorana bound state $n_\mathrm{M}=0$ are forbidden by the fermion-parity conservation (the two groundstates have a different fermion parity).
Moreover, the tunneling between contiguous Majorana modes is exponentially suppressed in their mutual distance, with tunneling rates $\propto\ee^{-L/\xi_\mathrm{M}}$~\cite{bonderson_quasi-topological_2013,das-sarma_majorana_2015,beenakker_search_2020}.
Furthermore, transitions to excited higher-energy states are thermally suppressed by the particle-hole gap, which separates the Majorana bound state from bulk excitations, with transition rates $\propto\ee^{-\Delta/k_\mathrm{B}T}$~\cite{cheng_topological_2012,bonderson_quasi-topological_2013,das-sarma_majorana_2015,beenakker_search_2020}.
However, in the real world, fermion parity conservation breaks down as soon as the Majorana wire is coupled to an external system, which is practically unavoidable in any realistic setting.
In a typical setup, the Majorana wire is proximitized by a bulk superconductor, capacitively coupled to external gates, and via tunneling junctions to metallic leads used to probe or manipulate Majorana modes.
In this case, the conservation of fermion parity is broken due to the so-called quasiparticle poisoning~\cite{leijnse_scheme_2011}, i.e., the tunneling of superconducting quasiparticles from the bulk superconductor~\cite{rainis_majorana_2012} or a metallic lead~\cite{budich_failure_2012}.
These processes allow fermionic states to leak into the superconducting condensate and flip the fermion parity of the superconducting condensate.
Furthermore, decoherence may arise at finite temperatures due to transitions into gapped states~\cite{goldstein_decay_2011} mediated, e.g., by charge fluctuations in the metallic gates~\cite{schmidt_decoherence_2012,lai_exact_2018}, the residual charge of partially overlapping Majorana modes in finite wires~\cite{knapp_dephasing_2018}, nonuniform chemical potentials~\cite{aseev_lifetime_2018}, or electron-phonon interactions~\cite{aseev_degeneracy_2019}.
Other sources of decoherence appear as soon as one tries to dynamically control the Majorana modes to perform, e.g., braiding operations, due to nonadiabatic processes~\cite{cheng_nonadiabatic_2011,karzig_boosting_2013,scheurer_nonadiabatic_2013,karzig_optimal_2015,amorim_majorana_2015,knapp_the-nature_2016,sekania_braiding_2017,bauer_dynamics_2018} or thermal noise~\cite{pedrocchi_majorana_2015}.
All these processes contribute to a finite coherence time of Majorana modes. 

\subsection{Majorana bound states vs other near-zero energy localized states}

Apart from topologically-protected Majorana bound states, realistic Majorana nanowires may exhibit localized states with zero or near-zero energy below the bulk gap, induced by random disorder~\cite{liu_zero-bias_2012,pikulin_a-zero-voltage_2012,bagrets_class_2012,roy_topologically_2013,stanescu_disentangling_2013,pan_physical_2020,pan_generic_2020,das-sarma_disorder-induced_2021,pan_disorder_2021,pan_crossover_2021,pan_three-terminal_2021,pan_quantized_2021}, impurities~\cite{stanescu_nonlocality_2014,pan_crossover_2021}, strong interband coupling~\cite{bagrets_class_2012,woods_zero-energy_2019}, or finite-size effects~\cite{cayao_confinement-induced_2021}, smooth potentials~\cite{kells_near-zero-energy_2012,prada_transport_2012,stanescu_nonlocality_2014,liu_andreev_2017,penaranda_quantifying_2018,moore_two-terminal_2018,avila_non-hermitian_2019,vuik_reproducing_2019,stanescu_robust_2019,woods_zero-energy_2019,pan_physical_2020,pan_crossover_2021,pan_quantized_2021,pan_three-terminal_2021}, quantum dots~\cite{kells_near-zero-energy_2012,prada_transport_2012,stanescu_disentangling_2013,liu_andreev_2017,setiawan_electron_2017,ptok_controlling_2017,penaranda_quantifying_2018,moore_quantized_2018,moore_two-terminal_2018,liu_distinguishing_2018,reeg_zero-energy_2018,huang_metamorphosis_2018,stanescu_robust_2019,avila_non-hermitian_2019,vuik_reproducing_2019,lai_presence_2019,zeng_analytical_2019,sharma_hybridization_2020,pan_physical_2020,pan_three-terminal_2021,pan_quantized_2021,hess_local_2021}, or partial proximization of the nanowire~\cite{kells_near-zero-energy_2012,prada_transport_2012,chevallier_mutation_2012,stanescu_disentangling_2013,cayao_sns-junctions_2015,liu_andreev_2017,setiawan_electron_2017,penaranda_quantifying_2018,moore_quantized_2018,moore_two-terminal_2018,liu_distinguishing_2018,reeg_zero-energy_2018,huang_metamorphosis_2018,fleckenstein_decaying_2018,avila_non-hermitian_2019,vuik_reproducing_2019,stanescu_robust_2019,lai_presence_2019,sharma_hybridization_2020,pan_physical_2020,pan_three-terminal_2021,hess_local_2021,liu_majorana_2021}.
These states may exhibit signatures virtually indistinguishable from those produced by Majorana bound states~\cite{liu_andreev_2017,moore_quantized_2018,moore_two-terminal_2018,liu_distinguishing_2018,cao_decays_2019,lai_presence_2019,sharma_hybridization_2020,pan_physical_2020,pan_crossover_2021,pan_generic_2020,das-sarma_disorder-induced_2021,pan_quantized_2021,hess_local_2021}.
In contrast with topologically-protected Majorana bound states, however, these states may appear in the topologically trivial phase, not necessarily at the opposite ends of the nanowire, but typically near inhomogeneities or impurities, and are not exponentially localized~\cite{penaranda_quantifying_2018,moore_two-terminal_2018,avila_non-hermitian_2019,stanescu_robust_2019,pan_quantized_2021}.
Distinguishing between topologically nontrivial Majorana bound states and other localized states is the focus of intense research. 
A thorough account of the topic can be found in Ref.~\onlinecite{prada_from_2020}.

An emblematic example of such near-zero localized states are the so-called quasi-Majorana bound states (also called trivial Majorana bound states, partially-separated or partially-overlapping Majorana bound states), which localize near smooth variations of the chemical potential~\cite{kells_near-zero-energy_2012,prada_transport_2012,stanescu_nonlocality_2014,liu_andreev_2017,penaranda_quantifying_2018,moore_two-terminal_2018,avila_non-hermitian_2019,vuik_reproducing_2019,stanescu_robust_2019,woods_zero-energy_2019,pan_physical_2020,pan_crossover_2021,pan_quantized_2021,pan_three-terminal_2021} or superconducting order parameter~\cite{kells_near-zero-energy_2012,prada_transport_2012,chevallier_mutation_2012,stanescu_disentangling_2013,cayao_sns-junctions_2015,liu_andreev_2017,setiawan_electron_2017,penaranda_quantifying_2018,moore_quantized_2018,moore_two-terminal_2018,liu_distinguishing_2018,reeg_zero-energy_2018,huang_metamorphosis_2018,fleckenstein_decaying_2018,avila_non-hermitian_2019,vuik_reproducing_2019,stanescu_robust_2019,lai_presence_2019,sharma_hybridization_2020,pan_physical_2020,pan_three-terminal_2021,hess_local_2021,liu_majorana_2021}.
To illustrate one of the possible mechanisms which can produce near-zero energy quasi-Majorana bound states, let us consider the case where the chemical potential $\mu(x)$ is not uniform but smoothly drops from a finite value $\mu(x)=\mu_0$ to zero $\mu(x)=0$ at one end of the wire, as shown in the inset of \cref{fig:MajoDensity}(f).
In the case of inhomogeneous chemical potential, the Majorana nanowire is in the topologically nontrivial phase only if the magnetic field becomes $b>\sqrt{\mu(x)^2+\Delta^2}$ along the whole wire.
Conversely, it is in the trivial phase if $b<\sqrt{\mu(x)^2+\Delta^2}$ along the whole wire.
Other than that, there may be values of the magnetic field such that one segment of the wire is in the nontrivial phase, and the rest remains in the trivial phase.
In the specific example considered, for values of the field $b<|\Delta|\le\sqrt{\mu(x)^2+\Delta^2}$, the whole wire is in the topologically trivial phase, while for $b>\sqrt{\mu_0^2+\Delta^2}>\sqrt{\mu(x)^2+\Delta^2}$, the whole wire becomes nontrivial.
For intermediate values $|\Delta|<b<\sqrt{\mu_0^2+\Delta^2}$ instead, the leftmost segment of the wire becomes nontrivial being $b>\sqrt{\mu(x)^2+\Delta^2}$, while the rest of the wire remains trivial, being $b<\sqrt{\mu(x)^2+\Delta^2}$, as shown in the inset of \cref{fig:MajoDensity}(f).
Thus, at these intermediate values, one may expect the presence of Majorana modes localized at one end of the wire and at the boundary between the nontrivial and nontrivial segments.
\Cref{fig:MajoDensity}(e) shows the resulting energy spectra as a function of the magnetic field.
At $b\approx|\Delta|$, an energy level detaches from the bulk and becomes almost zero.
Thus, a near zero-energy state may appear at low fields before the closing and reopening of the bulk gap, i.e., before the onset of the topological phase transition to the nontrivial phase.
\Cref{fig:MajoDensity}(f) shows the probability density of this lowest energy level.
This state is given by a quasi-Majorana bound state formed by two overlapping Majorana modes, one localized at the left end of the nanowire, and the other close to one-fourth of the wire, where the quantity $\sqrt{\mu(x)^2+\Delta^2}-b$ changes sign, as shown in the inset of \cref{fig:MajoDensity}(f).
This point represents the boundary between nontrivial and trivial segments of the wire.
Intuitively, this regime can be described as a topologically inhomogeneous phase, where the \emph{local} topological invariant $\sgn(\sqrt{\mu(x)^2+\Delta^2}-b)$ assumes different values on different segments of the wire~\cite{penaranda_quantifying_2018,stanescu_robust_2019,marra_majorana/andreev_2022}.
However, it is important to note that, since the transition between nontrivial and trivial phases is driven by a smooth variation of the chemical potential, this second Majorana mode is not exponentially localized.
Finally, at $b=\sqrt{\mu_0^2+\Delta^2}$, the bulk gap closes and reopens, and for $b>\sqrt{\mu_0^2+\Delta^2}$, the whole wire becomes nontrivial, with the subgap state coalescing into a Majorana bound state [similar to the case considered in \cref{fig:MajoDensity}(a) and (b)].
The hybridization between the two Majorana modes of the quasi-Majorana bound state induces a finite energy splitting, which is however very small, being comparable to the hybridization of the two Majorana modes of the nontrivial Majorana bound state.
Indeed, the subgap energy level in \cref{fig:MajoDensity}(f) exhibits near-zero energy well before the topological transition.

\section{Experimental signatures\label{sec:experiments}}

\subsection{Differential conductance\label{sec:DifferentialConductance}}

\subsubsection{Quantized zero-bias peaks}

\begin{figure*}[t]
\includegraphics[width=\textwidth]{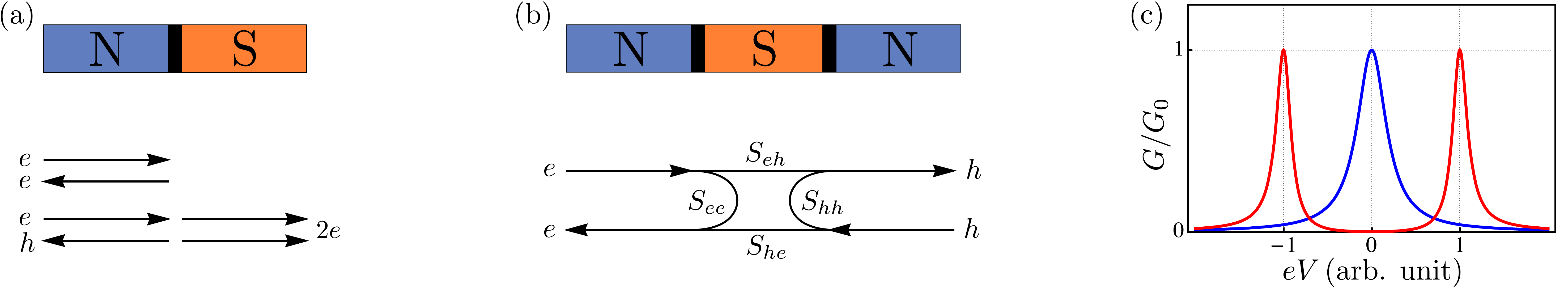}
\caption{
Quantized differential conductance in a Majorana nanowire.
(a)
In a typical conductance experiment, a proximitized nanowire (S) is connected to a metallic lead (N) through a tunneling barrier with a finite bias voltage.
When an electron from the lead arrives at the NS interface, the electron can be either normally-reflected with no net charge transferred or Andreev-reflected as a hole with two electrons transmitted through the barrier.
In this case, the incoming electron is transmitted through the barrier, coupling to another electron into a Cooper pair, and a hole propagates back into the metallic lead, with a net charge transfer of $2e$.
(b)
Andreev reflection can also be described as the transmission through a double barrier.
In this case, the metallic lead is separated into two leads, carrying electron and hole modes separately.
The incoming electron is reflected (normal reflection $S_{ee}$) or transmitted to the superconductor and then again transmitted as a hole into the hole lead (Andreev reflection $S_{eh}$).
(c)
In the presence of a Majorana mode at zero energy $E_\mathrm{M}=0$ and at zero temperature, Andreev reflections produce a zero-bias peak with a quantized height equal to the conductance quantum $G_0=2e^2/h$ at zero bias $eV=0$.
For a Majorana mode with a finite energy splitting $E_\mathrm{M}>0$, the zero-bias peak splits into two separate peaks at $eV=\pm E_\mathrm{M}$.
}
\label{fig:conductance}
\end{figure*}

A direct experimental signature of Majorana zero modes is the quantized peak in the differential conductance at zero bias~\cite{sengupta_midgap_2001,law_majorana_2009,flensberg_tunneling_2010,stanescu_majorana_2011,liu_detecting_2011,das-sarma_splitting_2012,fidkowski_universal_2012,prada_transport_2012,ioselevich_tunneling_2013,he_selective_2014}.
In a typical setup, the nanowire is connected to a metallic lead through a tunneling barrier with a finite voltage difference $V$, as in \cref{fig:nanowire}(a), and one measures the current $I$ through the barrier as a function of the applied voltage $V$.
This setup realizes a normal metal-superconductor (NS) junction between the metallic lead and the proximitized nanowire, as shown schematically in \cref{fig:conductance}(a).
In the presence of a Majorana mode at zero energy, the differential conductance exhibits a resonant peak at zero bias, which is quantized to the conductance quantum $G_0=2e^2/h$.

The differential conductance of an NS junction is the consequence of the so-called Andreev reflections.
When an electron from the lead arrives at the NS interface, two distinct physical processes may occur: normal reflection and Andreev reflection. 
In the case of normal reflection, the electron is simply reflected at the interface back into the metallic lead.
Normal reflection results in no net charge transferred through the barrier and zero current $I=0$.
In the case of Andreev reflection, the electron is transmitted through the barrier into the superconductor.
For energies above the particle-hole gap $E>\Delta E_\mathrm{ph}$, the transmitted electron produces a finite current into the superconductor.
Below the gap $E<\Delta E_\mathrm{ph}$, however, there are no propagating modes available in the superconductor. 
In this case, the transmitted electron is coupled to another electron with opposite spin to form a Cooper pair.
Charge conservation then mandates that a hole is created into the lead.
From the point of view of the normal lead, this process can be effectively described as an incident electron being reflected back as a hole with opposite momentum. 
The Andreev reflection from electron to hole results in a net charge transfer of $2e$ through the barrier.
Normal and Andreev reflection can be conceptually described as the transmission through a double barrier, as shown in \cref{fig:conductance}(b).
In this description, the metallic lead is separated into two leads, with each lead having only electron and hole propagating modes, respectively.
An incoming electron from the electron lead is then reflected back as an electron (normal reflection) or transmitted to the superconductor and then again transmitted as a hole into the hole lead, with the simultaneous creation of a Cooper pair into the superconductor (Andreev reflection).
This transmission process is described by
\begin{equation}\label{eq:transmission}
\begin{bmatrix}
\psi_{e}^+ \\
\psi_{h}^+
\end{bmatrix}
=
\begin{bmatrix}
S_{ee} & S_{eh} \\
S_{he} & S_{hh}
\end{bmatrix}
\cdot
\begin{bmatrix}
\psi_{e}^- \\
\psi_{h}^-
\end{bmatrix},
\end{equation}
where $\psi_e^-$ and $\psi_h^-$ are the incoming and $\psi_e^+$ and $\psi_h^+$ the outgoing electron and hole wavefunctions, respectively. 
For a nanowire with $M$ subgap modes and with an applied voltage $E=eV$, the scattering matrix is given by
\begin{equation}
S(E)=
\begin{bmatrix}
S_{ee} & S_{eh} \\
S_{he} & S_{hh}
\end{bmatrix},
\end{equation}
where the different $M\times M$ blocks $S_{\alpha\beta}$ describe the normal and the Andreev reflection processes.
The diagonal blocks $S_{ee}$ and $S_{hh}$ describe the normal reflection from electron to electron and from hole to hole, which results in no net charge transfer through the barrier.
The off-diagonal blocks $S_{he}$ and $S_{eh}$ describe instead the Andreev reflection from hole to electron and from electron to hole, which results in a net charge transfer of $2e$ through the barrier.
Below the gap $E<\Delta E_\mathrm{ph}$, assuming homogeneous and clean wires long enough to neglect transmission processes at the opposite end of the wire, the total amount of incoming electrons must equal the outgoing electrons (normally-reflected) and outgoing holes (Andreev-reflected).
Hence, the conservation of the probability current mandates that $S^\dag S=1$, i.e., the scattering matrix is unitary (see Refs.~\onlinecite{beenakker_random-matrix_1997,beenakker_random-matrix_2015}).

The differential conductance is given by the Landauer formula~\cite{blonder_transition_1982,takane_conductance_1992,fulga_scattering_2011,fulga_scattering_2012,beenakker_random-matrix_2015},
which at zero temperature and below the gap $E<\Delta E_\mathrm{ph}$ gives
\begin{equation}\label{eq:conductance}
	G(E)=\frac{\dd I}{\dd V}=
	\frac{2e^2}{h}\tr(S_{eh} S_{eh}^\dag)=
	\frac{2e^2}{h}\sum_{m=1}^M s_m,
\end{equation}
where $s_m$ are the eigenvalues of the matrix $S_{eh}S_{eh}^\dag$.
The factor 2 in the formula above is not due to the spin degeneracy but to the fact that the Andreev reflection of one electron $e$ amounts to the transfer of a charge $2e$ into the superconductor.
The eigenvalues $s_n$ are called the Andreev reflection eigenvalues and are equal to the transmission probabilities of each mode $m$, and are thus $0\le s_m\le1$, with $s_m=0$ and $s_m=1$ corresponding to perfect normal and perfect Andreev reflection, respectively.

At zero bias, the possible values of the differential conductance are entirely determined by symmetry considerations~\cite{fu_probing_2009,akhmerov_electrically_2009,akhmerov_quantized_2011,beenakker_random-matrix_2015}.
In a superconductor, the scattering matrix must exhibit particle-hole symmetry 
\begin{equation}\label{eq:SmatrixPH}
\mathcal{C}
S(E)
\mathcal{C}^{-1}
=
\tau_x 
S(E)^*
\tau_x
=
S(-E),
\end{equation}
which at zero bias $E=0$ mandates $\tau_x S(0)^*\tau_x=S(0)$, i.e.,
\begin{equation}
\begin{bmatrix}
S_{ee} & S_{eh} \\
S_{he} & S_{hh}
\end{bmatrix}
=
\begin{bmatrix}
S_{hh}^* & S_{he}^* \\
S_{eh}^* & S_{ee}^*
\end{bmatrix}
=
\begin{bmatrix}
S_{ee} & S_{eh} \\
S_{eh}^* & S_{ee}^*
\end{bmatrix}
,
\end{equation}
that is, $S_{ee}=S_{hh}^*$ and $S_{he}=S_{eh}^*$.
The possible values of the determinant of the scattering matrix at zero bias are constrained by symmetry:
Since the scattering matrix is unitary, the determinant satisfies $|\det(S(E))|=1$ and, due to particle-hole symmetry, it must be a real number for $E=0$, since \cref{eq:SmatrixPH} gives $\det(S(0)^*)=\det(S(0))$.
This mandates that $\det(S(0))=\pm1$.
Hence, one can define a topological invariant $\nu\in\mathbb{Z}_2$ of the scattering matrix~\cite{akhmerov_quantized_2011,fulga_scattering_2011,fulga_scattering_2012,beenakker_random-matrix_2015} as
\begin{equation}\label{eq:TIconductance}
	(-1)^\nu=\det(S(0))=\pm1,
\end{equation}
with trivial and nontrivial regimes corresponding to $\nu=0$ and $\nu=1$, respectively.
Another consequence of particle-hole symmetry is the Béri degeneracy~\cite{beri_dephasing-enabled_2009,beenakker_random-matrix_2015} of the Andreev reflection eigenvalues $s_m$:
Due to particle-hole symmetry, all Andreev reflection eigenvalues $s_m$ are degenerate, except for the values $s_m=0$ and $s_m=1$, corresponding respectively to perfect normal and perfect Andreev reflection.
Béri degeneracy enforced by particle-hole symmetry is the analogous to Kramers degeneracy enforced by time-reversal symmetry.

For a single mode below the gap, particle-hole symmetry gives $\det(S(0))=|S_{ee}|^2-|S_{eh}|^2=\pm1$, and unitarity gives $|S_{ee}|^2+|S_{eh}|^2=1$.
These two conditions fully determine the scattering matrix in the trivial and nontrivial regime 
and the differential conductance, which in this case is simply given by $G=(2e^2/h)|S_{eh}|^2$.
In the trivial case, one has $\det(S(0))=1$ and therefore $|S_{ee}|=1$ and $|S_{eh}|=0$. 
This corresponds to a regime of perfect normal reflection with zero conductance at zero bias 
\begin{equation}\label{eq:conductancetrivial}
S(0)=
\begin{bmatrix}
\ee^{\ii\theta} & 0 \\
0 & \ee^{-\ii\theta}
\end{bmatrix},
\quad
(-1)^\nu=1,
\quad
G(0)=0,
\end{equation}
as follows from \cref{eq:conductance,eq:TIconductance}, with $\theta$ a phase angle.
In the nontrivial case, one has $\det(S(0))=-1$ and therefore $|S_{ee}|=0$ and $|S_{eh}|=1$.
This corresponds to a regime of perfect Andreev reflection, with a quantized conductance at zero bias
\begin{equation}\label{eq:conductancenontrivial}
S(0)=
\begin{bmatrix}
0 & \ee^{\ii\theta} \\
\ee^{-\ii\theta} & 0
\end{bmatrix},
\quad
(-1)^\nu=-1,
\quad
G(0)=\frac{2e^2}h,
\end{equation}
as follows again from \cref{eq:conductance,eq:TIconductance}.
The conductance exhibits a resonant zero-bias peak which is quantized and equal to the conductance quantum $G_0$.
This result can also be understood in terms of the Béri degeneracy.
The Andreev reflection eigenvalue $s_1$ cannot be degenerate in the presence of one subgap mode.
Consequently, due to the Béri degeneracy, it must necessarily be $s_1=0$ or 1, which thus corresponds to $G=0$ or $G=G_0$ in \cref{eq:conductancetrivial,eq:conductancenontrivial}.
The regimes of perfect normal and perfect Andreev reflections are the only possible regimes allowed by particle-hole symmetry and by the conservation of the probability current (unitarity).
For an in-depth analysis of the interplay between symmetry and topology in the transport signatures of Majorana modes and for the scattering matrix approach in mesoscopic systems, see Refs.~\onlinecite{beenakker_random-matrix_1997,beenakker_random-matrix_2011,beenakker_random-matrix_2015}.

\subsubsection{Splitting and broadening of zero-bias peaks\label{sec:DifferentialConductanceSplitting}}

The explicit bias dependence of the resonant zero-bias peak in the presence of a single zero-energy Majorana mode can be calculated via Green's function methods~\cite{flensberg_tunneling_2010,stanescu_majorana_2011} or the Mahaux-Weidenmüller formula~\cite{beenakker_random-matrix_1997,nilsson_splitting_2008,vuik_reproducing_2019}, which yield~\footnote{
Note that this equation has been first derived to describe the zero-bias conductance peaks in $d$-wave high-temperature superconductors~\cite{tanaka_theory_1995,kashiwaya_origin_1995,kashiwaya_tunnelling_2000}
}
\begin{equation}
G(E)=\frac{2e^2}{h}\frac{\Gamma^2}{E^2+\Gamma^2},
\end{equation}
where $\Gamma$ is the tunneling rate through the barrier.
The peak at zero bias has a lorentzian shape with quantized height $G(0)=G_0$ and width $\Gamma$.
The values of the conductance at zero bias in the trivial and nontrivial regimes follow only from symmetry considerations and are thus not affected by the height of the barrier, disorder, or other perturbations:
The conductance can change from its quantized value to zero only by removing the Majorana bound state at zero energy, which requires the closing and reopening of the particle-hole gap and the concomitant change of the topological invariant.
Indeed, disorder may only affect the broadening of the peak, but not its height~\cite{akhmerov_quantized_2011,pientka_enhanced_2012}.
However, this is strictly true only at zero-temperature $T=0$ for Majorana bound states with exactly zero energy $E_\mathrm{M}=0$, and if the effect of perturbations breaking fermion-parity conservation, e.g., quasiparticle poisoning, can be neglected~\cite{leijnse_scheme_2011,rainis_majorana_2012,budich_failure_2012}.
Indeed, in finite-size wires, the Majorana modes localized at opposite ends hybridize, resulting in a finite energy splitting $E_\mathrm{M}>0$:
In this case, the quantized conductance becomes~\cite{flensberg_tunneling_2010,das-sarma_splitting_2012}
\begin{equation}
	G(E)=\frac{2e^2}{h} 
	\frac{ \Gamma ^2 E^2}{(E^2 - E_\mathrm{M}^2)^2 +\Gamma^2 E^2},
\end{equation}
which describes two peaks centered at $E=\pm E_\mathrm{M}$ with quantized conductance $G(\pm E_\mathrm{M})=G_0$, giving zero conductance at zero bias $G(0)=0$, as shown in \cref{fig:conductance}(c).
The splitting between the two peaks can be used to measure the Majorana energy splitting $E_\mathrm{M}$ and thus to estimate the localization length $\xi_\mathrm{M}$ via \cref{eq:MajoranaOscillations}.
For more complicated terminal geometries and in the presence of disorder, spatial variations of the parameters, or multiband wires, the differential conductance can be calculated numerically via the wavefunction matching method implemented in the Kwant code~\cite{groth_kwant:_2014}.

At finite temperatures $T>0$, the conductance is thermally broadened~\cite{lin_zero-bias_2012,setiawan_electron_2017} and can be written as~\cite{wimmer_quantum_2011}
\begin{equation}\thinmuskip=1mu\medmuskip=1.5mu\thickmuskip=2mu
	G(E,T)=
	\int_{-\infty}^\infty\dd E' \ G(E') \left( -\frac{\partial f(E'-E)}{\partial E} \right),
\end{equation}
where $f(E)=1/(\ee^{E/k_\mathrm{B}T}+1)$ is the Fermi function.
The effect of temperature is to lower the height and broaden the width of the zero-bias peak.
However, the conductance at zero temperature can still be obtained as the scaling limit of the peak at small temperatures~\cite{nichele_scaling_2017}.

\subsubsection{Other conductance signatures}

Quantized peaks in the differential conductance at zero or near-zero energy are the fingerprints of zero or near-zero subgap modes.
However, this is a necessary but not sufficient condition for the existence of Majorana modes: 
Several other mechanisms may produce subgap modes with nearly-quantized peaks at low energy, e.g., zero-bias Kondo resonances~\cite{lee_zero-bias_2012} or topologically-trivial Andreev bound states induced by disorder~\cite{liu_zero-bias_2012,pikulin_a-zero-voltage_2012,bagrets_class_2012,rainis_towards_2013,roy_topologically_2013,stanescu_disentangling_2013,woods_zero-energy_2019,pan_physical_2020,pan_generic_2020,das-sarma_disorder-induced_2021,pan_crossover_2021,pan_three-terminal_2021,pan_quantized_2021,pan_disorder_2021} or spatial inhomogeneities~\cite{kells_near-zero-energy_2012,prada_transport_2012,chevallier_mutation_2012,stanescu_disentangling_2013,roy_topologically_2013,stanescu_nonlocality_2014,cayao_sns-junctions_2015,liu_andreev_2017,setiawan_electron_2017,ptok_controlling_2017,penaranda_quantifying_2018,moore_two-terminal_2018,moore_quantized_2018,liu_distinguishing_2018,reeg_zero-energy_2018,huang_metamorphosis_2018,fleckenstein_decaying_2018,lai_presence_2019,avila_non-hermitian_2019,vuik_reproducing_2019,stanescu_robust_2019,lai_presence_2019,liu_conductance_2019,awoga_supercurrent_2019,zeng_analytical_2019,sharma_hybridization_2020,pan_physical_2020,pan_generic_2020,cayao_confinement-induced_2021,pan_crossover_2021,pan_quantized_2021,pan_three-terminal_2021,pan_quantized_2021,hess_local_2021,liu_majorana_2021,das-sarma_disorder-induced_2021,marra_majorana/andreev_2022} (see also Ref.~\onlinecite{prada_from_2020} for a review on the issue).
Moreover, zero and near-zero quantized peaks can also appear in $d$-wave high-temperature superconductors~\cite{tanaka_theory_1995,kashiwaya_origin_1995,kashiwaya_tunnelling_2000}, which were later proven to have a topological origin~\cite{sato_topology_2011}.
Hence, quantized peaks at zero bias do not necessarily correspond to the presence of topologically-protected Majorana bound states but may have alternative explanations.
This fact motivates the investigation of alternative, and possibly stronger, signatures and alternative setups.

For example, the simultaneous appearance of Majorana modes at the opposite ends of a nanowire can be probed in multiterminal setups~\cite{michaeli_electron_2017,rosdahl_andreev_2018,moore_two-terminal_2018,danon_nonlocal_2020,pan_three-terminal_2021,pikulin_protocol_2021,kejriwal_nonlocal_2022}, e.g., by measuring the differential conductance matrix 
\begin{equation}
	S=\begin{bmatrix}
	S_\mathrm{LL} & S_\mathrm{LR} \\
	S_\mathrm{RL} & S_\mathrm{RR} \\
	\end{bmatrix}=
	\begin{bmatrix}
	\dd I_\mathrm{L}/\dd V_\mathrm{L} & \dd I_\mathrm{L}/\dd V_\mathrm{R} \\
	\dd I_\mathrm{R}/\dd V_\mathrm{L} & \dd I_\mathrm{R}/\dd V_\mathrm{R} \\
	\end{bmatrix},
\end{equation}
where diagonal terms correspond to the local conductances (conductance on the same lead) and the off-diagonal terms to the nonlocal conductances (conductance between leads), with $I_\mathrm{L,R}$ and $V_\mathrm{L,R}$ respectively the voltages and currents measured at the left and right leads connected to the opposite ends of the wire.
This allows verifying the presence of zero-energy states simultaneously at the two ends of the nanowire and the concomitant closing and reopening of the bulk gap at the topological phase transition.

A direct signature of fractionalization can be obtained by measuring the differential conductance through a spinless quantum dot coupled to the end of a Majorana nanowire~\cite{liu_detecting_2011,vernek_subtle_2014}.
In the nontrivial regime, the Majorana mode at the end of the wire partially leaks into the quantum dot, inducing a resonant peak in the differential conductance through the dot with a quantized height given by 
\begin{equation}
	G=\frac{G_0}4=\frac{e^2}{2h},
\end{equation}
The value of the quantized peak $G_0/4$ has to be contrasted to the quantized peak of a single spinless electronic mode in the dot $G_0/2$ and that of a spin-degenerate electronic mode $G_0$.
This signature can be intuitively described as the resonant peak of ``half'' an electronic mode, which corresponds to the leaked Majorana mode in the dot.
Quantum dots coupled to nanowires can also be employed to probe the spin degree of freedom~\cite{leijnse_quantum_2011,prada_measuring_2017,chevallier_topological_2018,schuray_signatures_2020} and the nonlocality~\cite{liu_detecting_2011,prada_measuring_2017,schuray_fano_2017} of Majorana modes.

Another direct signature of Majorana bound states is the presence of quantized half-integer plateaus in the quantum point contact conductance in the ballistic regime~\cite{wimmer_quantum_2011}.
As a consequence of the Béri degeneracy, the ballistic point conductance in the topologically nontrivial phase exhibits plateaus with quantized conductance given by 
\begin{equation}
	G=(2n+1)G_0=(2n+1)\frac{2e^2}{h} \quad \text{with } n=0,1,\ldots,
\end{equation}
as a function of the bias voltage, with the first plateau $G=G_0$ being topologically robust against perturbations.
This has to be contrasted with the quantized integer plateaus $G=2nG_0$ in the topologically trivial phase and $G=nG_0$ in normal (nonsuperconducting) quantum point contacts, with first plateaus $G=0$ at zero bias.
The height of the first plateau in the nontrivial phase is thus halved compared to the trivial phase.
This signature as well can be seen as a consequence of fractionalization, i.e., a Majorana mode being ``half'' the degree of freedom of an electron in a superconductor.

Other conductance signatures which may distinguish between Majorana bound states and other trivial Andreev bound states can be obtained in alternative setups employing dissipative leads suppressing the contribution of trivial bound states~\cite{liu_proposed_2013,liu_universal_2022}, superconducting leads~\cite{peng_robust_2015}, or spin-polarized leads~\cite{sticlet_spin_2012,he_selective_2014,haim_signatures_2015,szumniak_spin_2017}.

\subsection{Josephson current}

\subsubsection{Fractional Josephson effect}

\begin{figure}[t]
\includegraphics[width=\columnwidth]{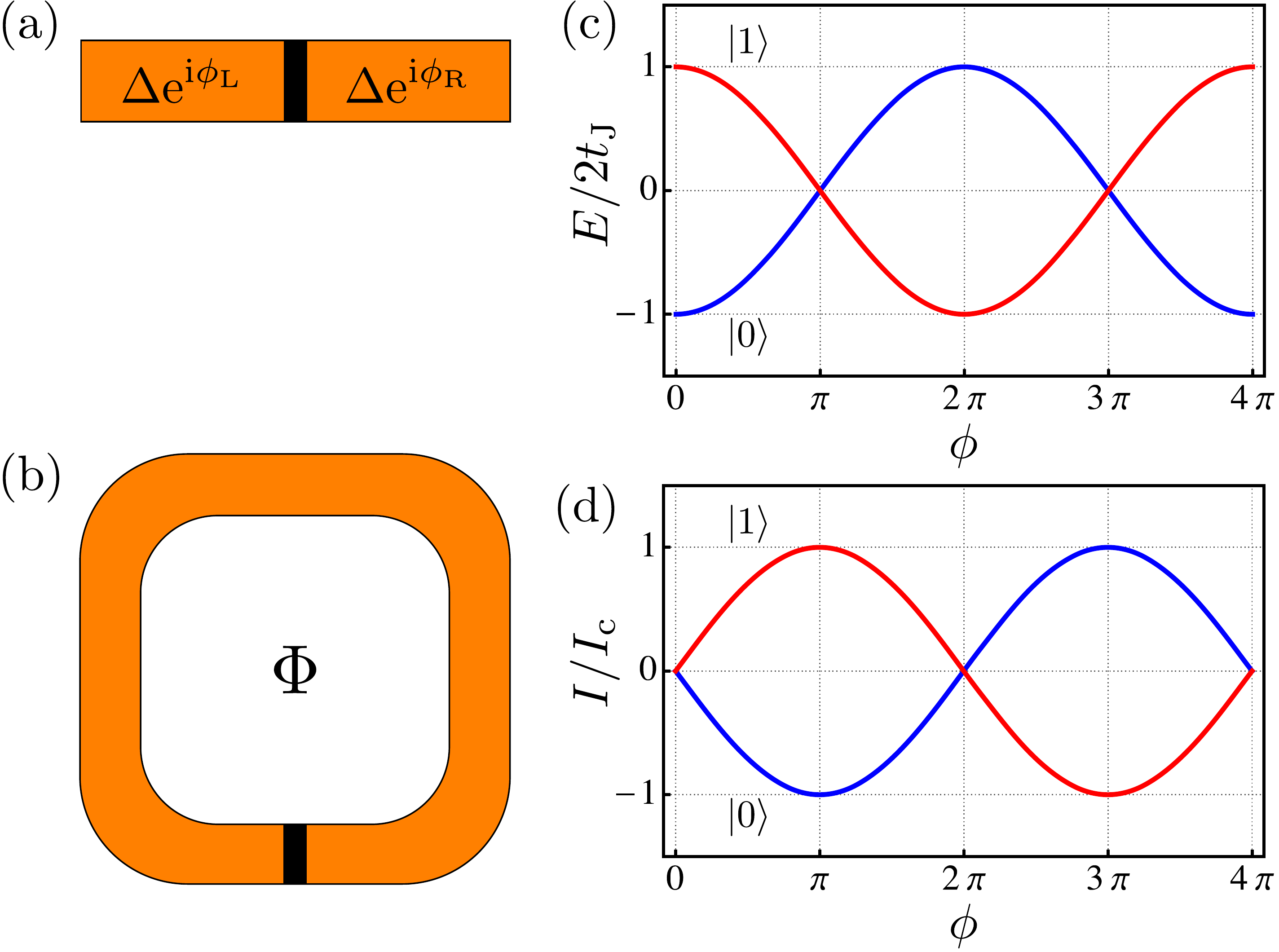}
\caption{
Fractional Josephson effect in a Majorana nanowire.
(a)
Josephson junction realized by a proximitized by two superconducting leads modeled as two 1D superconducting segments coupled by a tunneling barrier, with different superconducting phases $\phi_\mathrm{L}$ and $\phi_\mathrm{R}$ on the left and right sides.
(b)
A 1D superconductor in a ring geometry with a single Josephson junction enclosing a magnetic flux $\Phi$.
In this setup, the superconducting phase difference becomes $\phi=\phi_\mathrm{L}-\phi_\mathrm{R}=2\pi\Phi/\Phi_0$ due to the Aharonov-Bohm effect, with $\Phi_0=h/2e$ the superconducting flux quantum.
(c)
Energy levels of the Josephson junction as a function of the phase $\phi$.
The two states $\ket{0}$ and $\ket{1}$ have a finite energy splitting except for $\phi=\pi$, where the two states become degenerate at zero energy.
Transitions between the two energy levels at $\phi=\pi$ are suppressed by fermion parity conservation.
(d)
Josephson current as a function of the phase $\phi$.
The contributions of the two states $\ket{0}$ and $\ket{1}$ to the current have opposite signs.
The energy levels and the current are periodic with period $\phi=4\pi$.
}
\label{fig:Josephson}
\end{figure}

Another direct experimental signature of Majorana zero modes and their fractionalization is the so-called fractional Josephson effect, i.e., the $4\pi$ periodicity of the current-phase relation of Josephson junctions in the topologically nontrivial phase~\cite{kitaev_unpaired_2001,kwon_fractional_2004,fu_superconducting_2008,fu_josephson_2009,lutchyn_majorana_2010,oreg_helical_2010,ioselevich_anomalous_2011,van-heck_coulomb_2011,law_robustness_2011,pikulin_phenomenology_2012,san-jose_ac-josephson_2012,badiane_ac-josephson_2013}.
A Josephson junction can be realized in a hybrid semiconducting-superconducting setup where a nanowire is proximitized by two separate superconducting sections, as in \cref{fig:nanowire}(b).
This setup corresponds to a linear junction of two superconductors separated by a semiconducting barrier, as shown schematically in \cref{fig:Josephson}(a).
In a normal Josephson junction (i.e., topologically trivial), the Josephson current~\cite{josephson_possible_1962,josephson_supercurrents_1965} is described by the Josephson equations
\begin{equation}\label{eq:JosephsonEquations}
	I=I_\mathrm{c} \sin\phi,
	\qquad
	\partial_t \phi=\frac{2eV}\hbar,
\end{equation}
where $I_\mathrm{c}$ is the so-called critical current, $\phi=\phi_\mathrm{L}-\phi_\mathrm{R}$ the (gauge-invariant) phase difference, and $V$ is the voltage bias between the two superconductors.
In a tunnel junction, the critical current is given by the Ambegaokar–Baratoff formula~\cite{tinkham_introduction_2004} $I_\mathrm{c}=(\pi\Delta/2e R_\mathrm{N})\tanh(\Delta/2k_\mathrm{B}T)$, where $R_\mathrm{N}$ is the normal state resistance of the junction, given by $1/R_\mathrm{N}=(e^2/h)T_\mathrm{N}$, where $T_\mathrm{N}$ is the transparency of the junction. 
Typically, Josephson junctions are driven either by direct or alternating currents.
In the DC Josephson regime, the junction is current-driven, i.e., the current is kept constant in time with a value $I\in[-I_\mathrm{c},I_\mathrm{c}]$, corresponding to a zero voltage bias $V=0$ and constant phase $\phi$.
In the AC Josephson regime, instead, the junction is voltage-driven with a finite voltage bias $V>0$.
This results in the phase $\phi$ increasing linearly in time and a current with amplitude $I_\mathrm{c}$ and frequency $\omega_\mathrm{J}=2eV/\hbar$.
Alternatively, in the inverse AC Josephson regime, the junction is driven by microwave radiation $\omega_\mathrm{R}$, which induces a sinusoidal current in the junction.
This results in a staircase pattern in the current-voltage dependence, known as Shapiro steps~\cite{shapiro_josephson_1963}, due to the resonances between the microwave frequency and the phase winding induced by the voltage bias, occurring for $\omega_\mathrm{J}=n\omega_\mathrm{R}$, with $n$ integer (see Ref.~\onlinecite{poole_superconductivity_2014_p466}).
In all these setups, the phase difference is determined by the current and the voltage bias via \cref{eq:JosephsonEquations}.
However, the phase difference of the Josephson junction can be directly controlled by embedding the junction in an Aharonov-Bohm ring, i.e., a superconducting ring threaded by a magnetic flux, as shown schematically in \cref{fig:Josephson}(b).
Due to the Aharonov-Bohm effect, the magnetic flux $\Phi$ through the ring induces a finite phase difference $\phi=2\pi\Phi/\Phi_0$ where $\Phi_0=h/2e$ is the superconducting flux quantum.
For an overview of the Josephson effect, see Refs.~\onlinecite{tinkham_introduction_2004,poole_superconductivity_2014,nazarov_quantum_2009}.

The Josephson current can be directly calculated from the BdG Hamiltonian of the junction.
Since the energy variation in a time interval is equal to the work done $IV$, from \cref{eq:JosephsonEquations} follows that the energy increase over time is $\int{E\dd{t}}=\int{IV\dd{t}}=(\hbar/2e)\int{I\dd\phi}$.
Therefore the current is given by $I=(2e/\hbar)\partial_\phi E$, where $E$ is the total energy of the junction, up to constant terms.
This energy can be identified with the groundstate energy of the BdG Hamiltonian, which is the state with all quasiparticle hole states occupied, and is thus given by the sum of all occupied energy levels (with negative energy) $E=-\sum_n E_n$.
Hence, the Josephson current is
\begin{equation}\label{eq:JosephsonCurrent}
	I=\frac{2e}\hbar \frac{\partial E}{\partial\phi}
	= -\frac{2e}\hbar \sum_n\frac{\partial E_n}{\partial\phi},
\end{equation}
where the sum is extended over all occupied levels.
Since the Hamiltonian is invariant up to a gauge transformation $\phi\to\phi+2\pi$, one may naively expect that the energy levels $E_n$, and consequently the current, are always periodic in the phase with periodicity $2\pi$.
However, this is not the case in the presence of Majorana zero modes.
The phase periodicity of the Josephson current doubles in the presence of Majorana modes, resulting in a $4\pi$ periodic dependence.

Let us consider a concrete example, e.g., a nanowire proximitized by two superconducting leads, as shown in \cref{fig:nanowire}(b), modeled as two 1D superconducting segments coupled by a tunneling barrier, as shown in \cref{fig:Josephson}(a).
If both segments are in the nontrivial phase, they will both exhibit Majorana zero modes localized at their boundaries, with two inner Majorana modes localized across the junction and two outer Majorana modes localized away from the junction, one for each segment.
If the Majorana delocalization length is smaller than the length of the wire, the effect of the outer Majorana modes localized away from the junction can be neglected.
Hence, the low-energy effective Hamiltonian is just
\begin{equation}\label{eq:HeffW}
	\mathcal{H}_\mathrm{eff}(\phi)=
	\frac\ii2 E_\mathrm{M}(\phi)\gamma_\mathrm{L}\gamma_\mathrm{R}=
	E_\mathrm{M}(\phi) \left(n_\mathrm{M}-\frac12\right),
\end{equation}
where $\gamma_\mathrm{L}$ and $\gamma_\mathrm{R}$ are the Majorana modes localized at the left and right sides of the junction, $n_\mathrm{M}=(1+\ii\gamma_\mathrm{L}\gamma_\mathrm{R})/2$ is the corresponding number operator [cf.~\cref{eq:Majorananumberoperator}]
of the fermionic operator $d_\mathrm{M}=(\gamma_\mathrm{L}+\ii\gamma_\mathrm{R})/2$, 
and $E_\mathrm{M}(\phi)$ is the coupling strength, which depends on the phase difference across the junction.
This is the most general Hamiltonian of two coupled Majorana modes [see, e.g., \cref{eq:HeffW0}].
It has two energy eigenvalues $\ket{0}$ and $\ket{1}$ which are also eigenvalues of the number operator $n_\mathrm{M}$ with a well-defined fermion parity:
the even state with $n_\mathrm{M}\ket{0}=0$ and the odd state with $n_\mathrm{M}\ket{1}=\ket{1}$ with energies $-E_\mathrm{M}(\phi)/2$ and $E_\mathrm{M}(\phi)/2$.
The unique signatures of the Majorana zero modes can be understood from the Hamiltonian above using symmetry considerations.
Let us slowly (adiabatically) change the superconducting phase $\phi_\mathrm{L}\to\phi_\mathrm{L}+2\pi$ on the left side.
Since the phase of the superconducting pairing can be absorbed by a unitary transformation $c_n\to\ee^{-\ii\phi/2}c_n$ [see \cref{eq:H-pwave}], this change corresponds to $c_n\to-c_n$ and $\gamma_\mathrm{L}\to-\gamma_\mathrm{L}$, as one can verify from \cref{eq:Majorana}.
Consequently, we end up 
flipping the number operator $n_\mathrm{M}\to1-n_\mathrm{M}$ and the fermion parity
and reversing the sign of the Hamiltonian above.
However, since the Hamiltonian must be invariant under the gauge transformation $\phi\to\phi+2\pi$, one has that $\mathcal{H}_\mathrm{eff}(\phi+2\pi)=\mathcal{H}_\mathrm{eff}(\phi)$, which mandates $E_\mathrm{M}(\phi+2\pi)=-E_\mathrm{M}(\phi)$, i.e., the energy of the junction is not $2\pi$ periodic anymore.
This also gives that $E_\mathrm{M}(\phi+4\pi)=E_\mathrm{M}(\phi)$, i.e., the energy $E_\mathrm{M}(\phi)$ is $4\pi$ periodic. 
Moreover, \cref{eq:JosephsonCurrent} yields
\begin{equation}
	I(\phi)=-\frac{e}\hbar\partial_\phi E_\mathrm{M}(\phi) \left(2n_\mathrm{M}-1\right).
\end{equation}
As one can see, the Josephson current depends explicitly on the occupation number of the Majorana bound state at the junction.
If the fermion parity is conserved, the occupation number of the Majorana bound states at the junction is conserved as well: 
Consequently, the current $I(\phi)$ is $4\pi$ periodic and with $I(\phi+2\pi)=-I(\phi)$, due to the periodicity of the coupling $E_\mathrm{M}(\phi)$.
Notice that the Josephson current is $4\pi$ periodic, even though the Hamiltonian $\mathcal{H}_\mathrm{eff}$ is $2\pi$ periodic due to gauge invariance.
This apparent contradiction is resolved when one realizes that the fermion parity eigenstates $\ket{0}$ and $\ket{1}$ are not invariant under the gauge transformation $\phi\to\phi+2\pi$, which indeed flips the fermion parity transforming one state into the other.
Therefore, the $4\pi$ periodicity of the Josephson current is a direct consequence of fermion parity conservation.

The doubling of the current-phase periodicity can be understood again as the effect of fractionalization:
in the trivial phase, the Josephson current is carried by Cooper pairs (with charge $2e$); in the nontrivial phase, the current is carried by the Majorana modes localized at the junctions, with each one being ``half'' the degree of freedom of an electron.
In the AC Josephson junction setup (voltage-driven junction), the doubling of the phase periodicity corresponds to the halving of the frequency of the Josephson current~\cite{kwon_fractional_2004}, which is now 
$\omega_\mathrm{J}/2=eV/\hbar$.
The halving of the Josephson frequency can be probed by measuring the microwave radiation of the junction~\cite{badiane_ac-josephson_2013,virtanen_microwave_2013,vayrynen_microwave_2015,peng_signatures_2016}, the current noise spectrum~\cite{badiane_ac-josephson_2013,houzet_dynamics_2013}, or from the doubling of the width of the Shapiro steps in the nontrivial phase~\cite{jiang_unconventional_2011,dominguez_dynamical_2012,badiane_ac-josephson_2013,houzet_dynamics_2013,sau_detecting_2017,dominguez_josephson_2017} when the junction is driven by a current with both DC and AC components (induced by microwave radiation).
Other signatures of the topological phase transition are the discontinuities in the derivatives of the Josephson critical current~\cite{san-jose_multiple_2013,pientka_signatures_2013}, the discontinuous jumps in the current-phase relation~\cite{marra_signatures_2016,nesterov_anomalous_2016,murthy_energy_2020}, and the presence of nonzero Josephson current at $\phi=0$ when the magnetic field has a finite component $b_y>0$ in the direction of the spin-orbit coupling field~\cite{marra_signatures_2016,nesterov_anomalous_2016}.

\subsubsection{Josephson current-phase relation}

One can easily obtain the analytical expression of the current-phase relation of a topological Josephson junction using the Kitaev model~\cite{kitaev_unpaired_2001} in \cref{eq:H-pwave} in the Majorana chain regime $\Delta=t$.
Let us consider two 1D chains coupled by a tunneling barrier with transmission coefficient $t_\mathrm{J}$.
This system is described by the Hamiltonian~\cite{pientka_signatures_2013}
\begin{align}
\mathcal{H}=
-\!\sum_{n=1}^{2N}
\mu c^\dag_{n} c_{n}
-\!\sum_{n=1}^{N-1}&
\left(
t c^\dag_{n} c_{n+1} 
-
\Delta\ee^{\ii\phi_\mathrm{L}} c_{n} c_{n+1} + \text{h.c.}
\right)+
\nonumber\\
-\!\sum_{n=N}^{2N-1}&
\left(
t c^\dag_{n} c_{n+1} 
-
\Delta \ee^{\ii\phi_\mathrm{R}} c_{n} c_{n+1} + \text{h.c.}
\right)+
\nonumber\\
-t_\mathrm{J} &\left( c^\dag_{N} c_{N+1} + c^\dag_{N+1} c_{N}\right),
\end{align}
where the lattice sites $n=1,\ldots,N$ and $n=N+1,\ldots,2N$ correspond to the left and right sides of the tunneling barrier, with superconducting phases $\phi_\mathrm{L}$ and $\phi_\mathrm{R}$, respectively.
The superconducting phases of the left and right sides of the junction can be absorbed by the gauge transformation defined by
$c_{n}\to\ee^{-\ii\phi_\mathrm{L}/2}c_{n}$ for $n=1,\ldots,N$ and
$c_{n}\to\ee^{-\ii\phi_\mathrm{R}/2}c_{n}$ for $n=N+1,\ldots,2N$,
which gives
\begin{align}
\mathcal{H}=
-\sum_{n=1}^{2N} \mu c^\dag_{n} c_{n}
-\sum_{n=1}^{N-1} & \left( t c^\dag_{n} c_{n+1} 
-
 \Delta c_{n} c_{n+1} + \text{h.c.} \right)+
\nonumber\\
-\sum_{n=N}^{2N-1} & \left( t c^\dag_{n} c_{n+1} 
-
 \Delta c_{n} c_{n+1} + \text{h.c.} \right)+
\nonumber\\\label{eq:JosephsonGaugeInvariant}
-t_\mathrm{J} &
\left(\ee^{\ii\phi/2}c^\dag_{N}c_{N+1}+\text{h.c.}\right),
\end{align}
where $\phi=\phi_\mathrm{L}-\phi_\mathrm{R}$ is the phase difference across the junction.
As one can see from the equation above, this phase cannot be gauged away (it is gauge invariant), and thus the energy levels of the Hamiltonian will depend explicitly on $\phi$.
In the nontrivial phase $|\mu|<2t$, and if the two sides of the junction are completely decoupled $t_\mathrm{J}=0$, each side exhibit Majorana zero modes localized at the boundaries, i.e., at the lattice sites $n=1$ and $n=N$ (left side) and $n=N+1$ and $n=2N$ (right side).
In the Majorana chain regime with $\Delta=t$ and $\mu=0$ [see \cref{eq:H-pwave-majo-spatial}], the Majorana modes are fully localized and described by the operators
\begin{subequations}\begin{align}
	d_\mathrm{L}&=\frac12(\gamma_{B,N}
-
	\ii\gamma_{A,1}),
	\\
	d_\mathrm{R}&=\frac12(\gamma_{B,2N}
-
	\ii\gamma_{A,N+1}),
\end{align}\end{subequations}
as in \cref{eq:MajoranaEdgeStates}.
For $t_\mathrm{J}\neq0$, the Majorana modes localized at $n=N$ and $n=N+1$ hybridize via tunneling, inducing a finite energy splitting between the modes. 
Since the Majorana zero modes at $n=1$ and $n=2N$ are not affected, this regime is described by an effective Hamiltonian $\mathcal{H}_\mathrm{eff}$ obtained by projecting on the Majorana modes $\gamma_{B,N}$ and $\gamma_{A,N+1}$, i.e., by taking 
$c_N=(\gamma_{A,N}+\ii\gamma_{B,N})/2\to{\ii}\gamma_{B,N}/2$ and
$c_{N+1}=(\gamma_{A,N+1}+\ii\gamma_{B,N+1})/2\to\gamma_{A,N+1}/2$
in \cref{eq:JosephsonGaugeInvariant}, which gives~\cite{pientka_signatures_2013}
\begin{align}
	\mathcal{H}_\mathrm{eff}
	=
	&
	\frac14 t_\mathrm{J} \left( \ii \ee^{\ii\phi/2} \gamma_{B,N} \gamma_{A,N+1} + \text{h.c.}\right)
	=\nonumber\\
	=
	&
	\frac\ii2 t_\mathrm{J} \cos\left(\frac{\phi}2\right) \gamma_{B,N} \gamma_{A,N+1} 
	=\nonumber\\
	=
	&
	t_\mathrm{J} \cos\left(\frac{\phi}2\right) \left(n_\mathrm{M}-\frac12\right),
	\label{eq:JosephsonHamiltonian}
\end{align}
where 
$n_\mathrm{M}=(1+\ii\gamma_{B,N}\gamma_{A,N+1})/2$ is the number operator 
corresponding to 
the Majorana bound state 
$d_\mathrm{M}=(\gamma_{B,N}+\ii\gamma_{A,N+1})/2$
across the junction [cf.~\cref{eq:Majorananumberoperator}].
The Hamiltonian above, which is analogous to \cref{eq:HeffW} when 
$E_\mathrm{M}(\phi)=t_\mathrm{J}\cos(\phi/2)$, 
has two energy eigenvalues that are also eigenvalues of the number operator.
The state $\ket{0}$ has even fermion parity and energy 
$-(t_\mathrm{J}/2)\cos(\phi/2)$, 
while the state $\ket{1}$ has odd fermion parity with opposite energy 
$(t_\mathrm{J}/2)\cos(\phi/2)$.
Even though the Hamiltonian $\mathcal{H}_\mathrm{eff}$ is $2\pi$ periodic, the energies of the two eigenstates are $4\pi$ periodic.
Indeed, the gauge transformation $\phi\to\phi+2\pi$ flips both the fermion parity and the sign of the energy eigenvalues, leaving the Hamiltonian $\mathcal{H}_\mathrm{eff}$ unchanged.
The energies of the two states cross at zero energy for $\phi=\pm\pi$: 
However, transitions between the two states are suppressed by fermion parity conservation.
Using \cref{eq:JosephsonCurrent}, the contributions to the Josephson current of the two states are given by
\begin{equation}\label{eq:JosephsonCurrentNontrivial}
	I(\phi)
	=
	\frac{e t_\mathrm{J}}{2\hbar}\sin\left(\frac{\phi}2\right)\left(2n_\mathrm{M}-1\right),
\end{equation}
which is $4\pi$ periodic in the phase $\phi$.
Since the transparency of the junction is $T_\mathrm{N}\propto|t_\mathrm{J}|^2$, the nontrivial contribution to the Josephson current has magnitude $I_\mathrm{c}={et_\mathrm{J}}/{2\hbar}\propto\sqrt{T_\mathrm{N}}$, which has to be contrasted with the Ambegaokar–Baratoff formula~\cite{tinkham_introduction_2004} which gives $I_\mathrm{c}\propto T_\mathrm{N}$ for a trivial junction.
Therefore the nontrivial contributions are larger than trivial ones in tunnel junctions $T_\mathrm{N}\ll1$.

\Cref{eq:JosephsonHamiltonian,eq:JosephsonCurrentNontrivial} have been derived using the Kitaev model in the Majorana chain regime in \cref{eq:H-pwave-majo-spatial}, i.e., assuming that the Majorana edge modes are perfectly localized at a single lattice site with a localization length $\xi_\mathrm{M}=0$.
One can expect these equations to be valid, up to exponentially small corrections, also in the case of finite Majorana localization lengths $\xi_\mathrm{M}$.
Furthermore, since the low energy sector of the Oreg-Lutchyn Hamiltonian is equivalent to a spinless $p$-wave superconductor (as discussed in \cref{sec:band-inversion}), these results are still valid in the case of Josephson junctions realized by Majorana nanowires, as long as the Majorana localization length $\xi_\mathrm{M}$ is small compared with the wire length.

Figures~\ref{fig:Josephson}(c) and~\ref{fig:Josephson}(d) show the energy levels and Josephson current as a function of the phase $\phi$.
At $\phi=0$, there is a finite energy splitting between the two states $\ket{0}$ and $\ket{1}$, with energies 
$E=\mp t_\mathrm{J}/2$, 
respectively, while the Josephson current carried by both states is zero.
For $\phi>0$, the energy splitting between the two states decreases and becomes degenerate at zero energy for $\phi=\pi$, while the current reaches the maximum values 
$I=\mp I_\mathrm{c}=\mp et_\mathrm{J}/2\hbar$.
Consequently, for $\phi>\pi$, the energy splitting increases again, and at $\phi=2\pi$, the energies of the two states 
$\ket{0}$ and $\ket{1}$ become 
$E=\pm t_\mathrm{J}/2$, 
respectively, while the current becomes zero.
Hence, an adiabatic evolution of the phase from $\phi=0$ to $\phi=2\pi$ does not bring back to the initial configuration: The energies of the two states at $\phi=2\pi$ are inverted, and the initial configuration is reached only when $\phi=4\pi$.
The crossing of the two energy levels is protected by fermion parity conservation: 
Since the states $\ket{0}$ and $\ket{1}$ have different fermion parity, transitions between the two states violate parity conservation.
Therefore, even when the two states are degenerate at $\phi=\pm\pi$, the transition between the two states is strongly suppressed.
Hence, if fermion parity is conserved, the Josephson current is $4\pi$ periodic in the phase.

In realistic systems, a direct probe of the $4\pi$ periodicity of the Josephson current is rather hard to achieve:
Spatially-separated Majorana modes have a finite, although exponentially small, overlap.
If the Majorana delocalization length is comparable with the length of the wire, this residual overlap cannot be neglected.
In this case, the effect of the Majorana modes localized away from the junction, neglected in \cref{eq:HeffW}, has to be taken into account.
This results in the hybridization of the Majorana modes with a finite energy splitting $\propto\ee^{-L/\xi_\mathrm{M}}$, where $L$ is the distance between contiguous Majorana modes.
This energy splitting produces an avoided crossing of the two energy sectors of \cref{eq:JosephsonHamiltonian} at $\phi=\pi$ restoring the $2\pi$ periodicity of the Josephson current in finite systems~\cite{van-heck_coulomb_2011,san-jose_ac-josephson_2012,pikulin_phenomenology_2012,cayao_majorana_2017,cayao_andreev_2018,cayao_finite_2018}.
The total fermion parity of the Majorana modes at the junction and away from the junction is still conserved:
However, the avoided crossing breaks the parity conservation of the two inner Majorana modes at the junction.
Another process that breaks fermion parity conservation and restores the $2\pi$ periodicity is quasiparticle-poisoning~\cite{fu_josephson_2009,leijnse_scheme_2011,rainis_majorana_2012,budich_failure_2012}.
For this reason, experimental probes of the fractional Josephson effect need to be performed by sweeping the phase $\phi$ on time scales shorter than the quasiparticle poisoning time $\tau_\mathrm{QP}$.
On the other hand, rapidly changing the phase $\phi$ breaks adiabaticity and drives the junction out of equilibrium.
This may result in nonadiabatic (Landau-Zener) transitions to higher energy levels with the same fermion parity~\cite{billangeon_ac-josephson_2007,san-jose_ac-josephson_2012,pikulin_phenomenology_2012}, which may produce $4\pi$ periodicity even in topologically trivial regimes.

\subsection{Teleportation\label{sec:CoulombBlockade}}

\subsubsection{Coulomb blockade in floating nanowires}

\begin{figure*}[t]
\includegraphics[width=\textwidth]{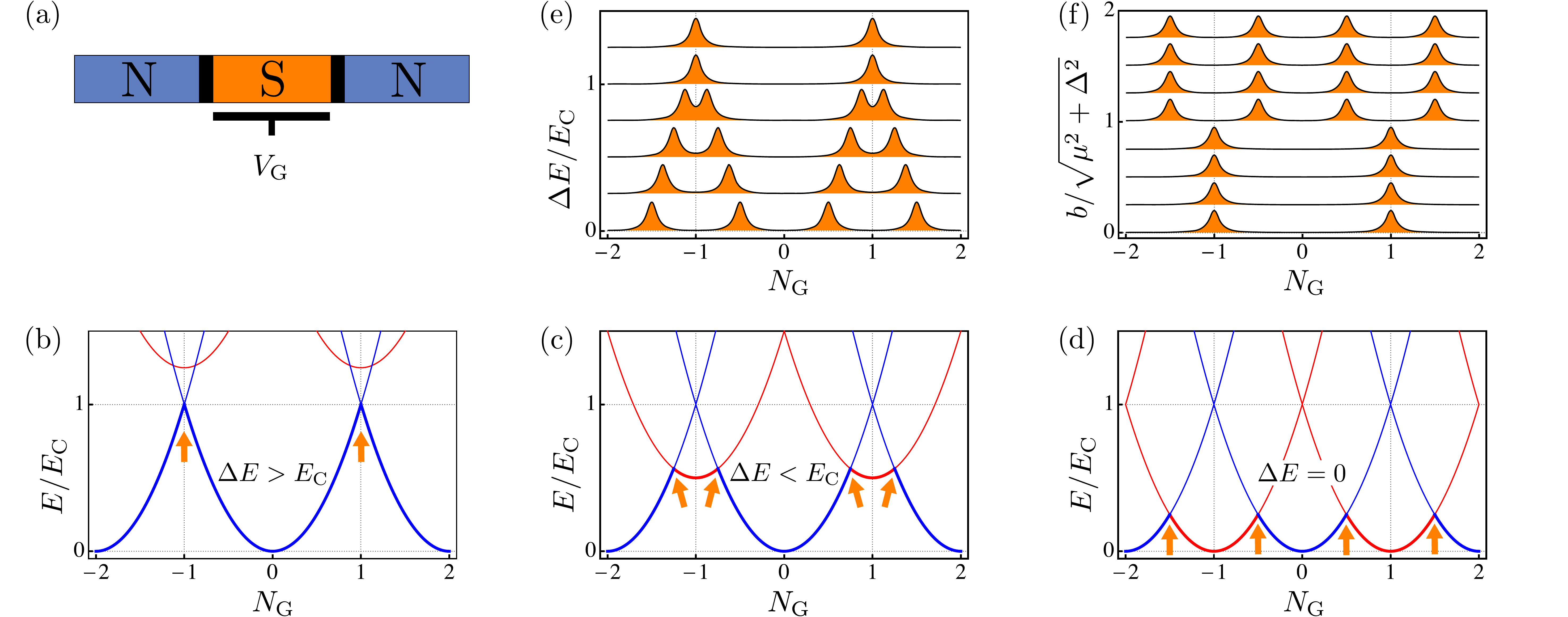}
\caption{ 
Coulomb blockade regime in a Majorana nanowire.
(a)
A proximitized nanowire connected to two metallic leads with a bias voltage $V$ and an applied gate voltage $V_\mathrm{G}$.
[(b)-(d)] 
Energy levels of the many-particle states as a function of the gate charge $N_\mathrm{G}$, with the degeneracy points indicated by arrows.
(b) 
In the trivial phase, the lowest single-particle energy level available coincides with the particle-hole gap $\Delta E=\Delta{E}_\mathrm{Z}$.
Assuming that the particle-hole gap is larger than the charging energy $\Delta E>E_\mathrm{C}$, the lowest energy levels have an even number of electrons.
The degeneracy points connect energy states with the same fermion parity, and the spacing between consecutive degeneracy points is $2e$.
(c) 
If $\Delta E<E_\mathrm{C}$ instead, energy levels with odd fermion parity may become energetically favorable for certain values of the gate charge.
The degeneracy points connect energy states with different fermion parity.
(d)
In the nontrivial phase, the lowest single-particle energy level available coincides with the energy of the Majorana bound state $\Delta E=E_\mathrm{M}\approx0$.
The spacing between consecutive degeneracy points is $e$ in this case.
(e) 
Coulomb blockade peaks as a function of the gate charge and of the energy of the lowest single-particle level. 
The degeneracy points correspond to peaks in the differential conductance at zero bias.
The peaks are evenly spaced and periodic in the gate charge $eN_\mathrm{G}$ with period $2e$ for $\Delta E>E_\mathrm{C}$ and period $e$ for $\Delta E=0$.
The halving of the period is a signature of the presence of a single-particle state at zero energy, e.g., a Majorana bound state.
(f) 
Coulomb blockade peaks as a function of the gate charge and magnetic field with $E_\mathrm{M}=0$, showing the topological transition at $b=\sqrt{\mu^2+\Delta^2}$ between trivial and nontrivial phases with periodicities $2e$ and $e$, respectively.
}
\label{fig:Blockade}
\end{figure*}

Another direct signature of Majorana zero modes, which fully reveal their nonlocal nature, is the so-called ``teleportation'', i.e., the tunneling through spatially-separated Majorana modes across the two ends of a Majorana nanowire in the Coulomb blockade regime~\cite{nilsson_splitting_2008,fu_electron_2010,zazunov_coulomb_2011,hutzen_majorana_2012}.
This regime is realized in mesoscopic systems capacitively coupled to the ground when the Coulomb repulsion between electrons becomes the dominant energy scale, suppressing electron tunneling and transport processes.
Let us consider a Majorana nanowire in a floating setup, i.e., isolated from the ground, connected to two metallic leads with a bias voltage $V$, and with an applied gate voltage $V_\mathrm{G}$, as shown schematically in \cref{fig:Blockade}(a).
In this setup, known as ``Majorana island'', the strong Coulomb repulsion between electrons becomes the dominant energy scale, and electron tunneling is energetically unfavorable. 
Consequently, the tunneling conductance is strongly suppressed at small temperatures and small bias voltages.
In the constant interaction model, one assumes that the total energy due to the Coulomb repulsions of the electrons in the nanowire only depends on the total charge.
This energy contribution is equal to the electrostatic energy $Q^2/2C=e^2N^2/2C$, where $Q=-eN=-e(N_\mathrm{e}-N_\mathrm{i})$ is the excess electronic charge, given by the sum between the ionic charge $eN_\mathrm{i}$ and the bare electronic charge $-eN_\mathrm{e}$, 
$N=N_\mathrm{e}-N_\mathrm{i}$ the number of excess electrons,
and $C$ the total capacitance of the wire.
Hence, the energy required to add one electron to a neutral system, i.e., the so-called charging energy, is given by $E_\mathrm{C}=e^2/2C$.
If one applies a finite voltage $V_\mathrm{G}$ by gating, the total energy of the island gains an additional term $-eNV_\mathrm{G}$, due to the Coulomb repulsion between the excess electronic charge and the gate charge $eN_\mathrm{G}=CV_\mathrm{G}$.
The sum of the electrostatic energy due to Coulomb repulsion between the excess charges $e^2N^2/2C=E_\mathrm{C}N^2$ and between the excess charges and gate charge $-eNV_\mathrm{G}=-2E_\mathrm{C}NN_\mathrm{G}$ amounts to the total energy of the island in the normal (i.e., nonsuperconducting) phase, given by
\begin{equation}
	E(N)=E_\mathrm{C}(N-N_\mathrm{G})^2,
\end{equation}
up to terms not depending on $N$.
In the superconducting phase, the total energy acquires an extra term that depends on the fermion parity of the condensate.
If the number of electrons is even, all electrons pair up into Cooper pairs with no extra energy cost, whereas if the number of electrons is odd, there is one extra unpaired electron that will occupy the lowest energy level available.
Hence, the total energy of the wire is given by~\cite{tuominen_experimental_1992,averin_single-electron_1992,eiles_even-odd_1993,lafarge_measurement_1993}
\begin{equation}\label{eq:CoulombParabolas}
	E(N)=E_\mathrm{C}(N-N_\mathrm{G})^2+n_\mathrm{F} \Delta E,
\end{equation}
with $n_\mathrm{F}=0,1$ for even and odd $N$,
and $\Delta E$ the energy of the lowest single-particle energy level available.
The energies $E(N)$ represent the energy of the many-particle states of $N$ electrons. 
If the spacing between these energy levels is nonzero $E(N+n)-E(N)>0$, there is a finite energy cost to add one or more electrons due to the Coulomb interaction.
In this case, there is no empty energy level available at the Fermi level, and the wire is insulating, i.e., the tunneling at zero bias is suppressed at small temperatures.
However, if two energy levels $E(N)$ and $E(N+n)$ become degenerate, the energy cost of adding one or more electrons is zero, and there is a finite transition probability between the two states.
Consequently, the wire has zero conductance with the exception of the degeneracy points, where the conductance exhibits the so-called Coulomb blockade peaks at zero bias.

\subsubsection{Period-halving of the Coulomb blockade peaks}

In a topologically trivial superconductor, the energy of the lowest single-particle energy level is equal to the particle-hole gap $\Delta E=\Delta E_\mathrm{ph}$.
If $\Delta E>E_\mathrm{C}$, the lowest energy configuration for a given gate voltage $N_\mathrm{G}$ is the many-particle state with an even number of electrons $N$, which minimizes $|N-N_\mathrm{G}|$ in \cref{eq:CoulombParabolas}.
In fact, due to the energy cost $\Delta E$, configurations with odd fermion parity are energetically unfavorable.
The energy spacing between two consecutive many-particle states with the same fermion parity is given by
\begin{equation}
E(N+2)-E(N)=4E_\mathrm{C}\left(N-N_\mathrm{G}+1\right),
\end{equation}
which follows directly from \cref{eq:CoulombParabolas}.
In this regime, the conductance at zero bias and small temperature $T\ll E_\mathrm{C}$ is zero, with the exceptions of the degeneracy points where $E(N+2)=E(N)$, i.e., when the gate charge is equal to an odd integer $N_\mathrm{G}=N+1$ in units of the electron charge
\begin{equation}\label{eq:CBtrivial}
	e N_\mathrm{G}=e (N	+ 1),
	\quad \text{for } N\in\mathbb{Z}\mathrm{\,\, even}.
\end{equation}
Hence, the Coulomb blockade peaks at zero bias are evenly spaced and periodic in the gate charge with periodicity $2e$.
\Cref{fig:Blockade}(b) shows the energies of the many-particle states with different electron numbers for $\Delta E>E_\mathrm{C}$ as a function of the gate charge, crossing at the degeneracy points.
In this regime, the conductance peaks periodically in the gate charge $eN_\mathrm{G}$ with period $2e$, as shown in \cref{fig:Blockade}(e) for $\Delta E/E_\mathrm{C}>1$. 

If $\Delta E<E_\mathrm{C}$ instead, configurations with an odd fermion parity may become energetically favorable for certain values of the gate charge.
Using \cref{eq:CoulombParabolas}, the energy spacing between two consecutive many-particle states is given by
\begin{align}
	E(N+1)-E(N)&=2E_\mathrm{C}\left(N-N_\mathrm{G}+\frac12\right)+
	(-1)^{N}
	\Delta E,
\end{align}
where the last term depends explicitly on the fermion parity 
of the island $P=(-1)^{N}$.
In this regime, the conductance at zero bias and small temperature $T\ll\Delta E$ is zero, with the exceptions of the degeneracy points where $E(N+1)=E(N)$.
Therefore, the conductance peaks at the degeneracy points are given by
\begin{equation}\label{eq:CoulombOscillations}
	e N_\mathrm{G}=e \left(N+\frac{1}2 +(-1)^{N} \frac{\Delta E}{2E_\mathrm{C}}\right),
	\quad \text{for } N\in\mathbb{Z}.
\end{equation}
In this case, the Coulomb blockade peaks are not evenly spaced but still follow a periodic pattern in the gate charge with periodicity $2e$.
\Cref{fig:Blockade}(c) shows the energies of the many-particle states for $\Delta E<E_\mathrm{C}$ as a function of the gate charge, crossing at the degeneracy points where energy levels with different fermion parity have the same energy.
The spacing between consecutive peaks with even and odd fermion parity is alternatively 
$e(1\pm \Delta E/E_\mathrm{C})$, 
as shown in \cref{fig:Blockade}(e) for $0<\Delta E/E_\mathrm{C}<1$. 

The Coulomb blockade regime allows one to distinguish the trivial and nontrivial phases of 
the Majorana island by measuring the conductance as a function of the gate charge~\cite{fu_electron_2010,hutzen_majorana_2012,chiu_conductance_2017,liu_proposal_2019} or putting the island into an Aharonov-Bohm interferometer~\cite{fu_electron_2010,sau_proposal_2015,vijay_teleportation-based_2016,hell_distinguishing_2018,liu_proposal_2019}.
In the trivial phase, the lowest single-particle energy level available coincides with the particle-hole gap $\Delta E=\Delta{E}_\mathrm{Z}$ in \cref{eq:EZ_gap}.
The gap closes at the critical value of the magnetic field $b=\sqrt{\mu^2+\Delta^2}$, driving the system into the nontrivial phase.
In this regime, the lowest single-particle energy level available is the Majorana bound state, $\Delta E=E_\mathrm{M}\propto\ee^{-L/\xi_\mathrm{M}}$, which is exponentially small in the wire length $L$.
If the Majorana localization length is much shorter than the wire length $\xi_\mathrm{M}\ll L$, one has that $\Delta E=E_\mathrm{M}\approx0$.
In this case, the conductance peaks at the degeneracy points given by
\begin{equation}\label{eq:CBnontrivial}
	e N_\mathrm{G}=e \left(N+\frac12\right),
	\quad \text{for } N\in\mathbb{Z},
\end{equation}
as follows from \cref{eq:CoulombOscillations} by taking $\Delta E=0$.
Hence, the Coulomb blockade peaks are now evenly spaced and periodic in the gate charge with periodicity $e$.
The Coulomb blockade peaks correspond to the tunneling in and out of the two spatially-separated Majorana modes at the opposite ends of the island, which effectively amounts to the tunneling through a single fermionic state, i.e., the Majorana bound states formed by the combination of the two Majorana modes.
This single-electron tunneling process thus corresponds to a peak with a maximum height quantized to $G=G_0/2=e^2/h$~\cite{fu_electron_2010,hutzen_majorana_2012} in the limit of large charging energies $E_\mathrm{C}$, which is half the conductance of the resonant zero-bias peak through a Majorana mode in the noninteracting case discussed in \cref{sec:DifferentialConductance}.
Remarkably, this tunneling of single electrons mediated by the Majorana zero modes through two opposite terminals is independent of the length of the Majorana island, i.e., independent of the distance between the spatially-separated Majorana modes.
Hence, this process is referred to as ``teleportation''.

\Cref{fig:Blockade}(d) shows the energies of the many-particle states for $\Delta E=0$ and the corresponding degeneracy points of energy levels with different fermion parity.
The spacing between consecutive peaks with even and odd fermion parity is $e$, i.e., the peaks are periodic in the gate charge $eN_\mathrm{G}$ with period $e$, as in \cref{fig:Blockade}(e) for $\Delta E/E_\mathrm{C}=0$. 
Therefore, at the topological phase transition, the Coulomb blockade peaks suddenly split, and their periodicity changes from $2e$ in the trivial phase [as in \cref{eq:CBtrivial}] to $e$ in the trivial phase [as in \cref{eq:CBnontrivial}].
The transition between the trivial phase $b<\sqrt{\mu^2+\Delta^2}$ with periodicity $2e$ to the nontrivial phase $b>\sqrt{\mu^2+\Delta^2}$ with periodicity $e$ is shown in \cref{fig:Blockade}(f).
The halving of the period of the peaks is thus a direct signature of the presence of a Majorana bound state at zero energy and a direct probe of its nonlocal nature.
Furthermore, in short wires, i.e., with a length comparable to the Majorana localization length $\xi_\mathrm{M}\sim L$, the energy $\Delta E=E_\mathrm{M}>0$ is not negligible. 
In this regime, the Coulomb peaks are not evenly spaced but follow a periodic pattern with periodicity $2e$ described by \cref{eq:CoulombOscillations}, where the fermion parity depends on the occupation number of the Majorana bound state $n_\mathrm{M}$, being $P=(-1)^N=(-1)^{n_\mathrm{M}}$.
Therefore, the spacing between consecutive peaks can be used to measure the Majorana energy splitting $E_\mathrm{M}$ and the oscillations of the Majorana energy~\cite{chiu_conductance_2017}, providing an estimate of the localization length $\xi_\mathrm{M}$ via \cref{eq:MajoranaOscillations}.

\subsection{Experimental advances}

The first experimental signatures consistent with Majorana bound states date back to 2012, with the zero-bias conductance peaks revealed via tunneling spectroscopy in a stacked
InSb/NbTiN nanowire heterostructure~\cite{mourik_signatures_2012}, similar to \cref{fig:nanowire}(a),
and the fractional Josephson effect in 
InSb/Nb nanowire-based junctions~\cite{rokhinson_the-fractional_2012}, similar to \cref{fig:nanowire}(b).
Similar devices were used in the following zero-bias peaks experiments on 
InAs/Al~\cite{das_zero-bias_2012}, 
InSb/Nb~\cite{deng_anomalous_2012},
InAs/NbN~\cite{finck_anomalous_2013}, 
InSb/Ti/NbTiN~\cite{churchill_superconductor-nanowire_2013} nanowires,
and Coulomb blockade spectroscopy in InSb/Nb~\cite{deng_parity_2014} devices.
Common issues of these early experiments were the zero-bias peak being much lower than the theoretically predicted quantized value and the presence of a so-called soft gap, i.e., a finite conductance below the superconducting gap even at zero magnetic fields, indicating the presence of spurious quasiparticle states near zero energy.

An engineering breakthrough came in 2015 with the development of a new fabrication technique where a superconducting thin film is epitaxially grown on the surface of hexagonal nanowires~\cite{krogstrup_epitaxy_2015}, as in \cref{fig:nanowire}(c).
This led to a second generation of hybrid
InAs/Al~\cite{krogstrup_epitaxy_2015,chang_hard_2015},
InSb/NbTiN~\cite{zhang_ballistic_2017,gul_hard_2017}, and
InSb/Al~\cite{shen_parity_2018} nanowire devices.
Due to an epitaxially clean and homogeneous superconducting-semiconducting interface, these partially-covered nanowires exhibit hard gaps~\cite{chang_hard_2015,zhang_ballistic_2017,gul_hard_2017}, i.e., a very low conductance below the superconducting gap with a strong suppression of quasiparticle states near zero energy, in contrast with earlier experiments.
Crucially, this led to more convincing experimental signatures, e.g., 
resolving separate trivial and nontrivial regions in the phase diagram in InSb/NbTiN~\cite{gul_ballistic_2018,chen_experimental_2017},
larger zero-bias peaks in InSb/Al~\cite{zhang_large_2021} devices, 
a direct observation of the nonlocality of Majorana bound states via tunneling spectroscopy with a quantum dot in InAs/Al nanowires~\cite{deng_majorana_2016,deng_nonlocality_2018},
the fractional Josephson effect via the Josephson radiation frequency in InAs/Al nanowire-based junctions~\cite{laroche_observation_2019}.
Other signatures were provided by Coulomb blockade spectroscopy
in isolated InAs/Al~\cite{higginbotham_parity_2015,albrecht_exponential_2016,albrecht_transport_2017,sherman_normal_2017,van-zanten_photon-assisted_2020} and InSb/Al~\cite{shen_parity_2018} Majorana islands.
A second breakthrough was the fabrication of multiterminal devices
in InAs/Al~\cite{grivnin_concomitant_2019,anselmetti_end-to-end_2019,menard_conductance-matrix_2020,puglia_closing_2021}
and InSb/Al~\cite{heedt_shadow-wall_2021} nanowires.
This setup allows to directly probe the nonlocal signatures of Majorana bound states by investigating the presence of zero-bias peaks at both ends of the wire~\cite{anselmetti_end-to-end_2019,menard_conductance-matrix_2020,puglia_closing_2021,heedt_shadow-wall_2021}, measuring the full conductance matrix~\cite{menard_conductance-matrix_2020,puglia_closing_2021} and simultaneously detecting the closing and reopening of the topological gap~\cite{grivnin_concomitant_2019,puglia_closing_2021}.

An alternative setup is realized by the so-called full-shell nanowires, i.e., semiconducting nanowires fully coated by an epitaxially grown superconducting shell~\cite{vaitiekenas_flux-induced_2020}, shown in \cref{fig:nanowire}(d).
These devices present several advantages: 
The full-shell covering protects the wire from surface impurities and, moreover, the topological phase may be reached at relatively low magnetic fields.
Signatures of Majorana bound states in InAs/Al full-shell nanowires at low magnetic fields have been reported via tunneling and Coulomb blockade spectroscopy~\cite{vaitiekenas_flux-induced_2020}.
However, these Majorana bound states may not be fully topologically protected against the mixing between subbands with different angular momenta, which arise when considering realistic experimental conditions, such as broken axial symmetry (e.g., due to the hexagonal section of the wire) and interface disorder~\cite{penaranda_even-odd_2020}.
Another possible issue is that the spin-orbit coupling induced in the wire may be rather small, requiring a fine-tuning of the nanowire radius to reach the topologically nontrivial phase~\cite{woods_electronic_2019}.
Moreover, these systems may also present zero-bias peaks unrelated to topological superconductivity~\cite{valentini_nontopological_2021,valentini_majorana-like_2022}.

Another promising setup is realized by hexagonal nanowires epitaxially covered by both superconducting and ferromagnetic layers, as in InAs/EuS/Al hybrid nanowire devices~\cite{liu_coherent_2020,liu_semiconductorferromagnetic_2020}, shown in \cref{fig:nanowire}(e).
In these doubly-covered nanowires, the time-reversal symmetry is broken by the exchange field induced by proximity with the ferromagnetic insulator~\cite{sau_generic_2010,alicea_majorana_2010}, allowing for the emergence of a stable zero-bias peak at zero applied magnetic field, as observed in a recent experiment~\cite{vaitiekenas_zero-bias_2021}.

Other 1D platforms that exhibit signatures compatible with the presence of Majorana bound states are metallic nanowire heterostructures~\cite{manna_signature_2020} and effectively 1D systems obtained in 2D semiconductor-superconductor planar heterostructures~\cite{nichele_scaling_2017,fornieri_evidence_2019,whiticar_coherent_2020,dartiailh_phase_2021} or in proximitized topological insulators (HgTe)~\cite{wiedenmann_4varphi-periodic_2016,bocquillon_gapless_2017,deacon_josephson_2017,ren_topological_2019}.

Very recently, evidence of a topologically nontrivial phase and Majorana bound states have been reported in an InAs/Al hybrid device, where a 1D quantum wire has been obtained by deposition of a narrow strip of Al over an InAs quantum well~\cite{aghaee_inas-al_2022} in a three-terminal setup~\cite{rosdahl_andreev_2018,danon_nonlocal_2020,pan_three-terminal_2021,pikulin_protocol_2021}.
As a first step, the experiment determined the regions of the parameter space where zero-bias peaks occur at both ends of the wire by measuring the local conductances at both ends.
As a second step, the full local and nonlocal conductance matrix was measured in these regions, including the nonlocal cross-conductance between the two ends, as a function of the applied magnetic field and gate voltage.
Since the nonlocal conductance can reveal the closing and reopening of the bulk gap, this two-step protocol allows finding the regions of the parameter space with zero-bias peaks at both ends of the wire, which are stable for a reasonably extended range of the applied magnetic field, and which appear after the closing and reopening of the bulk gap.
Ideally, these regions correspond to the nontrivial phases of the device, where the bulk is gapped with stable Majorana bound states. 
As claimed by the authors~\cite{pikulin_protocol_2021}, this procedure removes the subjectivity and the possible biases in the experimental search and reduces possible mistakes in the identification of Majorana bound states which, however, cannot be ruled out completely.

\section{Outlook\label{sec:outlook}}

To date, the many experiments on zero-bias conductance peaks~\cite{mourik_signatures_2012,das_zero-bias_2012,deng_anomalous_2012,finck_anomalous_2013,churchill_superconductor-nanowire_2013,lee_spin-resolved_2014,deng_majorana_2016,chen_experimental_2017,deng_nonlocality_2018,gul_ballistic_2018,bommer_spin-orbit_2019,grivnin_concomitant_2019,menard_conductance-matrix_2020,puglia_closing_2021,heedt_shadow-wall_2021,zhang_large_2021,aghaee_inas-al_2022}, fractional Josephson current~\cite{rokhinson_the-fractional_2012,laroche_observation_2019}, and Coulomb blockade spectroscopy~\cite{deng_parity_2014,higginbotham_parity_2015,deng_majorana_2016,albrecht_exponential_2016,albrecht_transport_2017,shen_parity_2018,van-zanten_photon-assisted_2020} are consistent with the existence of topologically-nontrivial Majorana bound states in proximitized semiconducting nanowires at nonzero magnetic fields, although some experiments could not be fully replicated~\cite{frolov_quantum_2021,castelvecchi_evidence_2021}.
Signatures of Majorana bound states have been reported in several other nanowire-based setups, e.g., semiconducting nanowires fully coated with a superconducting shell (full-shell nanowires)~\cite{vaitiekenas_flux-induced_2020} or partially covered by superconducting and ferromagnetic layers~\cite{vaitiekenas_zero-bias_2021}, and in metallic nanowires sandwiched between a superconductor and a ferromagnetic insulator~\cite{manna_signature_2020}, as well as in other setups such as atomic chains on superconductor substrates~\cite{nadj-perge_observation_2014,ruby_end-states_2015,pawlak_probing_2016,feldman_high-resolution_2017,jeon_distinguishing_2017,kim_toward_2018,schneider_topological_2021,schneider_precursors_2022}, superconducting planar heterostructures~\cite{nichele_scaling_2017,suominen_zero-energy_2017,fornieri_evidence_2019,whiticar_coherent_2020,dartiailh_phase_2021}, proximitized topological insulators (HgTe)~\cite{wiedenmann_4varphi-periodic_2016,bocquillon_gapless_2017,deacon_josephson_2017,ren_topological_2019}, in helical hinge states~\cite{jack_observation_2019}, and in the vortex core of topological insulator-superconductor heterostructures~\cite{xu_experimental_2015,menard_isolated_2019}.
However, despite the significant advances in the fabrication of the experimental heterostructures~\cite{krogstrup_epitaxy_2015,chang_hard_2015,zhang_ballistic_2017,gul_hard_2017}, the improvement of the quality of the interfaces, and the reduction of nonmagnetic and magnetic disorder, definitive proof of the existence of Majorana bound states in Majorana nanowires is still lacking.
Indeed, many of the reported signatures may have alternative explanations, including zero-bias Kondo resonances~\cite{lee_zero-bias_2012}, disorder~\cite{liu_zero-bias_2012,pikulin_a-zero-voltage_2012,bagrets_class_2012,rainis_towards_2013,roy_topologically_2013,stanescu_disentangling_2013,pan_physical_2020,pan_generic_2020,das-sarma_disorder-induced_2021,pan_disorder_2021,pan_crossover_2021,pan_three-terminal_2021,pan_quantized_2021,pan_on-demand_2022}, localized impurities~\cite{stanescu_nonlocality_2014,pan_crossover_2021}, strong interband coupling~\cite{bagrets_class_2012,woods_zero-energy_2019}, or finite-size effects~\cite{cayao_confinement-induced_2021}, which may produce zero-energy or near zero-energy modes even in the topologically trivial phase.
In particular, subgap Andreev modes~\cite{prada_from_2020} may appear as zero or near-zero energy excitations in the trivial phase induced by smooth potentials, accidental quantum dots, partial proximization of the nanowire, finite-size effects, or disorder~\cite{kells_near-zero-energy_2012,prada_transport_2012,chevallier_mutation_2012,liu_zero-bias_2012,pikulin_a-zero-voltage_2012,stanescu_disentangling_2013,roy_topologically_2013,stanescu_nonlocality_2014,cayao_sns-junctions_2015,liu_andreev_2017,setiawan_electron_2017,ptok_controlling_2017,penaranda_quantifying_2018,moore_two-terminal_2018,moore_quantized_2018,liu_distinguishing_2018,reeg_zero-energy_2018,huang_metamorphosis_2018,fleckenstein_decaying_2018,lai_presence_2019,avila_non-hermitian_2019,vuik_reproducing_2019,stanescu_robust_2019,woods_zero-energy_2019,lai_presence_2019,liu_conductance_2019,awoga_supercurrent_2019,yavilberg_differentiating_2019,chiu_fractional_2019,zeng_analytical_2019,sharma_hybridization_2020,pan_physical_2020,pan_generic_2020,cayao_confinement-induced_2021,pan_disorder_2021,pan_crossover_2021,pan_quantized_2021,pan_three-terminal_2021,pan_quantized_2021,hess_local_2021,liu_majorana_2021,das-sarma_disorder-induced_2021,marra_majorana/andreev_2022,zeng_partially_2022}, not necessarily localized at opposite ends and with experimental signatures which may mimic those of Majorana bound states~\cite{lee_spin-resolved_2014,suominen_zero-energy_2017,de-moor_electric_2018,chen_ubiquitous_2019,anselmetti_end-to-end_2019,junger_magnetic-field-independent_2020,yu_non-majorana_2021,valentini_nontopological_2021,shen_full_2021,bjergfelt_superconductivity_2021,kuster_non-majorana_2022,valentini_majorana-like_2022,estrada-saldana_excitations_2022,poschl_nonlocal_2022}.
These modes, known as quasi-Majorana bound states, trivial Majorana bound states, partially-separated or partially-overlapping Majorana bound states, materialize as a superposition of a pair of Majorana modes, partially-separated in space but with a sizable overlap, typically appearing without a sharp quantum phase transition~\cite{prada_transport_2012,stanescu_to-close_2012,stanescu_disentangling_2013,mishmash_approaching_2016,huang_metamorphosis_2018,marra_majorana/andreev_2022}.
However, whereas Majorana bound states are exponentially localized at the nanowire ends, quasi-Majorana bound states are typically localized near inhomogeneities or impurities and do not necessarily exhibit exponential localization~\cite{penaranda_quantifying_2018,moore_two-terminal_2018,avila_non-hermitian_2019,stanescu_robust_2019,pan_quantized_2021}.
Note that the Majorana modes forming a quasi-Majorana bound state may still have a large degree of separation with a small overlap and almost zero energy~\cite{penaranda_quantifying_2018,vuik_reproducing_2019} and, in this case, exhibit a full Majorana character, being self-adjoint and particle-hole symmetric with nonabelian braiding statistics~\cite{vuik_reproducing_2019,zeng_feasibility_2020}. 
Indeed, there exists a smooth crossover between quasi-Majorana bound states and Majorana bound states in nonuniform nanowires~\cite{moore_two-terminal_2018,stanescu_robust_2019,avila_non-hermitian_2019,vuik_reproducing_2019,pan_crossover_2021,marra_majorana/andreev_2022,zeng_partially_2022}.
In this case, subgap states typically detach from the bulk excitation spectrum and gradually approach zero energy without a sharp topological phase transition~\cite{prada_transport_2012,stanescu_to-close_2012,stanescu_disentangling_2013,mishmash_approaching_2016,huang_metamorphosis_2018,marra_majorana/andreev_2022}.
Another source of zero-energy states in topologically trivial nanowires is the coupling to external reservoirs (e.g., metallic leads, gates), which may induce the presence of exceptional points in the nonhermitian complex energy spectrum~\cite{san-jose_majorana_2016,avila_non-hermitian_2019} (see Ref.~\onlinecite{prada_from_2020}).

In light of the above, stronger evidence of topologically-nontrivial Majorana modes has been provided by the observation of their nonlocal signatures in the differential conductance in nanowires coupled with a quantum dot~\cite{leijnse_quantum_2011,leijnse_scheme_2011,liu_detecting_2011,vernek_subtle_2014,prada_measuring_2017,clarke_experimentally_2017,schuray_fano_2017,chevallier_topological_2018,schuray_signatures_2020,mishmash_dephasing_2020,ricco_topological_2021}, e.g., via tunneling spectroscopy~\cite{deng_majorana_2016,deng_nonlocality_2018}, or in multiterminal setups~\cite{michaeli_electron_2017,rosdahl_andreev_2018,moore_two-terminal_2018,danon_nonlocal_2020,pan_three-terminal_2021,pikulin_protocol_2021,kejriwal_nonlocal_2022} as in recent experiments~\cite{grivnin_concomitant_2019,anselmetti_end-to-end_2019,menard_conductance-matrix_2020,puglia_closing_2021,heedt_shadow-wall_2021,aghaee_inas-al_2022}, where the closing of the bulk gap and the presence of zero energy states at the two ends of the nanowire may be probed independently.
As mentioned, evidence of Majorana bound states has been reported in an InAs/Al hybrid device in a three-terminal setup~\cite{aghaee_inas-al_2022}, by mapping the regions of the parameter space where zero-bias peaks appear at both ends of the system with the concurrent opening and closing of the bulk gap.
Other conductance signatures of Majorana bound states may be obtained by suppressing the contribution of trivial Andreev bound states to the differential conductance via dissipative leads~\cite{liu_proposed_2013,liu_universal_2022}, as in a recent experiment~\cite{zhang_suppressing_2022}, measuring the half-integer plateaus in the quantum point contact conductance in the ballistic regime~\cite{wimmer_quantum_2011}, or probing the tunneling current through superconducting leads~\cite{peng_robust_2015} or spin-polarized leads~\cite{sticlet_spin_2012,he_selective_2014,haim_signatures_2015,szumniak_spin_2017}.
Direct evidence may also be obtained by probing the spin degree of freedom of Majorana modes~\cite{leijnse_quantum_2011,sticlet_spin_2012,he_selective_2014,haim_signatures_2015,prada_measuring_2017,szumniak_spin_2017,chevallier_topological_2018,schuray_signatures_2020}, their nonlocality~\cite{nilsson_splitting_2008,prada_measuring_2017,schuray_fano_2017,clarke_experimentally_2017,romito_ubiquitous_2017,cayao_distinguishing_2021,ricco_topological_2021,kejriwal_nonlocal_2022}, and by quantifying their lifetimes, coherence times~\cite{leijnse_scheme_2011,mishmash_dephasing_2020}, and degree of mutual overlap~\cite{das-sarma_splitting_2012,clarke_experimentally_2017,penaranda_quantifying_2018}.
Another approach is to directly probe the signatures of the topological phase transition~\cite{akhmerov_quantized_2011,stanescu_to-close_2012,mishmash_approaching_2016,chevallier_topological_2018}, e.g., in the Josephson critical current~\cite{pientka_signatures_2013,san-jose_multiple_2013,san-jose_mapping_2014,pientka_topological_2017} or spin susceptibility~\cite{pakizer_signatures_2021}, as experimentally measured in Majorana nanowire junctions~\cite{tiira_magnetically-driven_2017} and planar Josephson junctions~\cite{dartiailh_phase_2021}, or the fermion parity anomaly in the Coulomb blockade regime~\cite{liu_proposal_2019}.
Further evidence may be provided by the interplay between Kondo and Majorana physics~\cite{lee_kondo_2013,cheng_interplay_2014} or signatures of the topological Kondo effect stemming from the coupling between conduction electrons and the nonlocal Majorana bound state~\cite{beri_topological_2012,altland_multiterminal_2013,altland_multichannel_2014,galpin_conductance_2014,liu_topological_2021}.
Ideally, only the experimental observation on the same sample of multiple, independent, and mutually consistent signatures on a reasonably large parameter range may prove beyond reasonable doubt the existence of Majorana bound states in Majorana nanowires~\cite{frolov_topological_2020}, e.g., by performing both tunneling and Coulomb spectroscopy on the same device~\cite{vaitiekenas_flux-induced_2020,valentini_majorana-like_2022}. 
For an overview of the recent experimental advances in the field, see Refs.~\onlinecite{lutchyn_majorana_2018,zhang_next_2019,frolov_topological_2020,flensberg_engineered_2021,jack_detecting_2021,fu_experimental_2021,cao_recent_2022}.

The next milestones in the route to topological quantum computation with Majorana modes are the experimental demonstration of their nonabelian braiding statistics and the realization of scalable architectures employing Majorana-based topological qubits~\cite{aasen_milestones_2016,karzig_scalable_2017,plugge_majorana_2017} in networks of Majorana nanowires~\cite{sau_universal_2010,alicea_non-abelian_2011,sau_controlling_2011,clarke_majorana_2011,halperin_adiabatic_2012,van-heck_coulomb-assisted_2012,hyart_flux-controlled_2013,vijay_teleportation-based_2016,aasen_milestones_2016,karzig_scalable_2017,plugge_majorana_2017}.
A first indirect demonstration of the nonabelian statistics may be achieved by verifying the fusion rules by measuring the fermion parity of the qubit during a fusion experiment~\cite{aasen_milestones_2016,plugge_majorana_2017,steiner_readout_2020,zhou_fusion_2022,seoane-souto_timescales_2020,krojer_demonstrating_2022,souto_fusion_2022}.
A more direct demonstration of nonabelian braiding instead requires the adiabatical exchange of the Majorana modes in physical space~\cite{alicea_non-abelian_2011,clarke_majorana_2011,romito_manipulating_2012,halperin_adiabatic_2012} or in parameter space~\cite{flensberg_non-abelian_2011,sau_controlling_2011,hassler_the-top-transmon:_2011,van-heck_coulomb-assisted_2012,hyart_flux-controlled_2013,karzig_shortcuts_2015,aasen_milestones_2016}, or by 
measurement-only approaches via projective measurements of the fermion parity~\cite{bonderson_measurement-only_2008,bonderson_measurement-only_2009,bonderson_measurement-only_2013,vijay_teleportation-based_2016,plugge_majorana_2017,karzig_scalable_2017}. 
Nevertheless, braiding alone cannot realize universal quantum computation in a fully topological-protected way:
Majorana qubits need to be complemented by auxiliary topologically-unprotected quantum operations~\cite{bravyi_universal_2005,bravyi_universal_2006,sau_universal_2010,bonderson_implementing_2010,hassler_anyonic_2010,bonderson_topological_2011,leijnse_quantum_2011,jiang_interface_2011,aasen_milestones_2016,karzig_universal_2016,hoffman_universal_2016,karzig_scalable_2017,karzig_robust_2019}.
Furthermore, even the topological-protected braidings of Majorana modes are still prone to decoherence due to quasiparticle poisoning~\cite{leijnse_scheme_2011,goldstein_decay_2011,schmidt_decoherence_2012,budich_failure_2012,rainis_majorana_2012,aseev_lifetime_2018,knapp_dephasing_2018,lai_exact_2018,aseev_degeneracy_2019} and nonadiabatic processes~\cite{cheng_nonadiabatic_2011,karzig_boosting_2013,scheurer_nonadiabatic_2013,karzig_optimal_2015,pedrocchi_majorana_2015,amorim_majorana_2015,knapp_the-nature_2016,sekania_braiding_2017,bauer_dynamics_2018}.
Hence, to implement fault-tolerant quantum computation, Majorana-based qubits still need to be supplemented by quantum error correction codes~\cite{vijay_majorana_2015,landau_towards_2016,plugge_roadmap_2016,litinski_combining_2017,manousakis_majorana_2017,knapp_modeling_2018,litinski_quantum_2018}.

It is worth mentioning that the relevance of Majorana modes and topological superconductors goes well beyond their immediate applications in the field of topological quantum computing described here.
For example, the realization of hybrid Majorana-transmon qubits~\cite{hassler_the-top-transmon:_2011,hyart_flux-controlled_2013,ginossar_microwave_2014,schrade_majorana_2018,avila_majorana_2020,avila_superconducting_2020,sabonis_destructive_2020} may have an impact in the field of superconducting quantum computing, where qubits are implemented through superconducting circuits (see Ref.~\onlinecite{aguado_a-perspective_2020}).
Moreover, the braiding of Majorana modes and the corresponding topological quantum gates can be mapped into the time evolutions of a quantum simulator realized, e.g., by trapped ions, ultracold atoms, superconducting circuits, or photonic systems, as in recent experiments~\cite{xu_simulating_2016,xu_photonic_2018}.
Finally, the experimental and theoretical effort to realize Majorana modes may lead to the development of more advanced platforms for topological quantum computation:
Parafermions, a generalization of Majorana fermions, may provide a larger set of fully topologically-protected quantum gates, as opposed to topological Majorana qubits~\cite{lindner_fractionalizing_2012,cheng_superconducting_2012,burrello_topological_2013,motruk_topological_2013,clarke_exotic_2013,klinovaja_parafermions_2014,klinovaja_kramers_2014,klinovaja_time-reversal_2014,hutter_quantum_2016} (see Refs.~\onlinecite{alicea_designer_2015,alicea_topological_2016,schmidt_bosonization_2020} for a review).
On top of this, Majorana modes may be employed as building blocks of strongly-interacting models such as Kitaev spin models~\cite{kells_kitaev_2014,sagi_spin_2019}, the topologically-ordered Majorana toric code~\cite{xu_fractionalization_2010,terhal_from_2012,roy_quantum_2017,roy_charge_2018,ziesen_topological_2019}, and other stabilizer and error-correcting codes~\cite{vijay_majorana_2015,vijay_physical_2016,landau_towards_2016,plugge_roadmap_2016,manousakis_majorana_2017,litinski_combining_2017,litinski_quantum_2018}, which could lead to fault-tolerant topological quantum computation~\cite{kitaev_quantum_1997,kitaev_fault-tolerant_2003,kitaev_anyons_2006}.

Beyond quantum computation, Majorana modes are also relevant for studying supersymmetric theories in condensed matter, in particular emergent space-time supersymmetry~\cite{rahmani_emergent_2015,rahmani_phase_2015,ebisu_supersymmetry_2019,rahmani_interacting_2019,sannomiya_supersymmetry_2019} and emergent quantum-mechanical supersymmetry~\cite{hsieh_all-majorana_2016,huang_supersymmetry_2017,marra_1d-majorana_2022,marra_dispersive_2022}.
Finally, the Sachdev-Ye-Kitaev model~\cite{sachdev_gapless_1993,pikulin_black_2017,chew_approximating_2017,liu_quantum_2018}, realized by a collection of Majorana modes with
infinite-range (i.e., one-to-all) many-body interactions with random couplings, has been studied in connection with strongly-correlated non-Fermi liquids, quantum many-body chaos, quantum gravity, and black holes physics (see Ref.~\onlinecite{chowdhury_sachdev-ye-kitaev_2021} for a review).

The route from theoretical models to experiments and from experiments to technological applications will hopefully lead to the design of viable technological platforms for topological quantum computation, which will constitute a breakthrough in quantum information science and simultaneously contribute to fundamental advances in understanding quantum physics.

\begin{acknowledgments}
I thank Nicolas Laflorencie, Ananda Roy, and Yukio Tanaka for valuable feedback. 
This work is supported by the Japan Science and Technology Agency (JST) of the Ministry of Education, Culture, Sports, Science and Technology (MEXT), JST CREST Grant.~No.~JPMJCR19T2, and the Japan Society for the Promotion of Science (JSPS) Grant-in-Aid for Early-Career Scientists (Grant No.~20K14375).
\end{acknowledgments}

\appendix

\section{Discretization via finite-differences (lattice model)\label{sec:discretization}}

A continuum Hamiltonian can be discretized by approximating the fields and the derivatives with finite differences on a discrete lattice $x_n=na$, using
\begin{subequations}\label{eq:discretization}
\begin{align}
\Psi(x)&\to c_n,\\
p_x\Psi(x)&\to-\frac{\ii\hbar }{2a} \left(c_{n+1}-c_{n-1}\right),\\
p_x^2\Psi(x)&\to-\frac{\hbar^2}{a^2} \left(c_{n+1}-2c_{n}+c_{n-1}\right),
\end{align}
\end{subequations}
where $c_n=\Psi(na)$, and approximating the integral with discrete sums.
Conversely, one can derive the continuum Hamiltonian in momentum space from the discrete (tight-binding) Hamiltonian by expanding the momentum dispersion at small momenta
\begin{equation}
\sin{(ka)}\approx ka,\qquad
\cos{(ka)}\approx 1-\frac{(ka)^2}2.
\end{equation}

\section{The pfaffian of antisymmetric matrices\label{sec:pfaffian}}

The pfaffian of a $2n\times2n$ antisymmetric matrix $A=-A^\intercal$ with matrix entries $a_{ij}$ is defined as~\cite{wimmer_algorithm_2012}
\begin{equation}
\pf(A)=\frac{1}{2^n n!}\sum_{\sigma\in S_{2n}}\sgn(\sigma)\prod_{k=1}^n 
a_{\sigma(2k-1),\sigma(2k)},
\end{equation}
where the sum is over the symmetric group $S_{2n}$ of permutations of the set $\{1,2,\ldots,2n\}$, and $\sgn(\sigma)$ is the sign of the permutation $\sigma$.
An equivalent definition is
\begin{equation}\label{eq:pfaffian}
\pf(A)=\sum_\sigma\sgn(\sigma)\prod_{k=1}^n 
a_{\sigma(2k-1),\sigma(2k)},
\end{equation}
where the sum is over all permutations in the form
\begin{equation}
\sigma=\begin{bmatrix}1 & 2 & 3&\dots & 2n\\i_1&j_1&i_2&\dots& j_n\end{bmatrix},
\end{equation}
such that $i_k<j_k$ and $i_k<i_{k+1}$ for all indices $k$.
The pfaffian of a $2\times2$ antisymmetric matrix is simply
\begin{equation}
\pf\begin{bmatrix} 0 & a_{12} \\ -a_{12} & 0 \end{bmatrix}=a_{12},
\end{equation}
while the pfaffian of a $4\times4$ antisymmetric matrix is 
\begin{equation}
\pf\begin{bmatrix} 
0		& a_{12}	& a_{13}	& a_{14} \\ 
-a_{12}	& 0			& a_{23}	& a_{24} \\ 
-a_{13}	& -a_{23}	& 0			& a_{34} \\
-a_{14}	& -a_{24}	& -a_{34}	& 0
\end{bmatrix}=a_{12}a_{34} - a_{13}a_{24} + a_{14}a_{23}.
\end{equation}

The pfaffian of an antisymmetric matrix is a square root of its determinant, i.e., its determinant is the square of its pfaffian $\det(A)=\pf(A)^2$.
If the matrix $A$ is real, then the pfaffian is a real number $\pf(A)=\pm\sqrt{\det(A)}$ and have a well-defined sign $\sgn\pf(A)=\pf(A)/\sqrt{\det(A)}$ if $\pf(A)\neq0$.

\section{Zero-energy end modes of the tight-binding (lattice) Hamiltonian\label{sec:endmodes}}

We want to find the zero-energy modes of \cref{eq:H-pwave-majo} satisfying open boundary conditions.
Let us start considering a generic fermionic state
\begin{equation}
d^\dag_\mathrm{M}=
\sum_{n=1}^N \left( \psi_{A,n} \gamma_{A,n} + \psi_{B,n} \gamma_{B,n} \right).
\end{equation}
To be an eigenstate of the Hamiltonian with zero energy, this fermionic state must satisfy $\mathcal{H}d^\dag_\mathrm{M}\ket{0}=0$, which mandates $[\mathcal{H},d^\dag_\mathrm{M}]=0$ or equivalently 
\begin{equation}\label{eq:Motion1}
\sum_{n=1}^N
	\psi_{A,n} [\mathcal{H}, \gamma_{A,n}]+\psi_{B,n} [\mathcal{H},\gamma_{B,n}]=0.
\end{equation}
A direct calculation of the commutators above yields the equation of motion for the wavefunction~\cite{semenoff_stretching_2006}
\begin{subequations}\label{eq:Motion2}\begin{align}
	\mu\psi_{A,n} + (t+\Delta)\psi_{A,n+1} + (t-\Delta)\psi_{A,n-1} = 0,
	\\
	\mu\psi_{B,n} + (t-\Delta)\psi_{B,n+1} + (t+\Delta)\psi_{B,n-1} = 0,
\end{align}\end{subequations}
for $n=2,\ldots,N-1$, which can be arranged in matrix form as
\begin{equation}\label{eq:MotionEquation}\thinmuskip=1mu\medmuskip=1.5mu\thickmuskip=2mu\renewcommand{\arraystretch}{0.8}\begin{bmatrix}
\mu &0\\
0 & \mu
\end{bmatrix}
\bm{\psi}_n
+
\begin{bmatrix}
t-\Delta & 0\\
0 & t+\Delta
\end{bmatrix}
\bm{\psi}_{n
-
1}
+
\begin{bmatrix}
t+\Delta & 0\\
0 & t-\Delta
\end{bmatrix}
\bm{\psi}_{n
+
1}
=0,
\end{equation}
where the Majorana spinor is defined as $\bm{\psi}_n=[\psi_{A,n},\psi_{B,n}]^\intercal$.
To extend the above equations to all lattice sites $n=1,\ldots,N$, one can include two additional lattice sites at the beginning and the end of the chain, respectively $n=0$ and $n=N+1$, and impose that the Majorana wavefunctions have zero amplitude on these additional sites.
Thus, we search for the zero-energy eigenstates 
and require that the wavefunctions vanish at the ends of the chain.
Since we are interested in localized modes, we make the following ansatz 
\begin{equation}
	\bm{\psi}_n= z^n \begin{bmatrix} 
	\alpha \\ \beta 
\\ 
	\end{bmatrix},
\end{equation}
and impose the boundary conditions
\begin{equation}\label{eq:BoundaryCondition}
	\bm{\psi}_{0}=\bm{\psi}_{N+1}=0.
\end{equation}
Therefore, the equation of motion in \cref{eq:MotionEquation} becomes
\begin{equation}\label{eq:MotionEquationMatrix}\thinmuskip=1mu\medmuskip=1.5mu\thickmuskip=2mu\begin{bmatrix}\renewcommand{\arraystretch}{0.8}
\mu + (t+\Delta)z+(t-\Delta)/z & 
\hspace{-10pt}
0\\
0 & 
\hspace{-10pt}
\mu+(t-\Delta)z+(t+\Delta)/z
\end{bmatrix}\cdot
\begin{bmatrix}
\alpha\\\beta
\end{bmatrix}
=0.
\end{equation}
This equation has nonzero solutions 
($\alpha\neq0$ and $\beta\neq0$) 
if and only if the determinant of the matrix is zero, i.e., if one of the diagonal elements is zero.
The 
first and second
diagonal elements are zero, respectively, if $(t+\Delta)z^2+\mu z+(t-\Delta)=0$ and $(t-\Delta)z^2+\mu z+(t+\Delta)=0$.
The solutions of these quadratic equations are respectively $z_\pm$ and $z'_\pm=1/z_\pm$, given by
\begin{subequations}\label{eq:zpm}\begin{align}
z_\pm=
\frac{-\mu\pm\sqrt{\mu^2-4(t^2-\Delta^2)}}{2(t+\Delta)},
\\
1/z_\mp=
\frac{-\mu\pm\sqrt{\mu^2-4(t^2-\Delta^2)}}{2(t-\Delta)}.
\end{align}\end{subequations}
Using Vieta's formula for quadratic equations, one also has
\begin{equation}\label{eq:Vieta}
z_+ z_-=\frac{t-\Delta}{t+\Delta},
\end{equation}
which mandates $|z_+ z_-|<1$.
If $z=z_\pm$, the first diagonal element is 
zero 
while the second is 
nonzero,
and therefore the solutions of the eigenvalue equation in \cref{eq:MotionEquationMatrix} are the vectors 
$[\alpha,\beta]\propto[1,0]$.
If $z=1/z_\pm$, conversely, the second diagonal element is 
zero 
 and the first is 
nonzero,
and therefore the solutions of the eigenvalue equation are 
$[\alpha,\beta]\propto[0,1]$.
Therefore there are two orthogonal zero-energy solutions given by
\begin{subequations}\label{eq:solutions}\begin{align}
\bm{\psi}^A_{n}=&
\left(
A_+\, z_+^n
\ \, + \ \,
A_-\, z_-^n
\,
\right)
\:
\begin{bmatrix}
1\\0
\end{bmatrix},
\\
\bm{\psi}^B_{n}=&
\left(
B_+\, z_+^{-n}
+
B_-\, z_-^{-n}
\right)
\begin{bmatrix}
0\\1
\end{bmatrix},
\end{align}\end{subequations}
where the coefficients $B_\pm$ and $A_\pm$ need to be determined by the boundary conditions in \cref{eq:BoundaryCondition}.
If $|z_+|,|z_-|<1$, then the spinor wavefunction $\bm{\psi}^B_{n}$ increases, and $\bm{\psi}^A_{n}$ decreases along the chain.
In this case, the boundary conditions $\bm{\psi}^B_0=0$ and $\bm{\psi}^A_{N+1}=0$ can be satisfied in the limit $N\to\infty$, whereas the boundary conditions $\bm{\psi}^B_{N+1}=0$ and $\bm{\psi}^A_{0}=0$ give $B_+z_+^{-N-1}=-B_-z_-^{-N-1}$ and $A_+=-A_-$ respectively.
Taking $A_+=-A_-=B_+z_+^{-N-1}=-B_-z_-^{-N-1}$, one gets
\begin{subequations}\label{eq:ZEmodes}\begin{align}
\bm{\psi}^A_n=&
\frac1Z\left(
z_+^n
-
z_-^n
\right)
\begin{bmatrix}
1\\0
\end{bmatrix},
\\
\bm{\psi}^B_n=&
\frac1Z
\left(
z_+^{N+1-n}
-
z_-^{N+1-n}
\right)
\begin{bmatrix}
0\\1
\end{bmatrix},
\end{align}\end{subequations}
where $1/Z$ is a normalization factor.
In the particle-hole basis, these wavefunctions yield \cref{eq:MESgeneralNambu}.
These zero-energy modes correspond to two Majorana end modes localized at the opposite ends of the chain in \cref{eq:MESgeneral}.
In the Majorana chain regime $\Delta=t$, one has that $z_+=0$ or $z_-=0$ and formally can substitute $z_\pm^0\to1$ in the above equations for $n=0$, formally recovering the open boundary conditions in \cref{eq:BoundaryCondition}.

In order to obtain the zero-energy modes in \cref{eq:ZEmodes} satisfying the boundary conditions in \cref{eq:BoundaryCondition}, one needs to assume that $|z_+|,|z_-|<1$.
In this case, indeed, 
the boundary conditions in \cref{eq:BoundaryCondition} are satisfied for $N\to\infty$, and approximately up to exponentially small terms for $N<\infty$.
I will now show that this assumption is equivalent to assuming that $|\mu|<2t$.
This will guarantee the existence of the Majorana end modes.
First, we start to notice that \cref{eq:Vieta} mandates that at least one of the two solutions $z_\pm$ is smaller than one, i.e., one has $z_+<1$ or $z_-<1$, or both $|z_+|,|z_-|<1$.
If only one solution is larger than one, i.e.,
$|z_+|>1$ or $|z_+|>1$, then $\bm{\psi}^B_n$ and $\bm{\psi}^A_n$ are both a sum of an exponentially increasing and exponentially decreasing component. 
For this reason, they cannot satisfy the boundary conditions, not even approximately.
Let us consider separately the cases where $z_\pm$ in \cref{eq:zpm} are real and distinct [i.e., $\mu^2>4(t^2-\Delta^2)$], real and degenerate [i.e., $\mu^2=4(t^2-\Delta^2)$], or complex conjugates [i.e., $\mu^2<4(t^2-\Delta^2)$].
If $z_\pm$ are complex conjugate numbers, i.e., for $\mu^2<4(t^2-\Delta^2)<(2t)^2$, \cref{eq:Vieta} mandates $|z_+z_-|=|z_+|^2=|z_-|^2<1$.
If $z_\pm$ are real numbers instead, i.e., for $\mu^2>4(t^2-\Delta^2)$, \cref{eq:Vieta} mandates $|z_+z_-|=|z_+||z_-|<1$.
In order to have both $|z_+|<1$ and $|z_-|<1$, one may impose $(|z_+|^2-1)(|z_-|^2-1)={(4t^2-\mu^2)}/{(t+\Delta)^2}>0$.
Thus, one has $|z_+|,|z_-|<1$ if $|\mu|<2t$.
In the limiting case $\mu^2=4(t^2-\Delta^2)$, the two real roots are degenerate $z_+=z_-$, and the zero-energy modes $\bm{\psi}^A_n$ and $\bm{\psi}^B_n$ can be obtained by taking the limit $\mu\to\pm\sqrt{4(t^2-\Delta^2)}$.
Hence, whether $z_\pm$ are real or complex numbers, one has $|z_+|,|z_-|<1$ as long as $|\mu|<2t$. 
In this case, the Hamiltonian allows the existence of Majorana modes localized at the ends of the chain.
On the other hand, for $|\mu|>2t$, one has $|z_+|>1$ or $|z_-|>1$.
In this case, there are no zero-energy end modes. 
Finally, at the closing of the particle-hole gap $|\mu|=2t$, one has either $|z_+|=1$ or $|z_-|=1$.
In this case, the zero-energy modes $\bm{\psi}^A_n$ and $\bm{\psi}^B_n$ become extended along the whole chain.

The normalization factor is obtained by imposing the normalization condition one has $Z^2=\sum_{n=0}^N|z_+|^{2n}+|z_-|^{2n}-2\Re((z_+z_-^*)^n)$.
Using the well-known formulas for the geometric sums, in the limit of large system sizes $N\rightarrow\infty$, one gets
\begin{equation}\thinmuskip=1mu\medmuskip=1.5mu\thickmuskip=2mu
Z=\sqrt{
\dfrac{1}{1-|z_+|^2}+\dfrac{1}{1-|z_-|^2}-2\Re\left(\dfrac{1}{1-z_+ z_-^*}\right)
},
\end{equation}
which yields \cref{eq:MajoranaNormalizationInfty} using \cref{eq:zpm,eq:Vieta}.

\section{Change of Nambu basis\label{sec:nambu}}

The Hamiltonian density $H(k)$ in \cref{eq:H-swave-kspaceC-BdG} can be rewritten as
\begin{align}
	H(k)&=\left(\varepsilon_k+b_x\sigma_x+b_z\sigma_z\right)\tau_z+b_y \sigma_y+
	\nonumber\\\label{eq:H-swave-kspaceC-BdG-oldnambu}
	&+\alpha k\sigma_y\tau_z+\Delta\sigma_y\tau_y,
\end{align} 
where $\sigma_{xyz}$ and $\tau_{xyz}$ are the Pauli matrices respectively in spin and particle-hole space.
In this form, the $\mathrm{SU}(2)$ symmetry of the magnetic field term is somewhat hidden.
For this reason, it is useful to perform a change of basis into the Nambu spinor
$\bm\Psi^\dag(k)=[\Psi^\dag_{\up}(k),\Psi^\dag_{\down}(k),\Psi_{\down}(-k),-\Psi_{\up}(-k)]$
via the unitary transformation
$U=\diag(\id,\ii\sigma_y)$
which yields
\begin{align}
\begin{bmatrix}
\id & 0\\
0 & \ii\sigma_y\\
\end{bmatrix}
\cdot
&
\begin{bmatrix}
h(k) & -\Delta\ii\sigma_y \\
\Delta\ii\sigma_y & -h(-k)^* \\
\end{bmatrix}
\cdot
\begin{bmatrix}
\id & 0\\
0 & \ii\sigma_y\\
\end{bmatrix}^\dag
=\nonumber\\=
&
\begin{bmatrix}
h(k) & -\Delta\\
-\Delta & -\sigma_y h(-k)^*\sigma_y \\
\end{bmatrix},
\end{align}
which gives \cref{eq:H-swave-kspaceC-BdG-newnambu}.
The symmetry operators in the new Nambu basis in \cref{eq:CTSnewnambu} are obtained from \cref{eq:CTS} via the unitary transformation defined above.
Notice that the time-reversal symmetry operator $\mathcal{T}=\ii\sigma_y\mathcal{K}$ is not affected by the change of basis.

\end{document}